\documentstyle[11pt,aaspp4]{article} 
\newcommand{\kmsM} {km~s$^{-1}$~Mpc$^{-1}$}
\newcommand{\subsun}{\mbox{$_{\odot}$}}
\newcommand{\etal}{{\it et al.\/}}

\lefthead{Cohen {\it et al.} }
\righthead{HDF Redshift Survey}

\begin{document}

\title{Caltech Faint Galaxy Redshift Survey XVI:\\
The Luminosity Function for Galaxies in the Region of the HDF-North \\
to $z$ = 1.5\altaffilmark{1}}

\author{Judith G. Cohen\altaffilmark{2} }

\altaffiltext{1}{Based in large part on observations obtained at the
	W.M. Keck Observatory, which is operated jointly by the California 
	Institute of Technology, the University of California
        and NASA,}
\altaffiltext{2}{Palomar Observatory, Mail Stop 105-24,
	California Institute of Technology, Pasadena, CA \, 91125}

\begin{abstract}

We have carried out a study of the luminosity function (henceforth LF) of galaxies
in the region  of the HDF-North using our very complete redshift catalog.
We divide the sample into five redshift bins covering the range
$0.01<z<1.5$ and consider three primary galaxy spectral classes.
We solve for the LF at four rest frame wavelengths from 0.24 to 2.2$\mu$.
We find that the LFs for quiescent galaxies have shallow faint end slopes, 
while those of 
galaxies with detectable emission lines have steeper
faint end slopes.  Furthermore these slopes appear to be independent 
of redshift out to $z=1.05$ for each galaxy spectral grouping
and agree well with comparable local determinations.
We then fix $\alpha$ to obtain values of $L^*$ for
each galaxy spectral grouping as a function of redshift. 
We find that galaxies with strong absorption lines
become brighter with $z$ with $Q \sim 0.6$ at all
rest frame bands studied here, 
where $Q = \Delta{\rm{log}}[L^*(z)]/\Delta{z}$, while galaxies
with detectable emission lines (i.e. star forming galaxies) 
show a smaller change in $L^*$ with
redshift at all bands, $Q \sim 0.3$, with $Q$ becoming significantly larger
at rest frame 2400\AA.

Passive evolution models of galaxies are in reasonable agreement
with these results for absorption line dominated galaxies, while
plausible star formation histories can reproduce the behavior
of the emission line galaxies.
% The major discrepancy with the specific
% set of galaxy spectral synthesis models
% we adopt, those of Poggianti (1997), is the prediction
% of much more luminosity fading at rest frame $U$ for
% galaxies with a brief single initial burst of star formation than 
% is actually  from our analysis of quiescent galaxies.
%  New Poggianti models fit much better, private communication
%

We find a constant co-moving number density and
stellar mass in galaxies 
out to $z\sim1.05$.  By stretching all the correction factors applied to the
galaxy counts in the  highest
redshift bin to their
maximum possible values, we can just barely achieve this between $z=1.05$
and $z \sim 1.3$.   The
major epoch(s) of star formation and of galaxy formation
must have occurred even earlier.

The UV luminosity density, an indicator of the
star formation rate, has increased by a factor of $\sim$4 over the
period $z=0$ to $z=1$.

\end{abstract}


\section{Introduction}

The goal of the Caltech Faint Galaxy Redshift Survey is to understand
the evolution of galaxies to $z\gtrsim1$.  In this paper we present
the luminosity function (henceforth the LF)
for galaxies in the region of the Hubble Deep
Field-North (henceforth the HDF) (Williams \etal\ 1996) from
our very complete redshift survey.

First we discuss the completeness of the observed sample
in $U$, $R$ and $K$.  We 
describe the derivation of the rest frame luminosity function in $R$ in
some detail in \S\ref{analysis}. We then briefly indicate the differences 
in procedure for
rest frame 2400\AA, $U$ and $K$.
We discuss the evolution with redshift of $L^*$ (the characteristic
luminosity of the LF), the co-moving number
density and the co-moving luminosity density  for
various galaxy spectral groupings from $z=0.01$ to $z=1.5$.  We then 
compare our results to Poggianti's (1997)
models for galaxy spectral evolution in \S\ref{poggianti_comp}.
In \S\ref{mass} we compute the total stellar mass in galaxies as a function
of redshift.  
Comparisons with the results obtained by previous redshift surveys 
are given in \S\ref{compare}.

Throughout, our goal, given the sample size in hand, 
is the simplest possible analysis, even if at some sacrifice
of accuracy, since errors arising from the small sample size will dominate
over any limitations imposed here through use of a
relatively unsophisticated analysis.
An appendix demonstrates that the derived luminosity densities
are very robust.

As in earlier papers in this series, we 
adopt the cosmology H$_0 = 60$ \kmsM, $\Omega_M = 0.3$, 
${\Omega}_{\Lambda} = 0$.
Over the redshift interval of most interest, a flat universe with
$\Omega_{\Lambda} = 0.7$ and a Hubble constant of H$_0 = 67$ \kmsM\
gives galaxy luminosities very close to those derived below.
Additional comments regarding modifying the adopted cosmology
are given in \S\ref{cosmology}.

We use luminosities 
rather than magnitudes.  By luminosity, we mean the quantity
$\nu L_{\nu}$ with units of W
at a particular wavelength in the rest frame.  We express luminosities
as the base 10 log of this quantity.  See Cohen \etal\ (2000) for
more details.

\section{The Sample}

The redshift survey of galaxies in the region of the HDF by
Cohen \etal\ (2000) (henceforth C00), 
supplemented in Cohen (2001), is the sample used
here.  This survey strives for a complete set of redshifts for
all objects with $R < 24.0$ in the HDF itself, and for objects with
$R < 23.5$ in the Flanking Fields within a diameter of 8 arcmin
centered on the HDF.  Extremely high completeness (95\% for the HDF and
${\gtrsim}93$\% for the Flanking Fields) was achieved through a 
spectroscopic program that extended over five years.

The primary source of photometry is Hogg \etal\ (2000) (henceforth H00), 
who present four color catalogs for
galaxies in the region of the HDF.  They used the Sextractor package 
(Bertin \& Arnouts 1996) and relied on it to extrapolate the 
measured magnitudes to 
a large aperture.  The spectral energy distributions
of these galaxies were derived,
parameterized in the rest frame, by Cohen (2001).
Those fits are adopted to calculate all luminosities used in this paper.
SED parameters for 6 galaxies in the Flanking Fields with redshifts
taken from from Dawson \etal\ (2001) are given in appendix.

At least for the HDF itself, the issue of possibly missing
low surface brightness galaxies is moot in the redshift range
under consideration here since the HST/WFPC2
imaging  of Williams \etal\ (1996)
is extremely deep with respect to the limiting magnitude of the 
redshift survey (see also  Driver 2001). We assume that the H00
photometric catalog in the Flanking Fields is not seriously
affected by the potential loss of low surface brightness galaxies.

%  630 with 5 from Dawson et al.
Our redshift catalog for the region of the HDF 
contains 735 objects of which 631 are galaxies with $0 < z < 1.5$.
Ten galaxies (all with $0.25 < z < 1.5$)
are located just outside the 8 arcmin diameter sample boundary and
are excluded. The two AGNs with $z<1.5$ are also excluded
from the sample used for the LF analysis. 

%  Q    0.960000    20.3200
%  Q    0.962000    21.6900

We use the
galaxy spectral classification scheme defined in Cohen \etal\ (1999), which
basically characterizes the strength of the strongest emission lines,
particularly [OII] at 3727 \AA, [OIII] at 5007 \AA\ and H$\alpha$,
relative to the strong absorption features,
H and K of CaII and the normal absorption
in the Balmer lines.  To review briefly,
``${\cal E}$'' galaxies have spectra dominated by emission lines,
``${\cal A}$'' galaxies have spectra dominated by absorption lines,
while  ``${\cal I}$'' galaxies are of intermediate type.   Galaxies
with broad emission lines are denoted as spectral class ``${\cal Q}$''.
Starburst galaxies showing the higher Balmer lines 
(H$\gamma$, H$\delta$, etc.) in emission 
are denoted by ``${\cal B}$'', but for such faint
objects, it was not always possible to distinguish them from 
``${\cal E}$'' galaxies.
These classifications were assigned for the galaxies our sample in
the region of the HDF in C00 (Paper X) (see also Cohen 2001).

While the galaxy spectral classes that we use are very broad,
the possibility remains that galaxies may evolve between them
over time.  We must draw a distinction between
evolution of the LF with redshift 
and evolution of a particular galaxy with time.

\subsection{Completeness at Observed $U$, $R$ and $K$ Bands}

The calculation of completeness for our redshift survey
in the region of the HDF at observed $R$ is straightforward
as the sample for our redshift survey was selected from
the very deep $R$-band photometric catalog of H00.
We evaluate the completeness at $R$ by matching the redshift catalog
and the $R$-band photometric catalog of H00.
This was done in C00, but has to be reevaluated here
as additional redshifts, primarily from Cohen (2001), further increase the
completeness.  We require not the cumulative
completeness, but the completeness of the redshift survey per bin
in $R$ mag, $C(R)$, which we determine in 0.25 mag bins.
While the former is very high, the completeness
per magnitude bin is falling fairly rapidly at the faintest bins.

We cut off the sample for calculating the LF
where the completeness in a 0.25 mag bin falls to
$\sim$40\% for the Flanking Fields, which occurs at $R$ = 23.5 and cut at $R=24$ for the HDF
itself. 
Of the 631 galaxies with $0.01 < z < 1.5$, 
12 in the HDF have $R > 24$, and
50 in the Flanking Fields have $R > 23.5$.  This leaves a total sample of
553 galaxies.

Thus the sample actually analyzed here for the LF at rest-frame $R$
consists of 553 galaxies with $0.01<z<1.5$, 
93 of which are in the HDF and 460 are
in the Flanking Fields.  There are 21 galaxies in the
HDF with $23.5 < R < 24$, between the limits of the Flanking Fields 
cutoff in R and that of the HDF itself. 

The photometric catalog of H00 is not sufficiently deep at
observed $K$ or $U$.  
A study of the counts for the H00 catalog as a function of magnitude
suggests that it becomes significantly incomplete at $U > 24.5$
and at $K > 20.5$.  Hence there are many objects in the redshift catalog
with no detection at $K$ in the H00 database.
We therefore augment 
this primary photometric database by adding photometry at 
$K$ and $U$ for objects
missing such from the unpublished catalog of Barger, described
in Barger \etal\ (1998).   If there is still no observation at $K$,
we assume $R-K = 3.1$, the median value for the sample
at $z>0.5$.   If there
is still no observation at $U$, we use $U-R$ = 1.1 mag, again
the median for the sample at $z>0.5$,  except
for $\cal{A}$ galaxies, where we assume $U-R$ = 5.0 mag.  There
are only 24 galaxies in the redshift sample with
$0.01<z<1.5$ for which $U$ is missing, many of which are either
close pairs or represent matching problems between the various catalogs.
The completeness for the Flanking Fields
is then calculated in the same way described above as was used for $R$.

For the HDF itself, we augment the H00 photometric database 
with photometry from the Hawaii group, then 
use the catalog of Fernandez-Soto \etal\ (1999) as the master list against
which the completeness is evaluated for $K > 20$.
For brighter magnitudes, the H00 photometric catalogs as supplemented
with the Hawaii catalogs are used.

The resulting values for the HDF and for the Flanking Fields are
listed in Table~\ref{table_complete}.  All objects 
in this field brighter
than the the brightest entry in the table are Galactic stars.
 
The values used to calculate weights for the
LF analysis are slightly smoothed from those given in the table.
The observed $K$ completeness corrections were used for the rest frame
$K$ analysis, the observed $R$ completeness corrections were used for the 
rest frame $R$ and $U$ analysis, while the observed $U$
completeness corrections were used for the
analysis at rest frame 2400 \AA.

\subsection{The Two Weighting Schemes Used}

Two weighting schemes are applied in the following luminosity function
analysis.  The first is the conventional
weighting scheme used by many previous such analyses,
$W(G(R, z))$ = 1/$C(R)$, where $G(R,z)$ denotes a sample galaxy with with
redshift $z$ and observed $R$ mag.
Here the weight is a reflection of the completeness at a given magnitude
of the redshift catalog.  The general practice is to assign the missing
galaxies (i.e. the weight in excess of unity) to the redshift of
the observed galaxy.  We refer to this scheme as the ``R-mag weights''.
Similar weights have been calculated for $K$ and for $U$.

In addition, because of the difference in magnitude cutoff between
the HDF and the Flanking Fields, 
we multiply the weight of those objects in the HDF
with $23.5 < R < 24$ by the ratio of the solid angle subtended by the 
Flanking Fields to that of the HDF itself, which is 9.47.

These weights are valid if, and only if, the probability of determining
a redshift is a function only of magnitude.  However,
in our sample the range of redshift is very large.  The 
technical difficulties associated with the ``redshift desert'', the
regime $1.2 \lesssim z \lesssim 2$, between
where the standard optical emission lines shift out of the optical
band and the strong UV lines and Lyman limit move in, make redshift
determinations in this $z$-regime extremely difficult.  
The weighting scheme used ideally should take this into account.

Hence we introduce a second set of weights which we call
the high-$z$ weights.  The number of objects that should
be included in the sample based on their photometry but that are without 
redshifts is computed as before. However, {\it{all}} the missing objects are
put into the highest redshift bin.  Obviously use of the high-$z$
weights would be unacceptable in a sample with low completeness and 
with many bright (and hence presumably ``nearby'') objects without redshifts, 
but with the present sample we view it as a only a moderate 
extrapolation from reality.

\section{The Analysis Procedure \label{analysis} }

We emphasize that the galaxy SEDs are specified in the rest frame
(see Cohen 2001) and that all LFs are evaluated in the rest frame.
We begin with rest frame $R$, as the completeness corrections in the observed
$R$ band
are very accurately known to faint magnitudes due to the depth of
the photometry catalog of H00 and the SEDs are
very well defined there as well.  Note that $M_R = -22.00$ (rest frame)
$\equiv$log[$L(R)$(W)] = 36.97.

No extinction corrections for gas or dust within the galaxies 
themselves have been applied. 

\subsection{The Two Parameter Schechter Fits \label{2pshec} }

We use a standard maximum likelihood technique
to compute the luminosity functions (Sandage, Tammann \& Yahil 1979,
Efstathiou, Ellis \& Peterson, 1988, Lin \etal\ 1999).
We assume a Schechter (1976) function adequately describes the LF
at each $z$ for each set of galaxy spectral classes 
considered. The $R$ mag range is from 16 (there is nothing brighter
in this field except Galactic stars) to the faint end cutoff (24 in the HDF,
23.5 in the Flanking Fields).   
As a first reconnaissance of the problem, we break the sample into
the same four redshift ranges over the regime
from $z=0.25$ to 1.5 adopted in Cohen (2001).  We add  when feasible 
(i.e. for the rest frame $U$, $R$ and $K$ bands) 
the regime $0.01 < z < 0.25$ with some trepidation due to 
the selection of the HDF as a field devoid of bright galaxies
(Williams \etal\ 1996) as well as the limited
volume at low redshift of the narrow pencil beam probed here.
 
Table~\ref{table_zmean} lists the mean redshift of each
of the five bins as defined by the galaxies in the redshift sample
as well as the co-moving volume of the bins. The mean redshift of the observed sample is biased high for the low
redshift bins, where the increase in available volume is more rapid
than the drop off due to increasing distance and decreasing apparent
brightness.  Its biased somewhat low in the high redshift bins where the volume
is increasing slowly while the apparent brightness is dropping steadily.
The mean
redshift of the bins is also slightly affected by the presence and
exact redshifts of the most populous galaxy groups within some of the bins.
In accordance with our
philosophy  of simplicity in the analysis
described above, we ignore clustering. C00 have shown that
while there are no clusters (i.e. nothing comparable to an
Abell cluster) in this sample, there are many populous
groups of galaxies.
 
We carry out a standard two parameter solution for the 
characteristic luminosity $L^*$ and for the faint end slope $\alpha$
for the Schechter function 
$\phi(L) = dN/dL = \phi^*e^{-L/L^*} (L/L^*)^{\alpha}$. 
We are thus assuming that 
within each of the bins, these two parameters can be taken as 
roughly constant.  
The details of the analysis are similar to that described
by Lin \etal\ (1999).   We have five redshift bins and we
use six groupings of galaxy spectral classes.  The first three
are the primary galaxy spectral classes, $\cal{A}$,
$\cal{I}$,  $\cal{E+B}$ (henceforth just denoted as $\cal{E}$).
We also introduce two groupings of spectral classes to try to
take into account
the probable migration of galaxies between spectral classes due
purely to technical difficulties, $\cal{A+I}$ and $\cal{I+E}$, as well as
the class containing all galaxies (excluding AGN).

We require  a minimum sample of 15
galaxies in a bin.  The
$\cal{I}$ galaxy sample in the bin $0.8<z<1.05$ is too small
in the $R$-selected sample to be analyzed
independently. This is probably not an indication of a decline in
the number of such galaxies, but rather of a decline in our ability 
to discriminate between $\cal{I}$ galaxies and $\cal{E}$ galaxies
with spectra of a fixed exposure time and hence limited signal-to-noise
ratio.  The sample in the lowest redshift bin is very small, and can
only be analyzed as a whole, as is also true of the highest
redshift bin.  

There are several additional issues that affect the highest redshift bin.
First, the spectral types there
are not comparable in their meaning
to those for galaxies with $z < 1$.   Specifically 
galaxy spectral class $\cal{A}$ in the $z < 1$ sample
refers to galaxies with strong Ca II H+K lines, obvious Balmer
line absorption and a strong 4000 \AA\ break. At $z > 1.1$, however,
this galaxy spectral class designates the presence of absorption 
features and breaks characteristic
of the 2500 \AA\ mid-UV. This is not an issue in the present
work as the sample there is too small to subset in any way.

For the highest redshift bin, where the sample is dominated by galaxies with
strong emission lines, the range of $L(R)$ 
probed by our sample is small.  Hence the stability of the full solution
is delicate.  In the local Universe, samples of such galaxies show a very
steep faint end slope.  The solution for such a LF is much more difficult
than for one with the more modest faint end slope seen in the local Universe for
$\cal{A}$ galaxies.  There the luminosity function has a well defined
bend that can be used to separate $\alpha$ and $L^*$.  Such a
characteristic break is absent in local samples (and, as we will
see, the present
high redshift samples as well) of $\cal{E}$ galaxies, making the
separation of $\alpha$ and $L^*$ much more difficult.

If we consider the small sample in the highest redshift
bin, the fluctuations
introduced by a few extra galaxies in the HDF with $23.5 < R < 24.0$
are unacceptably large because the weight of each such galaxy in this grey zone is
high ($\ge 9.4$).  In a large sample, this can be tolerated, but
in this bin with $1.05<z<1.5$ the total sample is very small, and the excess weight
of the $\sim25$\% of the galaxies in the grey zone introduces an
instability.  The solution slides
towards very high values of $L^*$ and very steep faint end slopes,
even though no galaxies so luminous as the putative $L^*$ or anything near it
actually exist in the sample.  To overcome this, we found that
the best way to proceed was, for the highest redshift bin only, 
to cut the sample
back to limits of $R = 23.5$ in the HDF as well as in the 
Flanking Fields, thus making the
cutoff the same over the entire area of the survey.
A similar restriction of the maximum allowed weight
had to be imposed in all rest frame bands attempted
for the highest redshift bin only.

In the case of the highest redshift bin,
although
there are 38 galaxies in our redshift catalog for the
region of the HDF with $1.05<z<1.5$, 
seven of them have $R>24.0$ and thus are excluded 
from the $R$-selected sample irrespective of location.
Six more are located in the Flanking Fields with $R>23.5$ and hence
are eliminated.  Six more are in the HDF with $23.5 < R < 24.0$,
and also had to be eliminated from the $R$-selected sample.

The results of this classical analysis are given in Table~\ref{table2p}\
for the six galaxy spectral class groupings.  The errors given in 
this table and in Table~\ref{table2p_k} are 1 $\sigma$ uncertainties
for one parameter. The error is measured by projecting along the 
relevant axis the full range of the  two dimensional error contours.
% delta chi**2 = 1 below maximum  
These, for a fixed number of objects, 
depend on the shape of the 2D
uncertainty contours, which are shown 
Figures 1-6.  The 90\% confidence
level contours are indicated in these figures 
calculated in accordance with \S{14.5} of Press \etal\ (1986)  
for each of the galaxy spectral groupings for the three bins between
$z=0.25$ and 1.05.
% using an offset from the maximum probability of 4.61 (see Press..

Efstathiou,
Ellis \& Peterson (1988) evaluated the expected variance 
of $\alpha$ and $L^*$ analytically as a function of sample size
and verified their predictions with Monte Carlo calculations.  
Using their formulae,
we expect $\sigma(\alpha)$ = 0.25 for a sample of 100 galaxies.
Our measured errors are in reasonable agreement with their formulae.

Figures~\ref{figure_rlf_on_data_ie} and \ref{figure_rlf_on_data_a} 
show the resulting LF for two galaxy
spectral groupings ($\cal{A+I}$ and $\cal{E}$) 
for the redshift bin $0.5<z<0.8$.  The best fit LF from Table~\ref{table_fixa}
is also shown superposed on the data.
The lower panels in each of these figures
shows the errors in the LF due to Poisson statistics
as a function of luminosity for these two cases.  
The number of galaxies per step
in log($L$) is largest in the region
just below $L^*$.  Brighter than that, the factor of $e^{-L/L^*}$ rapidly cuts
the observed galaxy numbers in a small sample such as ours
to zero, while within a fixed redshift interval the faint end
magnitude cutoff of the redshift survey truncates the faint end
of the luminosity distribution of the observed galaxies.
When there is only one galaxy in a bin, $\Delta(\phi)/\phi$ becomes unity.

\section{The LF at Rest Frame $U$, $K$ and at 2400\AA}

Because of the substantial mean redshift of the sample,
the calculations for the LF parameters at rest frame $U$
are made using the observed $R$-selected sample and
use the observed $R$ weights, as do those at rest frame
$R$.  Hence they are quite similar to those for rest frame $R$ described
above.  The same treatment of the highest
redshift bin was required, eliminating all galaxies with $23.5<R<24$
in the HDF.  The LF parameters
for rest frame $U$ are given in Tables~\ref{table2p_u}
and \ref{table_fixa_u}, as well as interspersed in subsequent tables.

The calculations at rest frame $K$ used the observed $K$-selected
sample and the observed $K$ weights.
The number of galaxies in a redshift bin used to determine  the
LF at rest frame
$K$ depends on the faint end cutoffs adopted for observed $K$ and 
on the color of each 
galaxy, and hence often will not be exactly
the same as used for the $R$-selected LF, but should be fairly close.
Some of the bluer and fainter among the $\cal{E}$ galaxies 
which are members of the $R$-selected sample may be too faint to be
included in the 
$K$-selected sample.  Also, since the the observations
for the redshift catalog were defined by a observed $R$
photometric catalog,
there will be a more gradual decline in completeness
near the survey limit at any other observed band
due to the range of colors of galaxies.

For the LF at rest frame $K$, we
proceed as with the $R$-selected sample. We adopt a 
faint end cutoff for observed $K$
of 20.75 in the Flanking Fields and  21.25 in the HDF.  
To obtain convergence in the highest redshift bin, we had
to establish a maximum weight of 15, thus eliminating a small number
of galaxies.  With this modification, the highest $z$ bin
converged well with a total of 27 galaxies 
included with $1.05<z<1.5$.

The results of the computation of $\alpha$ and $L^*(K)$ 
at rest frame $K$ are given
in Tables~\ref{table2p_k} and \ref{table_fixa_k}. 
To guide our expectations, we note that based on the mean
galaxy SED parameters as a function of galaxy spectral
type given in Table~3 of Cohen (2001), we 
might expect that $L^*(K)-L^*(R)$ will be larger for
$\cal{A}$ galaxies than for $\cal{E}$ galaxies by about
0.2 dex.  This is in reasonable agreement with the results of
the calculations.  Also, as expected, since the  absorption line galaxies
are the reddest galaxies in the sample, their
the fractional contribution to the total luminosity density
is slightly larger at rest frame $K$ than it is at $R$.

There is excellent agreement in the mean value $\alpha$
we derive for each galaxy
spectral grouping between the rest frame $U$, $R$ and $K$ calculations.
The evolutionary rates over the interval $0.25<z<1.05$ 
(given in Table~\ref{table_q}) are very
similar for rest frame $U$, $R$ and $K$.

LF parameters were also calculated at 2400 \AA\ in the rest frame
for galaxies with $z>0.25$ where that regime shifts into the
ground based optical bandpass.  Observed $U$ is used to define the
sample, with cutoffs of $U=24.0$ in the Flanking Fields and 24.75 in the HDF.
Here the SED model is being stretched to the extreme blue limit
of its wavelength range of validity.
The results of the computation of $\alpha$ and $L^*(2400\AA)$ 
are given in Tables~\ref{table2p_q} and \ref{table_fixa_q}.
The $\cal{A}$ galaxies are very faint 
at observed $U$; many have $U>26$, well below
the sample cutoff.  They drop
out of the sample, leaving fewer $\cal{A}$ galaxies in many redshift bins
than the minimum adopted for a LF solution.   However, these
contribute very little to the total light at 2400\AA\ in the rest
frame, and hence a negligible error is introduced into the calculation
of the total luminosity density there.  Only the regime $0.25<z<0.8$ converged
well and could be used to define the mean values for $\alpha$ given
in Table~\ref{table_meana}.  

At 2400\AA,  
the faint end slope, even for $\cal{E}$ galaxies, may be
less steep than at longer wavelengths.   However, the larger SED errors
at this rest frame wavelength may somehow have affected the
analysis to produce this apparent flattening of the LF.

\section{The Behavior of $\alpha$ With Redshift}

As expected (see Figures~1-6),
the strong negative covariance between $L^*$ and $\alpha$ seen
in the analyses of local samples is also found here.  This 
complicates the issue of
comparing the resulting parameters among the various  redshift bins 
and galaxy spectral groupings.  Irrespective of this issue,
our analysis thus far has demonstrated
(see Tables~\ref{table2p}, \ref{table_meana},
\ref{table2p_u} and \ref{table2p_k})
that the faint end slope of the LFs in the redshift regime studied here
shows the same general dependence on
galaxy spectral type/morphological type found in numerous studies 
of the local LF reviewed in \S\ref{compare_alpha}.
Specifically, the faint end slope is
much steeper for late type spirals (strong emission line galaxies
in the  galaxy spectral classification scheme used here) than it is
for galaxies without detectable emission lines.
\footnote{We ignore the highest redshift bin, $1.05<z<1.5$,
where the determination of $\alpha$ is quite uncertain.
The solution for rest frame 2400\AA\ is also problematic.}
This is illustrated
in Figure~\ref{figure_alpha_comp}, 
to be discussed in more detail in \S\ref{compare_alpha}.

Furthermore, $\alpha$ appears to
be constant,  to within the uncertainties due to 
random
fluctuations given the number of galaxies in each bin,
for each galaxy spectral class group
over the redshift regime considered here.

\section{The Evolution of $L^*$ With Redshift \label{2pmodel} }

In the next set of fits, we continue to use the standard
two parameter Schechter function model with no
additional parameters.  To proceed, we need to make
an additional simplifying assumption.  We assume, based on the
results presented above, that 
$\alpha$ is in fact constant with redshift, and that the variations
in $\alpha$ within a given galaxy spectral grouping 
seen within Tables~\ref{table2p}, \ref{table2p_u}, \ref{table2p_k}
and \ref{table2p_q} are not real. 
An alternative approach, assuming that $L^*$ is constant with redshift
while $\alpha$ is allowed to vary, is explored in an appendix.

We adopt a constant value of $\alpha$ for each galaxy 
spectral grouping
independent of redshift.  These values for each galaxy
spectral grouping,  which are from 
the values of $\alpha$ found for each redshift bin given Tables~\ref{table2p},
\ref{table2p_u}, \ref{table2p_k}
and \ref{table2p_q}
with $0.25<z<1.05$ weighted by the number of galaxies in each sample, 
are given in Table~\ref{table_meana}.

The specific places where this assumption is most likely not to be valid
include the emission line galaxies in the
two higher redshift bins, where the galaxy spectral
types becomes less distinguishable due to the lower quality of the
spectra of such faint objects.  Many galaxies in that redshift regime which
might at lower redshift be considered to be of intermediate
galaxy spectral type tend to get classified as $\cal{E}$ galaxies.
The composite types
$\cal{A+I}$ and $\cal{I+E}$ were introduced precisely to explore this issue,
and indeed $\alpha$ for the second of these is closer to constant than
it is for the $\cal{E}$ galaxies considered alone.
Also the composite nature of the set consisting of all galaxies
may, at higher redshift, become more dominated by the galaxies
with strong emission lines, with their concomitant steeper faint end slope.
This would tend to produce a small decrease in $\alpha$ for this group
with increasing $z$, as may be seen in Table~\ref{table2p}, although
the possible effect is not larger than the errors.

We use the contours of constant $\Delta\chi^2$
for the two parameter solution of \S\ref{2pmodel} above (see
Figures \ref{figure_errors_a}-\ref{figure_errors_all}) to reevaluate $L^*$,
finding the value with the maximum likelihood for the appropriate fixed
value of $\alpha$ from Table~\ref{table_meana}. 
This must be done for each galaxy spectral grouping at each rest frame
wavelength and for each bin in redshift.
The results are given in
Table~\ref{table_fixa}, which is an updated version of
Table~\ref{table2p}\ assuming a fixed value of $\alpha$ holds for
each galaxy grouping, for rest frame $R$, and Tables~\ref{table_fixa_u},
\ref{table_fixa_k}, and \ref{table_fixa_q} for the remaining rest
frame bands analyzed here.

From the entries in
Table~\ref{table_fixa} combined with the median redshift for
each bin given in Table~\ref{table_zmean}, we can estimate 
directly the change in $L^*(R)$
with redshift. Similar calculations can be carried out
for each of the rest frame colors. These piecewise values of 
${\Delta}{\rm{log}}[L^*(R)]/\Delta{z}$ calculated over the redshift
regime $0.25<z<1.05$ are given in Table~\ref{table_qpiece}, where
the mean redshift of the first and last bin used defines
the range $\Delta{z}$, as are the values for the other three
rest frame bands.

With $\alpha$ forced to be constant,
we see an apparent increase in $L^*$ 
which we believe to be valid for the $\cal{A}$ galaxies
as well as for the  $\cal{A+I}$ composite group.
The $\cal{E}$ galaxies also show an apparent increase
in $L^*$ with redshift, but we ascribe this  
to the group having a $\alpha$ decreasing with increasing $z$
as the group becomes 
more composite, with an admixture of $\cal{I}$ galaxies
at high redshift.  Note that the biggest apparent 
increase in $L^*$ for the strong emission line galaxies
occurs at the highest redshift bin used.
Furthermore, the apparent brightening with redshift of $\cal{E}$ galaxies
is much less in Table~\ref{table2p} where $\alpha$ is allowed to vary
with $z$.
Also note that the solution for the composite
group $\cal{I+E}$ shows essentially no change in characteristic
luminosity with redshift for a fixed $\alpha$.  

The characteristic luminosity of the
sample as a whole is increasing slowly.  
In the highest redshift bin $1.05<z<1.5$ an uncertain
extrapolation was required to determine $L^*(R)$ from the $\alpha$ of the
maximum likelihood solution of Table~\ref{table2p} to the fixed
$\alpha$ adopted; little weight can be placed
on the value of $\alpha$ determined in this bin as the sample
does not reach deep enough and hence the results 
of the extrapolation must be viewed with some suspicion.

At $z \le 0.8$, at least between rest frame $U$ and $K$,
the absorption line dominated galaxies are the most luminous
of the primary galaxy spectral groupings considered, 
as was already stated in C00 and Cohen (2001).

\subsection{Formal Solution For The Evolution of $L^*$ with Redshift}

We attempt to derive in a more formal way the evolution of $L^*$ 
to verify
the results of the previous section using a global fit.  We 
introduce the parameter $Q$ used by Lin \etal\ (1999), among others,
where $L^*(z) = L^*(z=0) {\times} 10^{Qz}$.
Given the modest size of our
sample and the large redshift range, issues of the stability of 
any solution suggest that we need to simplify things somewhat and
avoid doing a full 3 parameter ($L^*$, $\alpha$, $Q$) fit for
each rest wavelength considered here.
To accomplish this, taking rest frame $R$ as an example, first
we fix $\alpha$ to the mean value given above
for each galaxy spectral grouping in Table~\ref{table_meana}.  
Second, we fix the value of
$L^*(R)$ at the mean redshift of the bin $0.5 < z < 0.8$
to the value obtained for that redshift bin in Table~\ref{table_fixa}. 
This redshift bin was chosen because it is more populated than the 
lower $z$ bins.
We are then left with only a single unknown, the parameter, $Q$.
We solve for the most likely value of $Q$.

In this solution, we have assumed a simple parametric form for the evolution
of $L^*$ with redshift and also assumed a value for $L^*$
at some particular redshift, as well as continued our assumption
of a fixed value for $\alpha$.  This is the price for avoiding the
use of individual redshift bins, each with their small samples and
large uncertainties, used in the previous section (\S~\ref{2pmodel})
to derive the evolution of $L^*$ in a piecewise fashion.
If we were sure we could match
our photometry and galaxy spectral groups onto those of local samples,
the much larger local samples could be used to provide the
benchmark luminosity $L^*(z=0)$.  However, this is not the case, and
we must use one of our own redshift bins as the benchmark.

The results of this exercise are given in Table~\ref{table_q}.  They 
confirm the simpler analysis given in the previous section.  Strong
evolution in $L^*(R)$ is found for the same specific galaxy spectral
groupings, $\cal{A}$, $\cal{I}$, $\cal{A+I}$ and $\cal{E}$ where $Q$
ranges from 0.6 to 1.2 as in \S\ref{2pmodel}.
As described above, the apparent high $Q$ found
for the $\cal{E}$ galaxies is believed to be spurious. 
Little/no evolution in the characteristic luminosity of the
same set of two galaxy spectral groupings ($\cal{I+E}$ and All)
as were approximately constant in the qualitative
test of the previous section is seen in this more quantitative test.
Similar results are obtained if $L^*(R)$ is held fixed to the value found
for the lowest redshift bin at that mean redshift.

The 1$\sigma$ single parameter errors are given in the last column of
Table~\ref{table_q}.  They suggest that while the overall trends of the
evolution of $L^*(R)$ with redshift can be discerned through analysis of
our limited dataset, a much larger sample of galaxies will be
needed to determine this to an accuracy of 10\%.

To verify the above results, we have carried out a solution allowing
all three parameters, $L^*$, $\alpha$ and $Q$, to vary for each of
the three rest frame bands $U$, $R$ and $K$.
We do this only for the most populous galaxy spectral
groupings, $\cal{I+E}$ and ``all'', and obtain
the results labeled ``(3p)'' in Table~\ref{table_q},
which are in 
agreement to within the uncertainties with the values 
given in the same table
for the somewhat more constrained solution procedure described above.
The values from the three parameter solution are to be preferred and
are those used subsequently.  The errors among these three
quantities show strong negative covariances, particularly
between $L^*$ and $Q$. 

With the improvement of a three parameter fit,
we are still forcing $\alpha$ to be a constant independent
of redshift.  Due to the small size of our sample,
even in the best case, our 1$\sigma$ uncertainty in $\alpha$ is $\pm0.15$.
Testing rest frame $R$ as an example,
we find that adoption of a value for $\alpha$ that is 0.15 too large 
will result in an underestimate of $L^*(R)$ by 0.35 dex for 
$\cal{E}$ galaxies,  but only by 0.05 dex for $\cal{A}$ galaxies.
This striking difference arising because of the
differing shapes of the error contours.  Very steep faint end slopes
are often accompanied by error contours which are much closer to vertical
in the $\alpha$-$L^*$ plane, as is apparent from comparing 
Figure~\ref{figure_errors_i} with Figure~\ref{figure_errors_e}.
Thus a small trend in $\alpha$ with $z$ which can be hidden within 
our uncertainties may still be enough to produce errors in $L^*$ of
0.2 dex for galaxy spectral groupings with very steep faint end slopes,
while $L^*$ determinations for galaxy spectral groupings with flatter 
faint end
slopes should be more accurate and less dependent on the
exact choice of $\alpha$.

\subsection{Overall Summary for the Behavior of $L^*$ \label{section_overall}}

The galaxy spectral groups $\cal{A}$, $\cal{I}$, $\cal{A+I}$
and $\cal{E}$
all show substantial brightening with redshift over the range
from 0.24 to 2.2$\mu$.  The composite group $\cal{I+E}$ and the
group consisting of all galaxies show a slower increase
in luminosity with redshift except at 2400 \AA.
The rise in $L^*$ found here for the galaxy spectral groupings
with flatter faint end slopes is believed to be real.
However, we suggest that the large apparent brightening seen for the
$\cal{E}$ spectral group is spurious, and is due to mixing
of $\cal{I}$ with $\cal{E}$ galaxies in the higher redshift bins
where these two galaxy classes become essentially indistinguishable. 
This mixing would produce almost undetectable changes in the
composite value of $\alpha$, given our uncertainties.

We thus view the subdivision of the sample into
galaxy spectral groupings whose results
are most likely to be valid as the $\cal{A}$ with the $\cal{I+E}$ groups.
Those, as well as the group containing all the galaxies,
will be the groups emphasized in the remainder of this work.

\section{The Evolution of Number Density with Redshift \label{number}}

The final step in this analysis is to compute the normalization
constant for the LF.  
The calculations
are carried out with $\alpha$ fixed at the mean value adopted
for each galaxy spectral grouping (given in Table~\ref{table_meana}) and the
resulting luminosities from Tables~\ref{table_fixa}.
For rest frame $R$ only, we give 1$\sigma$
errors as a multiplier to be applied to $\phi^*$. These were calculated
utilizing the $1\sigma$ error contours of the two parameter
LF solutions.  We take the $L^*$ and $\alpha$ from these contours
at the ends of the major and minor axes of the
error contour.  We then propagate these values through the codes
to calculate four values of $\phi^*$.  
We take the average of the absolute values of the fractional
deviations with respect to the nominal value as the $1\sigma$ error.
This is an approximation for the true $\sigma$, but should be close enough
for our purposes, given our small sample size.

It is important to stress that the errors given in Table~\ref{table_fixa}
include variations in $L^*$ and in $\alpha$, as well as the 
term arising from Poisson statistics.  Although the numbers
of galaxies in some of the bins are quite small, the
contribution to the total error of this last term is generally small.  These
errors have {\it{not}}
been calculated assuming fixed values of $\alpha$, but represent the
full range of likelihood with both parameters allowed to vary.

To understand the behavior of the uncertainty for 
$\phi^*$ shown in Table~\ref{table_fixa},
consider an idealized  sample of galaxies which 
is complete and which reaches as faint as $L(min)$, where
$L(min) \sim$0.1$L^*$.
The error in $\phi^*$ will depend on $\alpha$.  For
$\alpha$ large and positive, the LF approaches a $\delta$-function.
As long as the $\delta$-function
is located at $L^* > L(min)$, the error in $\phi^*$ will be zero.  
However, for the steep faint end slopes found for the LFs
of $\cal{E}$ galaxies, the integral of the LF will change significantly
for a small change in $\alpha$, and hence the uncertainty in
$\phi^*$ will be large.  
For a very deep sample where $L(min) \lesssim0.01L^*$,
a situation not achieved in the present sample, the changes in the 
the integral are small as $\alpha$ is varied even for 
steep faint end slopes.  For a fixed $\alpha$ and $L(min)$, as $L^*$
increases, $\phi^*$ decreases, and again the change in the integral
is larger for LFs with steeper faint end slopes.

Similar calculations were carried out to evaluate $\phi^*$
at each of the other
rest frame wavelength considered here; the results are
reported in the last column of 
\ref{table_fixa_k},
\ref{table_fixa_u}, and \ref{table_fixa_q}.

As an approximation to gain insight, we assume that 
at any rest frame wavelength and for a particular galaxy
spectral grouping, galaxy LFs
change with redshift only through variations of $L^*$.  With this
assumption, $N = \phi^* f(\alpha)$.
We can then directly compare the number densities as a function
of redshift within the
entries for a particular galaxy spectral grouping (i.e. a particular
choice of $\alpha$) simply by comparing the derived values of $\phi^*$.

We find that the co-moving number density of $\cal{A}$ galaxies is constant between
$z=0.4$ and $z=0.6$, but then
declines by 40\% at a mean $z=0.9$.   We have calculated using our SED
formalism the redshift at which a normal $\cal{A}$ galaxy with
a typical SED from Cohen (2001) becomes fainter than the cutoff
in the Flanking Field for a $L^*$ galaxy.  At observed $R$, this is
$z\sim0.9$ and at observed $K$ this is $z\sim1.25$.  Thus
it is only to be expected that almost all of 
the absorption line galaxies will
be fainter than than the faint end cutoff adopted at observed $R$
for $z>0.9$, and this produces a drastic drop in the  
number of $\cal{A}$ galaxies in the sample.  We emphasize that this
happens for the nominal SED of an absorption line galaxy. 

One way of checking that the co-moving number density of $\cal{A}$ galaxies is 
actually constant, at least out to $z=1.05$, is to calculate the luminosity
density from such galaxies in each redshift bin, but
including only such galaxies more luminous than $L_{min}(R)$.
Comparison of luminosity densities is much more robust than comparison
of individual LF parameters such as $\alpha$ or $L^*$.  We do
this by directly summing the $L(R)$ for the $\cal{A}$ observed
galaxies in the sample (no weighting is applied), 
adopting $L_{min} = 36.0 + 0.6z$.
This yields
values of 0.57, 1.16 and $0.86 \times 10^{36}$ W Mpc$^{-3}$
for the redshift bins $0.25-0.5, 0.5-0.8$ and $0.8-1.05$
respectively.  The selection criteria adopted for the HDF itself
(``no bright galaxies'') probably produces
the low value in the lowest redshift bin.  The values for the two higher
redshift bins agree to within 25\%.  Since luminous $\cal{A}$ galaxies
are highly clustered, clustering could easily
account for this difference.

Daddi, Cimitti, \& Renzini (2000) have evaluated the surface
density of EROs with colors corresponding to passively evolving
elliptical galaxies in the latest large area deep infrared
imaging surveys
(both their own and those of Thompson \etal\ 1999) and conclude
that there is good agreement in all respects between the
properties of the EROs and those predicted for ellipticals
at $z>1$.

The co-moving density of galaxies
with emission lines does not change by more than 20\%
out to $z=1.05$ based on the $\cal{I+E}$ and $\cal{E}$
galaxy spectral groupings.

The
co-moving number density of the entire sample of galaxies is dominated by
the $\cal{E}$ galaxies.  The lowest redshift bin ($z<0.25$) is low
by a factor of $\sim$2, which is not surprising considering the
deliberate choice of the HDF as a region devoid of bright (i.e. nearby)
galaxies.  Ignoring that bin, the volume density is constant
to within the uncertainties until the highest redshift bin, $1.05<z<1.5$, at which
point it falls drastically, by a factor of $\sim$10.  This is
in agreement with Figure~6 of C00, a plot of observed $R$ mag
versus $z$ for the galaxies in this sample, which gives the strong
visual impression that there is a deficit of galaxies at $z>1$.

The situation at the highest redshift bin is difficult to evaluate.
There is a litany of problems here, ranging from not seeing very
deep in the LF to having to extrapolate rest frame $R$ when it
corresponds to an observed wavelength of 2600\AA, thus requiring
trust in the  SED model perhaps beyond that warranted.
It must also be recalled that the grouping  of all galaxies is
a sum of galaxies of several different spectral groups whose
relative contributions change as a function of redshift.  It is 
thus the most likely galaxy spectral grouping of those we use here
to violate the assumptions made here about how galaxy LFs evolve
with redshift, specifically that $\alpha$ remains constant and that
LFs change only through variation of $L^*$.

\subsection{The Impact of the High-$z$ Weights\label{impact}}

We can estimate the number of objects that are missing
redshifts in the $R$-selected sample.  Overall our
redshift survey is $\sim$93\%
complete to the stated limits ($R=24$ in the HDF, $R=23.5$ in the 
Flanking Fields).  There are are about 735 objects with redshifts
in the region of the HDF, including Galactic stars, but we
need the number that fall within these photometric limits.  
If ignore the very small
number of higher-$z$ objects within these photometric
limits, we deduce that there 
are about 50 objects missing from the redshift
survey. In this weighting scheme, those objects are all put
into the highest redshift ($1.05 < z < 1.5$) bin, rather than into 
any of the lower $z$ bins.

For each additional restframe bandpass in which we calculate the LF, the
effective number of missing galaxies is derived by
examining the total number of objects used and the total weight $W$
under the mag-weight scheme.  We calculate the part of $W$ due 
to objects which 
are within the photometric limits of the survey but which do not have
spectroscopic redshifts (the ``missing'' galaxies).  
This calculation carried out at rest frame $K$ suggests that
the number of galaxies in the highest redshift bin, which is 
27 in this case, must be increased by $\sim$100.

The highest redshift bin is very sparsely populated.
Due to the completeness limits adopted,
the $R$-selected sample in this redshift bin contains only 
24 objects of the 38 galaxies with $1.05<z<1.5$
in the redshift catalog.  (The $K$-selected sample contains 27 galaxies
in this $z$ bin.)
Six of these 24 galaxies are in the HDF with
$23.5 < z < 24$, and hence have $W \sim 9.5$.  These were excluded
from the rest frame $R$ analysis, but not from the rest frame $K$ analysis.

Thus the impact of adopting the
high-$z$ weighting scheme is dramatic in the highest redshift bin,
increasing the number of objects by factors of $\sim$4.
This will provide the maximum possible correction factor to
$\phi^*$ for the highest redshift bin.
Since the redshift survey is very complete, the
impact on the lower $z$ bins of adopting this weighting scheme
is small.  Less than 15\%
of the total weight in the redshift bins up to $z=1.05$ in the
$R$-selected sample comes
from objects that are not observed.   The number
of objects in all lower redshift bins should be decreased at the same time
by $\sim$20\% for the
$0.8<z<1.05$ bin and by 8\% for the $0.25<z<0.5$ redshift bin
for the $R$-selected sample, with slightly higher values
prevailing for the $K$ selected sample.

\subsection{Correction Factors for the Number Density in the 
Highest Redshift Bin \label{correc_num}}

At first glance, as shown in \S\ref{impact}, the redshift regime
$1.05<z<1.5$ has a co-moving number density of galaxies drastically
smaller (by a factor of 10 or more) than that of the lower redshift bins. 
Our approach is to make our best effort estimate of the
correction factors that can be applied, leaning towards trying
to achieve a maximum co-moving number density, and to see if these
correction factors can realistically be made large enough to achieve
constant co-moving number density.  The biggest correction factor
will come from adopting the high-$z$ weighting scheme discussed above
for the galaxies missing spectroscopic
redshifts but within the adopted limits for
the photometric survey.  We adopt the values given in 
the previous section, i.e. the $R$ sample then has $\sim$100 galaxies
in the highest redshift bin, the $K$ sample $\sim$125 galaxies.
Note that this is a firm upper limit to the correction.

We can easily be missing another 10\% of the galaxies through problems 
in the photometry at these faint levels, overlapping images which have
not been properly resolved, etc.  Note that this is 10\% of {\it{all}}
galaxies in the sample, or perhaps 
more appropriately 10\% of all faint galaxies in the sample,
{\it{not}} 10\% of the galaxies in the highest redshift bin.
Just as for the galaxies missing redshifts, we apply this correction
assuming all these galaxies (55 galaxies) are at high redshift and
to be added to the highest redshift bin only.
This factor could be underestimated slightly, so since we are going
for the maximum believable correction, we apply
a 15\% correction, which is equivalent to adding 80 galaxies,
to the highest redshift bin.  (It is here that a sample defined 
exclusively through
HST imaging, with much better characteristic spatial resolution,
would be superior.  We have repeatedly requested HST imaging of the
HDF Flanking Fields.)

We must than add in all the $\cal{A}$ galaxies which
are mostly, at this high redshift, fainter than the photometric
cutoffs adopted.  Absorption line galaxies
constitute roughly 1/6 of the total $R$-selected sample  in
the lower redshift bins, and we assume this should apply
in the highest redshift bin as well.   This makes the membership
of the highest redshift bin be $\sim$200 galaxies and
brings the co-moving
number density in the bin $1.05<z<1.5$ to within
30\%  of the number density for lower redshifts.  We regard this
as close enough, given the errors.

Furthermore, at $z=1.5$, 
$R$ in the observed frame corresponds to a wavelength of
2600\AA, which is at the limit of validity at the blue end of our
SEDs.  (See Cohen 2001 for a more detailed discussion of the 
limitations of our SED model.)  The errors for $L(2400\AA)$
are larger than those of other bands considered here because of 
the uncertainty of the SED model there.

This combination of factors does just barely succeed in raising the population
of the highest redshift bin $1.05<z<1.5$ to the point where it is
consistent with constant number density of galaxies throughout
the sample volume.  We emphasize that we have pushed each factor
to its maximum to accomplish this.  We have perhaps pressed these
factors beyond the tolerance of the referee, to which the author
can only reply that after looking at the images and the spectra
and checking object by object over the course of the past few
years, she believes such large correction factors cannot be
ruled out.  

It is not possible for the co-moving
galaxy number density at luminosities near $L^*$ to have been 
significantly higher at $z\sim1.3$ than it is at present.

To summarize the conclusion of this discussion,
assuming that no large error has been introduced by fixing $\alpha$
for the highest redshift bin,
the apparent deficit of luminous galaxies with $z>1$ in our earlier
papers is probably $\it{not}$ real, but this requires
stretching the correction factors for various sample 
selection issues to their maximum credible values.  This result
needs to be verified with a deeper and much larger redshift sample 
accompanied by a photometric catalog based on better spatial
resolution images than those available to us.

\subsection{Comparison of Observed and Predicted Properties of the Sample}
 
As a check on the validity of our LF modeling,
we utilize the parameters determined above, together with the volume of 
the cone of our survey  through each of the redshift
bins, to predict the number of galaxies that
should have been observed in our sample in the various galaxy
spectral groupings in each of the redshift bins.  In these calculations,
we adopt a magnitude cutoff of $R=23.5$ ($R$ in the observed frame),
and find the corresponding value of $L(U)$ and $L(R)$ using the mean SED
parameters for each galaxy spectral type given in Cohen (2001).
Similar calculations are done for the rest frame $K$ sample
selected at observed $K$ with a cutoff of 20.75, and the
2400\AA\ sample selected at observed $U$ with a cutoff of 24.0.

These calculations are stringent tests of our SED model and
sample characterization.  The SED model is used with the mean
SED parameters of the various galaxy spectral groupings
from Table~3 of Cohen (2001) to convert between
rest frame luminosity and observed magnitudes.
This is required to determine 
the low luminosity cutoff at a particular rest frame wavelength
of the integral for the predicted co-moving number density
(and also the integral for the predicted mean observed luminosity
of the sample).  A small error in this value will produce, for galaxy
spectral groupings with steep faint end slopes, a substantial
error in the prediction ($\sim$30\% for a 0.2 mag error in $M$(cutoff)).

The results are presented in Table~\ref{table_num}.  As expected,
the LFs derived here predict the observed galaxy counts
to $R=23.5$ mag to within $\sim$30\% throughout.
The agreement is particularly good for galaxy spectral groups with
less steep faint end slopes.

Also illustrative is a calculation of the predicted mean $L(R)$ for the sample
of galaxies observed.  This calculation too is a stringent
test of our SED model
as the same procedure to define the lower limit in the integration
for the predicted mean $L(R)$ for the number densities
is used here as well.  We adopt the same cutoffs as 
above for the observed $U$, $R$ and $K$-selected samples.
The results are given in Table~\ref{table_lmean}. 
The entry in Table~\ref{table_lmean} for the group of ``all'' galaxies
include both the observed and predicted values for the mean
value of the rest-frame luminosity for the observed sample.  
These are in reasonable agreement throughout. 

The predicted values in Table~\ref{table_lmean} show the expected
very strong increase with redshift as the less luminous galaxies
drop out of the magnitude limited sample towards higher $z$.
Because of the steep faint end slope of the LF for the galaxies
with strong emission lines, their mean $L(R)$ within our sample increases by
a factor of $\sim$5 between $<z> = 0.4$ and $<z> = 0.9$ while the mean 
luminosity of the sample's absorption line galaxies is
predicted to increase only by a factor of $\sim$3.

\section{The Luminosity Density and its Evolution with Redshift \label{text_lumdens}}

We compute the luminosity density between 10$L^*$ and
$L^*/20$  by integration
using the parameters of the best fit LF at each rest frame
wavelength and for each galaxy spectral grouping. 
The resulting values for $\rho(L)$ using $\alpha$ and
$L^*$ from the two parameter
Schechter fits are given in
Table~\ref{table_lumdens}.  The appendix, where a different set
of assumptions are made to analyze our sample, establishes that these
values for $\rho(L)$ are very robust.  Note that these tabulated
values must still be corrected for the various selection issues
described in \S\ref{correc_num}, such
as the disappearance of $\cal{A}$ galaxies from the sample as they
become EROs.  We have not tried to adjust $\alpha$ for each of the
different rest frame wavelengths to be identical for each
galaxy spectral group; they are consistent to within the
uncertainties, but not identical.

For rest frame $R$ and $K$, we give 1$\sigma$
errors as a percentage of $\rho(L)$.  These were calculated
utilizing the $1\sigma$ error contours of the two parameter
LF solutions.  We take the $L^*$ and $\alpha$ from these contours
at the ends of the major and minor axes of the
error contour.  We then propagate these values through the codes
to calculate four values of $\phi^*$ and of $\rho(L)$.
We take the average of the absolute values of the fractional
deviations with respect to the nominal value as the $1\sigma$ error.
This is an approximation for the true $\sigma$, but should be close enough
for our purposes, given our small sample size.

Error estimates for $\phi^*$ can be calculated in
a similar manner, but are not very useful as
$\alpha$ is varying between the points selected on the error contour,
hence the results are not directly comparable.

We use the three parameter simultaneous fits for
$\alpha, L^*$ and $Q$, when available, to generate 
Figure~\ref{figure_lum_dens}, which shows the evolution of the luminosity
density from 0.24 to 2.2$\mu$ in the rest frame for the $\cal{A}$
and the $\cal{I+E}$ galaxy spectral groupings (closed and open circles
respectively), as well as for the group of all galaxies, indicated
by the larger symbols.  A correction for galaxies missing from the
sample calculated from the number densities in the relevant tables 
(e.g. Table~\ref{table_fixa} and its equivalent
at other rest frame wavelengths) has been applied to the
highest redshift bin $1.05<z<1.5$ and to the group of $\cal{A}$
galaxies for $0.8<z<1.05$, where they drop out of the sample as well.
This correction is for the highest redshift bin is large
(a factor of 3 -- 10) and uncertain.

Overall, we
see a slow increase in the luminosity density, with the 
increasing fractional contribution of the $\cal{A}$
galaxies at redder rest frame wavelengths, as expected. 
Correcting for the loss of $\cal{A}$ galaxies from the sample
in the highest redshift bins, Figure~\ref{figure_lum_dens}
suggests that the fractional contribution by the various galaxy
groupings does not change much with redshift for a particular
rest wavelength, at least over the redshift regime considered here. 

\subsection{The Star Formation Rate and Its Dependence on Redshift}

The flux in the ultraviolet continuum of galaxies is often used
as an indicator of the star formation rate, along with
emission at H$\alpha$, the 3727\AA\ line of [OII], other
photoionized optical emission lines as well as radio emission.
For $z>0.25$ we observe the mid-UV directly from the ground
for the galaxies in our sample.

We have computed the luminosity density at 2400\AA\ from
the data in our redshift survey.  There are three places relevant
quantities appear in the Tables.  The first is 
Table~\ref{table_fixa_q}, where $L^*(2400\AA)$ is 
seen to be rising rapidly with redshift for the $\cal{E}$
spectral grouping and for all the galaxies together.
This gives
rise to the entries for rest frame 2400\AA\ in 
the next table (Table~\ref{table_qpiece}), which are
computed directly from the above and hence also indicate
a rapid rise in $L^*(2400\AA)$ of a factor of 
5 -- 10 between $z=0$ and $z=1$.  The global solution
for $Q$ given in Table~\ref{table_q} also suggests
a brightening in $L^*(2400\AA)$ by a factor of 10 over
this redshift interval.  This value is in agreement with
the rapid increase in total emission in the 3727\AA\ [OII]
emission line found by Hogg \etal\ (1998) and by Lilly \etal\ (1996),
and disagrees with the value smaller by about a factor of 3
found by Cowie, Barger \& Songaila (1999)
who analyzed part of the present sample together with
additional data from the Hawaii Deep Fields to deduce the UV luminosity 
density. 

The luminosity density calculations for rest frame
2400\AA\ is given in Table~\ref{table_lumdens},  and at $0.25<z<0.5$ is 
very close to that determined recently for the local Universe by
Treyer \etal\ (1998).  The values in Table~\ref{table_lumdens}  support 
an increase in $\rho(L)(2400\AA)$ with redshift as well,
but by a somewhat smaller factor.  As shown in Figure~\ref{figure_lum_dens},
the increase in $\rho(L)$ at 2400\AA\ corresponds to $Q \sim 0.6$, i.e. a
factor of four increase of $\rho[L(2400\AA)]$ at 2400\AA\ between $z=0$ and $z=1$.

Because of the strong  negative
covariance between $\alpha$ and $L^*$, any estimate of the 
UV luminosity density based solely on values of $L^*$ 
which ignores completely the role of $\alpha$ must be 
viewed with suspicion. Hence
we prefer the estimate for the change of the UV luminosity density
with redshift based on the appropriate integrations for
$\rho[L(2400\AA)]$ described above, 
whose results are given in Table~\ref{table_lumdens}
and in Figure~\ref{figure_lum_dens}.  These suggest that
$\rho[L(2400\AA)]$ increases
by a factor of $\sim$4 between $z=0$ and $z=1$.
However, given the current controversy over this
issue and its overall importance,
we are planning to redo the measurement of the 3727\AA\ line 
strengths shortly as this should provide a more definitive value for the
star formation rate and one potentially less affected 
by extinction than a value based on the UV luminosity density.
Our preliminary analysis of the
spectra suggests that that
the  mean rest frame  equivalent width of the
3727\AA\ emission line of [OII] for our
observed sample of galaxies in the region of the HDF remains
approximately constant
with redshift, which requires the mean
observed equivalent width to increase $\propto(1+z)$.
Under these circumstances, the emitted flux in this emission 
line will increase substantially with
redshift due to the substantial rise in mean luminosity of the observed sample
given in Table~\ref{table_lmean}.
This issue will be discussed at length in the next paper in this series.

\section{Comparison with Galaxy Evolution Models \label{poggianti_comp} }

Once a model for the evolution of the integrated light for
a galaxy is adopted, a number of other calculations become
feasible.  We can, for example, extrapolate the luminosity densities
in each redshift bin
given in Table~\ref{table_lumdens} into luminosity
densities at $z=0$, taking out the expected trends due
to normal stellar evolution.  This is straightforward for
galaxies with little star formation.  For galaxies with
considerable star formation during the time (redshift) interval
in question, assumptions regarding the form of the
the SFR as a function of $z$ become more critical.
We adopt the galaxy evolution models of Poggianti (1997)
for this purpose.  In particular, her elliptical model
has an initial burst of star formation with an exponential
decay timescale of 1.0 Gyr and is 15 Gyr old.  Her model for Sc galaxies assumes
a star formation rate proportional to the gas fraction and
includes inflow.

Quiescent galaxies within which significant star formation has not occurred
during the entire redshift range spanned here will basically be
fading in luminosity strictly through stellar evolution.  
The detailed behavior of the fading as a function of redshift and
rest frame wavelength depends only on the adopted stellar evolutionary
tracks and on the conversion between time and redshift (the cosmology).
They are seen as $\cal{A}$ galaxies throughout the sample volume.
In actively star forming galaxies,
on the other hand, the fading with time  from passive stellar
evolution, although always occurring, can be overcome by the
additional luminosity input from young stars.

Figure~\ref{figure_lstar_urk_z} compares the luminosity evolution we find
for the region of the HDF
given in table~\ref{table_q} (filled circles) with those
predicted from the model.  The mean redshifts of the three bins
between $z=0.25$ and 1.05 are used, and the points
represent $L^*$ for rest frame $U$, $R$ and $K$ 
for $\cal{A}$ and $\cal{I+E}$ (i.e. relatively quiescent and
star-forming) galaxies.  These points are connected by
the solid lines. The three parameter solution for $Q$ is used.  Because here
we are focusing on $L^*$ only, the luminosities in each of the 
rest frame bands have been shifted to 
the value of $\alpha$ obtained for rest frame $R$ for each
galaxy spectral grouping.  This requires adding
a small constant offset to the result for to log(L*) in each redshift bin.
The constant is different for each galaxy spectral grouping and also
varies with rest frame wavelength.  (It is 0 at rest frame $R$.)

The predictions of Poggianti's models in this figure are represented
by the open circles, which for each rest frame color are
connected by dashed lines.

The agreement between the evolution in $L^*$ with redshift as a function
of rest frame color that we derive through our analysis of the LF
for galaxies in the region of the HDF with that predicted by
Poggianti's (1997) galaxy synthetic spectral evolution models
for the Sc model
with the star forming galaxies in our sample
is  reasonable at all rest frame wavelengths.  
The agreement for the more quiescent
galaxies is fair at $R$ and $K$, but poor at rest frame $U$.
The passively evolving model predicts a fading at rest frame
$U$ with increasing time that is significantly larger
than that inferred by our analysis.  The predicted fading
of $\sim$0.9 dex in log[$L^*(U)$] between 
$<z>=0.9$ and 0.6 is not present in the
data. 

One cannot argue that these galaxies at $z\gtrsim0.5$ are masquerading
as galaxies of any other spectral group as they
remain very red in the rest frame optical/near-IR.  Such red
galaxies very rarely, if ever, show emission lines
throughout the relevant redshift range (see Cohen 2001), and
hence have little or
no current star formation until beyond $z=1.5$. 
They therefore cannot have morphed out of
the $\cal{A}$ galaxy group.

It is undoubtedly true that an adjustment to the star formation
prescription used by Poggianti to generate the SED of the
quiescent galaxies could resolve these problems.  The general
form required is for a small tail of residual star formation
to continue with time beyond the initial burst.

At our request, Poggianti (private communication, 2001) has recalculated
the evolutionary corrections for the $\cal{A}$ galaxies using
our cosmology and also changing the e-folding time of the
initial burst from 1 Gyr to 0.5 Gyr.  The values published
in Poggianti (1997) are recovered with the appropriate inputs.
However, the new evolutionary
corrections at rest frame $U$ are much smaller 
at high $z$ than those
published in Poggianti (1997).  For example, at $z=1$, the new
evolutionary correction is $\sim$2.5 mag smaller.  These
new evolutionary corrections provide a much better fit to our
results.

Figure~\ref{figure_lstar_urk_z} recovers one of the key results
of Cohen (2001), that the major change in galaxy SEDs with
increasing redshift in this redshift range
is the rest frame UV becoming bluer for actively
star forming galaxies.  Here we add 
an evaluation of the trend of overall luminosity.

\subsection{Total Stellar Mass in Galaxies as a Function of Redshift \label{mass} }

One rationale for studying galaxies at $K$ is that
the integrated light there is much more representative 
of the total stellar mass of a galaxy than are the 
optical colors, where light from the most recent epoch
of star formation involving only a small fraction of the stellar mass
in the galaxy may dominate over that from the much more massive
older population as one moves towards the ultraviolet.
To convert the rest frame $K$ luminosity density we measure
as a function of redshift (given in Table~\ref{table_lumdens}) into total
stellar masses of the galaxies requires a model of evolving
galaxy spectral energy distributions
to evaluate their mass-to-light ratio as a function of look back time
and of their star formation history.  Because the evolutionary
corrections at $K$ are small and almost independent
of spectral type over the range of stars that contribute substantially
to the integrated light of a galaxy at this rest frame wavelength,
we are able accomplish this transformation with some degree of precision,
and hence to deduce the total stellar mass in galaxies as
a function of redshift.  Any variation in the total 
stellar mass must be a result
of star formation.

We use the models of Poggianti (1997)
for the evolution of galaxy integrated light to 
extrapolate the luminosity densities at rest frame $K$, 
$\rho(L/K)$ at $z_i$,
given in Table~\ref{table_lumdens} to values
at $z=0.0$, $\rho(L/K/z_i~{\rm{at}}~z=0.0)$, so that they can be directly
compared. This quantity 
is then converted into total stellar mass in galaxies assuming
a constant mass to light ratio at rest frame $K$ 
applies to all galaxies at $z=0.0$; we adopt 0.8 (in Solar units).

The results are given in Table~\ref{table_ml}.  The entries in the
table are given in the form of $L(K/z)$ extrapolated back to $z=0$
in units of W Mpc$^{-3}$. The
conversion factors required to obtain values in units of
$L\subsun$Mpc$^{-3}$ or $M\subsun$Mpc$^{-3}$ are given in the
notes to the table. The
uncertainties have been propagated using the values from
Table~\ref{table_lumdens} for $K$ and assuming a 15\%
uncertainty in the evolutionary correction that brings
the observed $L(K/z)$ back to $z=0$.  If the contribution
of the different galaxy spectral groupings to the integrated 
light at $K$ were constant with $z$, the $M/L$ adopted would
act strictly as a normalization factor.  Hence we ignore its
uncertainty in this calculation.  

There is
no sign in Table ~\ref{table_ml}
of any substantial increase in the total mass in stars between
$z=0.25$ and $z=1.05$.   The total stellar mass in galaxies
appears to be constant with redshift to within $1\sigma=\pm20$\%.
This statement is not inconsistent with
the claim made above in \S\ref{text_lumdens} that the 
UV luminosity density at 2400 \AA\
and hence the star formation rate
are increasing substantially between $z=0$ and $z=1$.  If we adopt
the conversion between UV flux and star formation rate given by
Kennicutt (1998) and integrate the UV luminosity density 
$\rho(L/2400\AA)$ as a function of
time between the present and $z=1.05$, we find that $\lesssim$20\%
of the total stellar mass currently in galaxies given Table~\ref{table_ml}
has been formed since $z=1.05$, depending on the exact choice of $Q$
for rest frame 2400\AA\ and on the mean extinction adopted.

\section{Changing the Cosmology \label{cosmology}}

Since the key parameters are proportional to $H_0^n$ ($n=2$
for $L$, $n=3$ for the co-moving number density, and
$n=1$ for the co-moving luminosity density), the consequences
of changes in $H_0$ are easily evaluated.
The effect of the choice of $\Omega$ and $\Lambda$ 
is more subtle, as both the luminosity distance
($L \propto D_L^{2}$),
and the cosmological co-moving volume, which
affects co-moving densities, are non-linear
in these parameters.  We compare the result
of adopting the 
flat model $\Omega_m = 0.3$ and $\Lambda=0.7$
suggested by the CMD fluctuation measurements
(de Bernardis \etal= 2001) instead of the
model adopted in this series of papers, $\Omega_m=0.3$, $\Lambda=0$,
which has the virtue that many cosmological
quantities of interest have analytical solutions.

The luminosity distance  for the flat model is 
$\sim$8\% larger than it is for the model we use, which difference can
be removed by modifying the adopted value of $H_0$ from 60 to 65 \kmsM.
More importantly, the ratio of luminosity distances
between  the two models  changes as a function of $z$ by only
about 10\% over the range $z=0.25$ to 1.5, with the flat model
having luminosity distances that increase more rapidly
with redshift.  Adoption of such a model would mean that
the more distant galaxies in our sample would become slightly
brighter by $\sim$0.07 dex in log($L$) compared to the nearest
galaxies.
These small changes are well within our errors.  Given the 
contribution from passive stellar evolution within
galaxies, which is larger than this cosmological difference,
there is no hope
of using our sample to learn anything about cosmology.

In the flat model with $\Omega_m=0.3$, the differential co-moving volume
between redshift $z$ and $z+\Delta(z)$, $V(z)$, is larger for the flat model
than for the cosmological model we use here.
Moreover, $V(z)$ for the flat model increases even more rapidly
at high redshift than does $V(z)$ for the cosmology adopted here.
[$V(z=1.0)/V(z=0.3)$ for $\Omega_m=0.3,\Lambda=0.7$]/
[$V(z=1.0)/V(z=0.3)$ for $\Omega_m=0.3,\Lambda=0.0$] = 1.23
for $\Delta(z)=0.10$.
This difference  exacerbates the problems of 
the low co-moving density we have inferred 
in our analysis for $z>1.1$ in \S\ref{number}.  A larger survey
with better understanding of the sample completeness for $z>1$ might
be sensitive to this difference in co-moving volume.

\section{Comparison with The Results of Other Surveys \label{compare}}

In comparing our characteristic luminosities
with those of other surveys, we convert
the cosmology to that adopted here.
\footnote{In practice, we just adjust for any difference in
the adopted values of $H_0$.}
We also report two
values of $L^*$.  The first is that obtained from the maximum
likelihood solution with fixed $\alpha$ given in the
appropriate one of Tables
~\ref{table_fixa}, \ref{table_fixa_u}, \ref{table_fixa_k}
or \ref{table_fixa_q} for the $z=0.25-0.5$ bin.  
This value is then extrapolated to 
$z=0.0$ (for comparison with local surveys) using 
the values for $Q$ from Table ~\ref{table_q}. 
Since all local redshift surveys have much larger galaxy
samples than we do, we assume their determinations of the
faint end slope of the LF are correct.
Therefore, a second
comparison value is obtained from our tables by finding $L^*$ at the
value of $\alpha$ of the local survey, which may not be the
same as our best value of $\alpha$ of Table~\ref{table_meana}.
Because of the covariance between $L^*$ and $\alpha$, this
second value of $L^*$ is perhaps the most valid for the
comparison.  Values of $\alpha$ are directly comparable; no
adjustments are required.

Local surveys determine the LF over a much wider range of
luminosity than is possible when very distant galaxies are involved.
Their luminosity density integrals are carried out to 
fainter lower limits in $L(R)$, and their total luminosity
densities are correspondingly higher.
To overcome this, we assume that all published local measurements
go to much fainter luminosities relative to $L^*$ than do ours
and that they should be corrected by the ratio 
appropriate to a Schechter function for the relevant value of $\alpha$,

$$   \Gamma(\alpha+2)/\int_{0.05}^{10}(L/L^*){\times}\phi(L/L^*) 
{\rm{d}}(L/L^*). $$

Table~\ref{table_lfcomp} gives the details for all the comparisons
described below.  In all such comparisons, one must remember that 
different survey teams
adopt slightly different definitions for their galaxy
spectral groupings.

\subsection{Comparison with Local Values for $\alpha$ 
and for $L^*(R)$\label{compare_alpha} }

The LCRS (Shectman \etal\ 1996), with $\sim$20,000
galaxies with a mean $z$ of 0.10, was the first local redshift
survey big enough to ensure that the LF determination would
be highly accurate.  Lin \etal\ (1996) carried out the LF analysis
in the (pseudo)-$R$ band 
of that survey. They find $\alpha=-0.70$ and,
adjusting to our cosmology, as we do for all such comparisons, 
log[$L^*(R)$] = 36.76 dex, including a small (0.1 mag) factor
to convert between the $R$ band photometric system of the CFRS and ours.
Separating the sample into emission and non-emission line objects, dividing 
at an equivalent width of 5 \AA\
for the 3727 \AA\ line of [OII], Lin \etal\
found variations of
the faint end slopes and characteristic luminosities shown in Figures
\ref{figure_alpha_comp} and \ref{figure_lstar_comp}, and given
in Table~\ref{table_lfcomp}.  In figure \ref{figure_lstar_comp}, we 
also superpose the predicted passive stellar evolution for an 
elliptical and for a Sc galaxy from Poggianti (1997), indicated 
by the thin solid curves.  These have an arbitrary vertical 
zero point offset, which was set for convenience here.

Blanton \etal\ (2001) report the preliminary LF from  the SDSS for
more than 11,000 galaxies  for five rest frame bands
3560 -- 9130 \AA.  Applying the conversion factor
between various $R$ band observational systems recommended
by Fukugita \etal\ (1996) leads to
$L^*(R) = 37.01$, with $\alpha= -1.16$.
It is very interesting that the faint end slope for the sample
as a whole does not vary noticeably over this wavelength range;
ours does not either, at least between 0.36 and 2.2$\mu$ in the rest
frame.

Initial results for the 2dF survey were presented in 
Folkes \etal\ (1999) in the $b_j$ band. Now that the 2dF
galaxy redshift sample is
in excess of 45,000, Cross \etal\ (2001) provide an
update on the galaxy LF, with improved photometry,
and  they introduce a more complex
complex bivariate form for the LF.
They deduce, for the entire sample, converting 
their photometry into $R$ using the mean color of galaxies
from Fukugita \etal\ (1996), a value of
$L^*(R)$ identical to within $\pm0.02$ mag with of that of Blanton \etal\
with $\alpha=-1.09$.

These large surveys provide a local calibration of the LF
to which our data is compared in Table~\ref{table_lfcomp}.

Our
value of log[$L^*(R)$(W)] at $z=0.0$ from the three parameter fit
using all galaxies is 37.20 dex.
When corrected to the same value of $\alpha$ as the 2dF 
results, this becomes log[$L^*(R)$(W)] = 36.96, in good agreement with their
values for L$^*$.  Overall the agreement shown in Table~\ref{table_lfcomp}
is reasonably good considering that a double extrapolation of
our $L^*$ values is often required, once in redshift and again in $\alpha$,
to obtain the entries given in the last column.

Our value of $\alpha$ ($-1.10\pm0.10$) for the entire sample of
is in  excellent agreement with these local measurements.

Given the high precision
photometry possible for bright extended nearby galaxies, 
Blanton \etal\ have carefully analyzed
the best way to extrapolate to true total galaxy brightness
measurements within very large radii.   They claim that
total galaxy luminosities were underestimated by a factor of
two by the LCRS survey through use of the isophotal aperture photometry
adopted by the LCRS group. However,
the luxury of a complex analysis using Petrosian (1976) magnitudes is beyond
our reach with such faint galaxies and ones so distant that spatial
resolution is very limited.

\subsection{Comparison with Local Values for $L^*(K)$\label{compare_klum} }

Cole \etal\ (2001) utilize the current overlap of
the 2MASS (Skrutskie \etal\ 1997)
all-sky infrared photometry campaign
and the 2dF redshift survey to find that $M^*(K_S) = -23.44$ 
and $\alpha(K_S) = -0.9$ for
$H_0 = 100$ \kmsM\ from a sample of $\sim$17,000
galaxies.  This translates into log[$L^*(K)$] = 36.88 dex
for $H_0 = 60$ \kmsM.  Kochanek \etal\ (2001) do the same
for the $\sim$4100 galaxies in common between 2MASS and the CFA
``$z$-cat''
redshift catalogs of nearby galaxies.  They find
that galaxies of different spectral types all have the same
faint end slope $\alpha \sim -0.9\pm0.1$, but that $M^*$ varies
from $-23.0$ to $-23.5$, being brighter for earlier morphological types.
Note that Marzke \etal\ (1994) has shown that
the CFA survey, when divided on the basis of galaxy morphology,
also yields LFs at $B$ whose faint end slopes are indistinguishable
among the various subsamples, except for the Sm-Im galaxies, which
show a much steeper faint end slope, but whose presence will be
minimal in $K$-selected samples of nearby galaxies.

\subsection{Comparison with Previous Results For Intermediate Redshift}

The largest surveys including at least part of the redshift range 
discussed here are the CNOC2 survey and the Canada-France
Redshift Survey (CFRS).
The CNOC2 survey (Yee \etal\ 2000) covers the regime $0.12 < z < 0.55$
and $17 < z < 21.5$ with $<z> \sim 0.3$, and contains about 2000
galaxies in four separate fields.  The LF analysis is given in Lin \etal\ (1999).  
Redshifts are only available for a subset of these;
SED parameters from their multi-color photometric
database are used to infer the redshifts of the remaining objects.
They use three SED based galaxy spectral classes, early, late and
intermediate, and calculate the LF in the rest frame $U$, $B$ and $R$ bands. 
They established that within this regime, the early and
intermediate type galaxies primarily show brightening at higher redshift,
with only modest density evolution, while the late type galaxies show
little luminosity evolution and strong density evolution.
The typical value of $Q$ they find for their early type galaxies corresponds
to $\sim$0.6, while for late type galaxies it is $\sim$0.1, which
are quite comparable to
the values we derive given in Table~\ref{table_q}.

Our results are in good agreement with their conclusions regarding early
type galaxies (which we call $\cal{A}$ and $\cal{I}$ galaxies),
extending them significantly higher in redshift.  However, we find
little density evolution among the late type galaxies, with moderate
luminosity evolution.

The values for the faint end slope of the LF given by 
Lin \etal\ (1996) are roughly independent of
redshift between their three bands ($U$ to $R$), and are quite
consistent with the values we derive given in Table~\ref{table_meana}.
The quantitative agreement
between our LF parameters and those determined by the CNOC2 survey
is quite good as is shown in Figures 6 and 7 and Table~\ref{table_lfcomp}.

The total luminosity density
of the CNOC2 survey at rest frame $R$, with a mean $z$ of 0.3, is about 
$13.8 \times 10^{34}$ W Mpc$^{-3}$.
When we correct our  value of $7.7 \times 10^{34}$ W Mpc$^{-3}$ by
the appropriate factor to compensate for the difference in the 
lower luminosity limit
of the integration between the CNOC2 survey and our work, we obtain
excellent agreement, to within $\sim$10\%.

The CFRS is described in 
Le F\'evre \etal\ (1995).  It contains 730 galaxies of which
591 have spectroscopic redshifts with median $z=0.56$
within five fields with $17.5<I<22.5$. 
The LF analysis of the CFRS is presented in Lilly \etal\ (1995).
They divide their
sample into two galaxy groups, ``red'' and``blue'' and solve for the
rest-frame $B$ luminosity evolution.
Here we do not agree as well
with regard to the evolution of the LF.  
As shown in figures~\ref{figure_alpha_comp} and \ref{figure_lstar_comp},
the CFRS tends to have LF solutions 
that run away as $z$ increases towards steeper faint end
slopes, which in turn requires very bright $L^*$ values.
While there are probably no galaxies so luminous in the data,
the fit is not rejected because the LF becomes so steep that the
probability of seeing them becomes very low.   Our
solutions do not show such extreme ranges in $\alpha$ for a given
galaxy spectral grouping, and thus we avoid the extremely high
$L^*$ values which they find in their highest $z$-bins.  As Lilly \etal\
state, because of the high coupling between the LF parameters,
the values they tabulate of these parameters 
should not be taken in isolation but rather
viewed together as the evolution of the LF, and in that spirit
our results can be viewed as being in somewhat better agreement.

We have
not carried out an analysis at rest frame $B$.  
The comparison shown in Figure~\ref{figure_lstar_comp}
assumes that log[$L^*(B)(W)]$ at $z=0.6$ is equivalent to
log[$L^*(R)(W)] = 37.2$ dex, while the conversions given in
Table~\ref{table_lfcomp} approximate $L^*(B)$ from our
results by using $0.5[L^*(U)+L^*(R)]$.  The comparison given in
the table is from a redshift regime ($z\sim0.6$) just before the CFRS
luminosity functions start becoming unexpectedly bright
as described above.

The luminosity density, since it is an integral over the luminosity
function, is a more robust measure to compare than the individual
parameters that are used to characterize the LF.
We evaluate the total luminosity density for the CFRS as
a function of redshift range and galaxy spectral grouping
by integrating the LFs
between 10$L^*$ and $L^*/20$ using
the parameters ($\alpha$, $L^*$ and $\phi^*$)
specified by the CFRS. We convert our results from Table~\ref{table_lumdens}
using the procedure above to derive rest frame $B$ values.
The total luminosity density
of the CFRS survey at rest frame $B$, with a mean $z$ of 0.3, is about 
$7.3 \times 10^{34}$ W Mpc$^{-3}$, which is very close to the value
$6.8 \times 10^{34}$ W Mpc$^{-3}$ we infer from Table~\ref{table_lumdens}.
At $z\sim0.9$, the comparison
is also very good, $13.1 \times 10^{34}$ W Mpc$^{-3}$ for the CFRS versus
$9.2 \times 10^{34}$ W Mpc$^{-3}$ for the total rest frame $B$ luminosity.
Comparisons of the total rest frame $B$ luminosity
for our $\cal{E}$ galaxy spectral class with the
CFRS  ``Blue'' galaxies are also very good.  
This is quite encouraging and suggests that fundamentally our
sample and the CFRS are in good agreement to $z\sim0.9$, but the
different
analysis procedures, sample sizes, sample uncertainties, etc. 
have led in some cases to quite different
values determined for $\alpha$ and $L^*$ between the two projects.

Im \etal\ (2001) have analyzed the LF for a sample of 145
red field E/S0 galaxies selected through their morphology
on deep HST images and their
colors to have $16.5 < I < 22$ and expected to
be at $z \lesssim 1$.  Of these, 44  have 
spectroscopic redshifts,
with photometric redshifts being used for the remaining galaxies.
Their analysis is similar in many ways to ours; for example,
in their preferred method, they fix $\alpha$ with redshift to $-1.0$,
a somewhat steeper faint end slope than that we find for early
type galaxies.
They find a brightening in rest-frame $B$ of
1.1--1.9 mag between $z=0$ and $z=0.8$, which is in good
agreement with our result.  Their constraint on the variation
of number density with redshift (constant to within a factor of 2
between $z=0$ and $z=0.8$) is considerably looser than our constraint
developed in \S\ref{number}.

\subsection{Comparison with Previous Results at $z \sim 3$ }

Steidel \etal (1999) give an analysis of the LF for $z\sim3$ Lyman
break galaxies.  Since theirs is a photometrically selected sample,
with spectroscopic confirmation
of only a fraction of these, they reach about as deep
into the LF at $z\sim3$ as we do in our highest redshift bins.
They find $\alpha = -1.60\pm0.13$, with a UV
luminosity density which when converted into the units used here
and with a correction factor applied for carrying the integration
as faint as $L^*/20$
becomes $\rho(L)$ at 1500\AA\ of 1.1 $\times 10^{34}$ W/Mpc$^3$.
This value is within a factor of two of the value we find for
$\rho(L)$ at 2400\AA\ at $z\sim1$.  

The co-moving number density
they derive for Lyman break galaxies at $z\sim3$ (i.e. their value
of $\phi^*$) is
very close to our value for $\cal{A}$ galaxies, and smaller
by a factor of $\sim$5 than that for our total sample.

Steidel \etal\ find a value of log[$L^*(W)]$ at 1500\AA\ of 37.59 dex.
Our value at 2400\AA\ for galaxies
at $z \sim 1.2$ is almost a factor of 10 smaller, but when
extrapolated to their very steep faint end slope, we find 
log[$L^*(2400\AA)(W)]\sim37.15$ dex, only a factor of $\sim3$
smaller.

\section{Comparison of Luminosity Evolution Determined Through Other Methods}

There are a number of other methods of determining the evolution of
luminosity with redshift.  
Studies of the fundamental plane of
elliptical galaxies by  Pahre, Djorgovski \& de Cavallho (1998)
and Jorgenson \etal\ (1999) have
determined with considerable precision the location of the fundamental
plane for nearby galaxies in the field and in nearby
rich clusters.  Kelson \etal\ (1999) and
van Dokkum \etal\ (1998) have explored rich clusters with deep
HST imaging (CL 1358+62 with $z=0.33$ and MS 1054$-$3 with 
$z=0.83$)
to analyze the evolution of early type galaxies, finding that
for rest frame $B$
$M/L_B \propto (-1.0\pm0.1)z$.
This value should be compared to that we obtain for
quiescent galaxies, $Q = 0.65$ (with a large uncertainty), which
corresponds to $-1.6$ mag.

Kochanek \etal\ (2000) have used the properties of a small sample
of gravitational lenses as determined from HST images and from
mass models of the systems (used to estimate the central velocity
dispersion of the galaxy) to constrain the evolution of the
fundamental plane of field elliptical galaxies.  They demonstrate
that to within the uncertainties of their analysis 
the lens galaxies appear to behave similarly to the 
cluster elliptical galaxies.

For spiral galaxies, the Tully-Fisher relation provides 
a way of determining galaxy masses.  Here the high redshift
galaxies have been attempted with LRIS at Keck by Vogt \etal\ (1996, 1997),
who have succeeded in measuring optical rotation curves
for a small sample of $0.1<z<0.8$ galaxies with emission
in the [OII] line at 3727\AA.  They find
only a small offset of $\le$0.4 mag in 
rest-frame $B$ with respect to the local relation,
which they ascribe to luminosity evolution in the field galaxy sample.
This value is consistent with the modest values of $Q$ found for
star forming galaxies in the present analysis.

There have been a number of attempts to study the LF in the HDF
using photometric redshifts, including the work of
Sawicki, Lin \& Yee (1997), Connolly \etal\ (1997), and 
Pascarelle, Lanzetta \& Fernandez-Soto (1998),
as well as the NICMOS based studies of Thompson, Weymann
\& Storrie-Lombardi (2001) and Dickinson (2000).  The area on
the sky of the HDF is very small, and hence only a broad brush
picture can thus be derived without insurmountable problems of small number
statistics. Since the number of galaxies
brighter than $R=24$ is very small ($\sim100$) and essentially
all of them have spectroscopic redshifts now, such efforts
are most valuable as a means of exploring the regime
beyond $z=1.1$, where our sample is small, as well as the
redshift ``desert'' between $1.5<z<2$  (once suitable
calibration of the photometric redshifts exists there).

\section{Summary}

This study of the LF of galaxies in the region of the HDF-North
relies on the large databases built up in earlier papers in this
series.  It has 
been made feasible by the joining of the very complete
redshift survey in that region
of C00, the photometric catalogs of H00, and the rest-frame SED
analysis of Cohen (2001).  We divide the sample into
six galaxy spectral class groupings, not
all of which are independent, and five redshift bins.
We use this data to carry out 
a classical LF analysis assuming the Schechter function with
fixed $L^*$ and $\alpha$
is an adequate description within each redshift bin and for each
galaxy spectral class grouping.
We evaluate the LF at rest frame 2400\AA, $U$, $R$ and $K$.

We find that the behavior of the faint end slope is 
consistent with that observed in the local Universe.  The LFs for
quiescent galaxies have shallow faint end slopes, while those of
galaxies with detectable emission lines have steeper
faint end slopes.  Furthermore, as shown in 
Table~\ref{table_meana}
these slopes are independent 
of redshift out to $z=1.05$ for each galaxy spectral grouping
and agree well with comparable local determinations.  The faint
end slopes are the same for rest frame $U$, $R$ and $K$ for
each galaxy spectral grouping.

We assume galaxy LFs for a particular galaxy spectral grouping
and rest frame wavelength
change with redshift only through variations
in $L^*$ (and $\phi^*$).  We fix $\alpha$,
choosing a mean value from the set of full two parameter LF fits,
 to obtain values of $L^*$ for
each galaxy spectral grouping in each of the redshift bins.
We find that $\cal{A}$
galaxies become brighter with $z$ with $Q \sim 0.6$ at all
rest frame bands studied here, 
where $Q = \Delta{\rm{log}}[L^*(z)]/\Delta{z}$, while galaxies
with detectable emission lines (i.e. star forming galaxies), which
dominate the total sample,
show a smaller change in $L^*$ with
redshift at all bands, $Q \sim 0.3$, becoming larger only at 2400\AA.  

Passive evolution models of galaxies are in reasonable agreement
with these results for absorption line dominated galaxies, while
plausible star formation histories can reproduce the behavior
of the emission line galaxies.  The major discrepancy with the specific
set of galaxy spectral synthesis models 
we adopt, those of Poggianti (1997), is the prediction
of much more luminosity evolution at rest frame $U$ for
galaxies with a brief single initial burst of star formation than 
is actually inferred from our analysis of quiescent galaxies.
This problem appears to be eliminated when Poggianti's very
recent unpublished (Poggianti 2001; private communication) models
are used.

Our naive view as illustrated in Figure 14 of Cohen \etal\ (2000)
of the $z\gtrsim1$ Universe is one sparsely populated with
bright blue galaxies, but we now see that this is too simplistic.
The co-moving number density of 
$\cal{A}$ galaxies drops rapidly beyond $z=0.8$, but the
SED analysis shows clearly that this is due to them
becoming EROs which are too faint to be included in our $R$-selected
sample.  Correcting for this, the co-moving number density
appears to be roughly constant for each of the various
galaxy spectral groupings, until the highest redshift bin
is reached.  There, for $1.05<z<1.5$, the apparent number
density is a factor of $\sim$10 low at all bands.
We correct for selection effects resulting from the
variation of our ability
to determine redshifts as a function of $z$ itself by throwing
all galaxies without redshifts within the magnitude cutoffs
of the sample into the highest $z$ bin.  We also make a generous
allowance for galaxies near the faint end of the sample that are 
missing from the ground based
photometric catalogs.  These, with several
other smaller factors, are just barely sufficient when
applied at their maximum possible values to allow constant
co-moving number density from $z=0.25$ to 1.5, with considerable
uncertainty at the highest redshift range.
If less extreme correction factors are used, the
co-moving number density of luminous galaxies begins to decline at $z>1$.

We calculate co-moving luminosity densities. We find
that the contributions to the total change as one would
expect.  $\cal{A}$ galaxies, being redder, contribute
a larger fraction of the luminosity density at longer
rest frame wavelengths.  We then use
galaxy evolution models to extrapolate 
back to $z=0.0$ the co-moving luminosity 
densities at $K$ for each redshift bin (see Table~\ref{table_ml}).
From this, we can then calculate the total
stellar mass in galaxies which appears to be constant
to within 15\% over this redshift range.

We confirm that the UV luminosity density, an indicator of
star formation, increased by a factor of $\sim$4 over the
period $z=0$ to $z=1$.

We find the co-moving  number density and the
stellar mass in galaxies to be approximately constant
out to $z\sim1.05$, and with more
uncertainty, to $z\sim1.3$.  The
major epoch(s) of star formation and of galaxy formation
must have occurred even earlier.

An examination of alternate possibilities for the
assumptions made here shows that while the values of the
LF parameters themselves (i.e. $\alpha$ and $L^*$)
may depend on the detailed assumptions made in the analysis, 
parameters involving
integration over the LF, such as the luminosity density,
are robust.

In spite of our best efforts, and with a total sample 
of 735 objects with redshifts in this field,
the numbers of galaxies are still small in the
highest redshift bin considered here with $1.05<z<1.5$.
To attempt a fainter magnitude limited
survey with existing telescopes and instrumentation
would be extremely expensive in terms of observing time.
Our plans for future work to extend and reinforce these
results will concentrate
on targeted surveys guided by photometric criteria for
candidate selection.

\acknowledgements The entire Keck/LRIS user community owes a huge debt
to Jerry Nelson, Gerry Smith, Bev Oke, and many other people who have
worked to make the Keck Telescope and LRIS a reality.  We are grateful
to the W. M. Keck Foundation, and particularly its late president,
Howard Keck, for the vision to fund the construction of the W. M. Keck
Observatory.  

We thank Roger Blandford and George Efstathiou 
for helpful discussions.
We thank Amy Barger and Len Cowie for access to their unpublished photometric
database for the region of the HDF.  

We than the referee for constructive and helpful suggestions.

This work was not supported by any federal agency.

\appendix

\section{SED Parameters For Six Galaxies from Dawson \etal\ (2001) }

We give the SED parameters for six galaxies 
in the Flanking Fields from the twelve new
redshifts from Dawson \etal\ (2001).  (The other six are
fainter than the cutoffs adopted here.) Five of these
are in the $R$-selected sample; all are in the $K$-selected
sample.  Based on the description of their spectra given by
Dawson \etal\, five are $\cal{E}$ galaxies while F36398\_1602,
is assigned a galaxy spectral class of $\cal{A}$.
The sBB SED model was used to derive the parameters given in 
Table~\ref{table_dawson}.

\section{An Alternative Derivation of the Luminosity Density}

Because of our small sample size and wide redshift range, we
needed to make several simplifying assumptions to achieve
a satisfactory LF analysis of our data.
In the main body of this paper, we followed conventional ideas
about the anticipated evolution of the LF, assuming that $\alpha$ remains
fixed with redshift, while $L^*$ changes.  Note that this does not
amount to assuming pure luminosity evolution, as $\phi^*$ is
determined from the data, and the number density is never constrained 
to be constant with redshift.

In this appendix, at the suggestion of the referee,
we explore the opposite extreme assumption.
We assume that $L^*(R)$ is fixed with redshift, and that
$\alpha$ may vary with $z$.  In this case, unlike in the
former, the values of $\phi^*$ for a specified galaxy
spectral grouping in the various redshift bins 
cannot be intercompared to  yield
the relative number density of galaxies.  Furthermore, any
discussion of the evolution of $L^*$ with redshift is meaningless,
as $L^*$ is fixed. 

We calculate the luminosity density, which we expect to be the
most robust parameter in our analysis, as it, together with the total
number of galaxies per unit volume brighter than some specified cutoff,
both represent integrals over the LF.  The results for the
modified assumptions are given in Table~\ref{table_fixlstar}.
We compare $\rho(L)$ with the luminosity densities at rest
frame $R$ given in Table~\ref{table_lumdens}.  The comparison
is very good; the two values of $\rho(L)$ dagree to within
20\% over the entire redshift range for each of the six galaxy
spectral groups.  Since these represent the extreme opposite
assumptions that can be made in the LF analysis, we conclude that
the determination of luminosity densities is very robust, as expected.

\begin{figure}
\epsscale{0.8}
% Comment in the following line to embed the postscript figure into the manuscript
% %\plotone{errors_a.ps}
\plotone{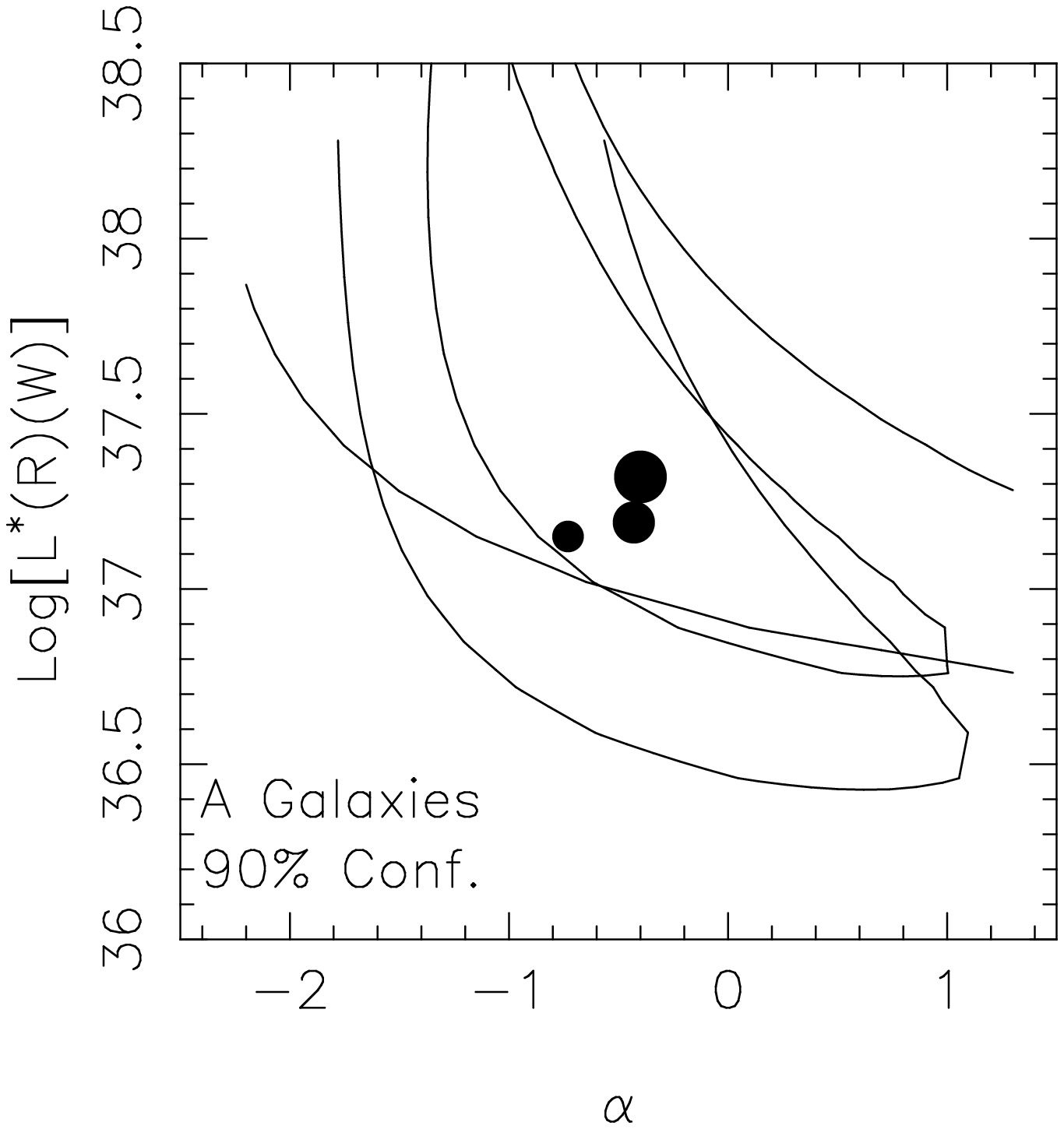}
\caption[figure1.ps]{The 90\% confidence level
contours for $\alpha$ and $L^*(R)$ for the $\cal{A}$ galaxies is
shown for the
three redshift bins between $z=0.25$ and 1.05.  The filled circle,
whose size increases with increasing redshift of the bin,
denotes the maximum likelihood solution. 
\label{figure_errors_a}}
\end{figure}

\begin{figure}
\epsscale{0.8}
% Comment in the following line to embed the postscript figure into the manuscript
%%\plotone{errors_i.ps}
\plotone{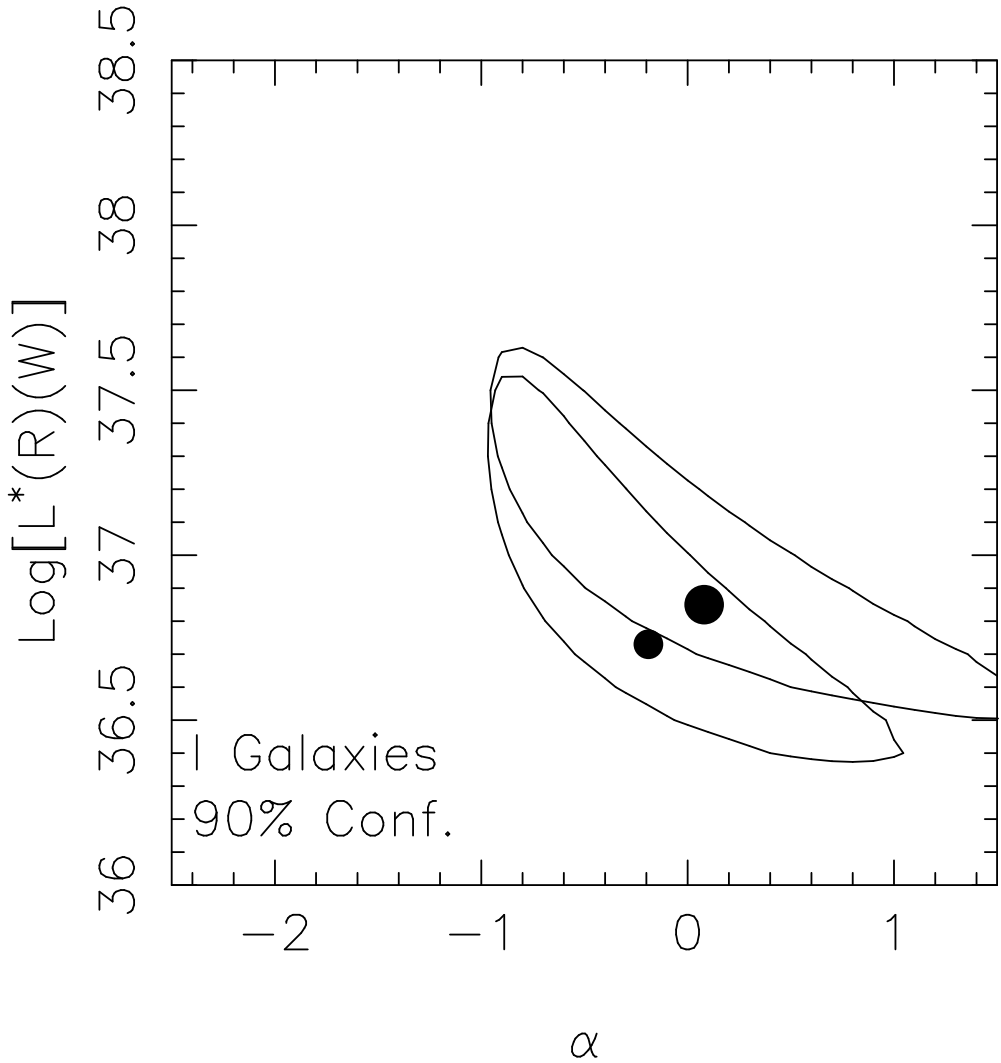}
\caption[figure2.ps]{The 90\% confidence level
contours for $\alpha$ and $L^*(R)$ for the $\cal{I}$ galaxies
is
shown for the redshift bins $0.25<z<0.50$ and $0.50<z<0.80$ only.
(The sample in this galaxy spectral grouping is too small
in the higher redshift bins.)
The filled circle,
whose size increases with increasing redshift of the bin,
denotes the maximum likelihood solution. 
\label{figure_errors_i}}
\end{figure}

\begin{figure}
\epsscale{0.8}
% Comment in the following line to embed the postscript figure into the manuscript
% %\plotone{errors_ai.ps}
\plotone{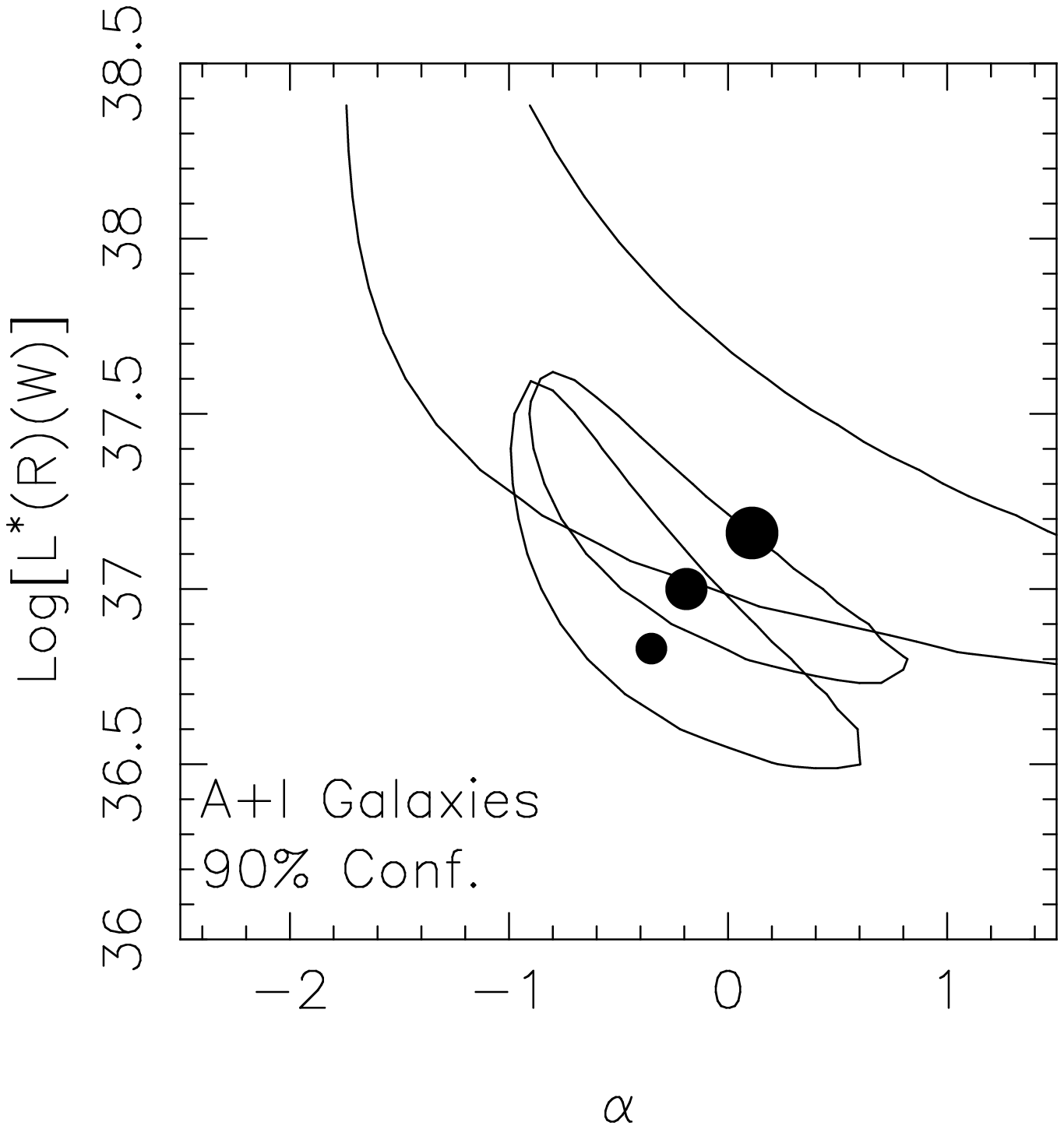}
\caption[figure2.ps]{The 90\% confidence level
contours for $\alpha$ and $L^*(R)$ for the $\cal{A+I}$ galaxies
is shown for the
three redshift bins between $z=0.25$ and 1.05.  The filled circle,
whose size increases with increasing redshift of the bin,
denotes the maximum likelihood solution. 
\label{figure_errors_ai}}
\end{figure}

\begin{figure}
\epsscale{0.8}
% Comment in the following line to embed the postscript figure into the manuscript
% %\plotone{errors_ie.ps}
\plotone{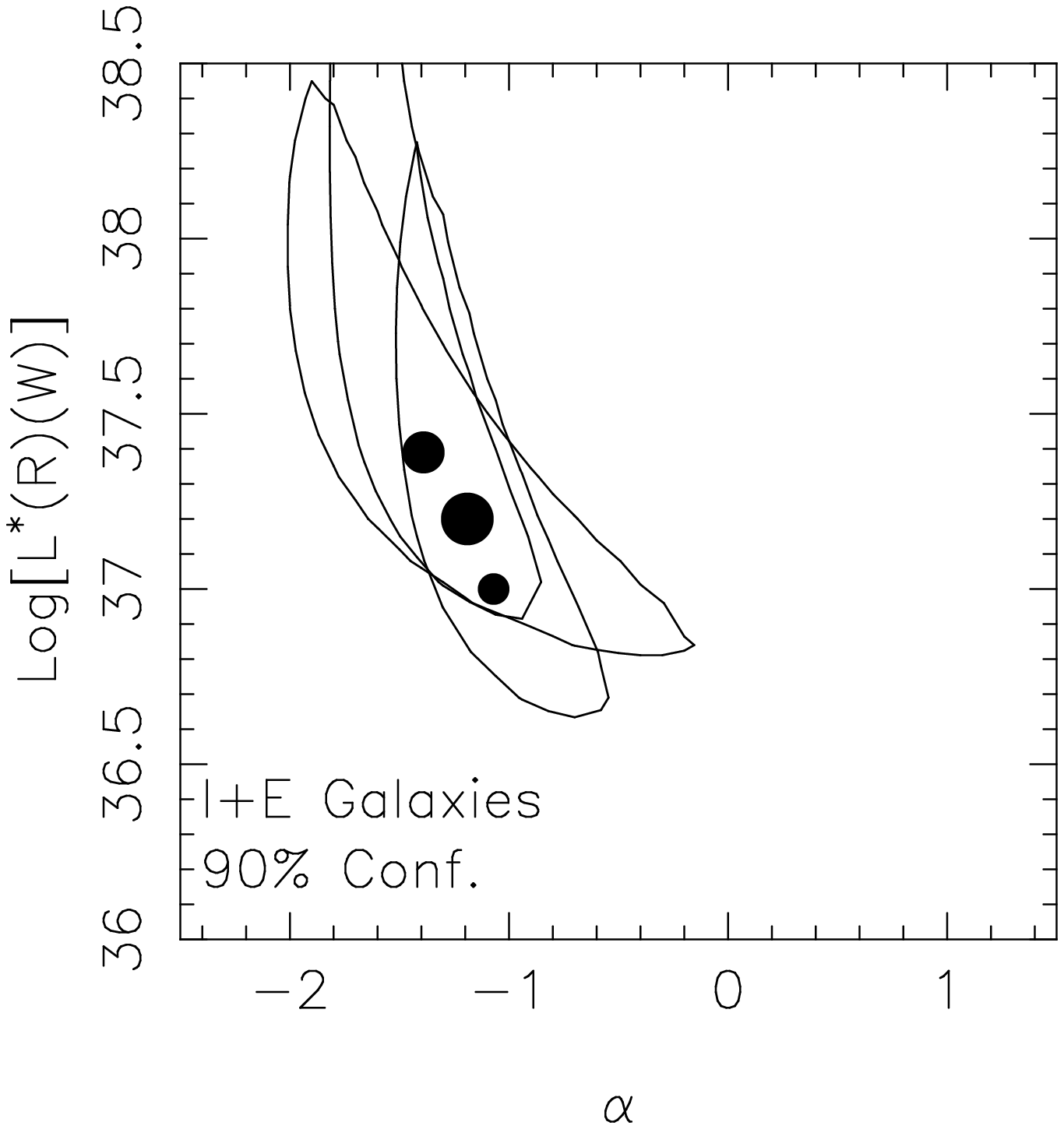}
\caption[figure2.ps]{The 90\% confidence level
contours for $\alpha$ and $L^*(R)$ for the $\cal{I+E}$ galaxies
is shown for the
three redshift bins between $z=0.25$ and 1.05.  The filled circle,
whose size increases with increasing redshift of the bin,
denotes the maximum likelihood solution. 
\label{figure_errors_ei}}
\end{figure}

\begin{figure}
\epsscale{0.8}
% Comment in the following line to embed the postscript figure into the manuscript
% %\plotone{errors_eb.ps}
\plotone{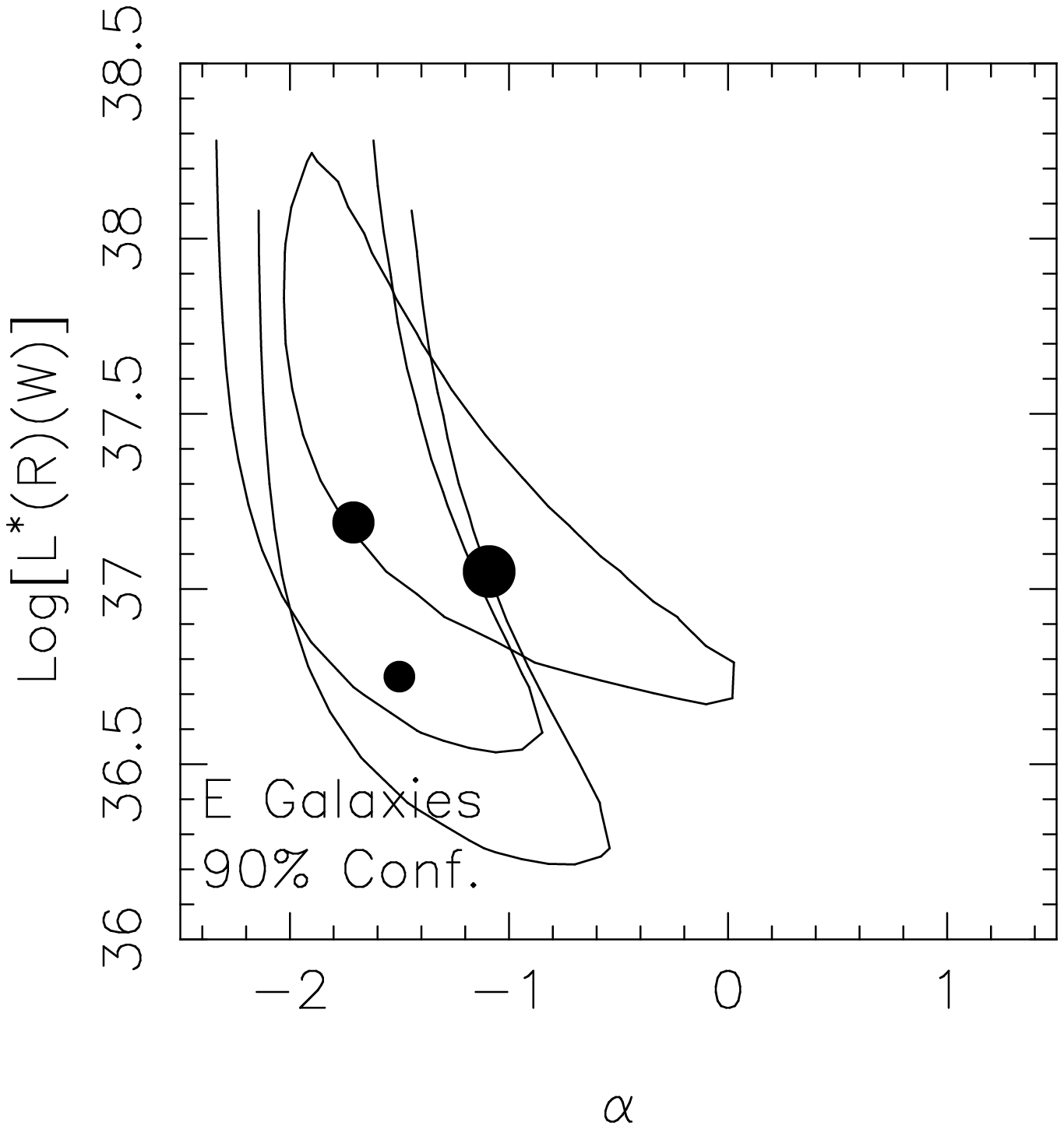}
\caption[figure2.ps]{The 90\% confidence level
contours for $\alpha$ and $L^*(R)$ for the $\cal{E+B}$ galaxies
is shown for the
three redshift bins between $z=0.25$ and 1.05.  The filled circle,
whose size increases with increasing redshift of the bin,
denotes the maximum likelihood solution. 
\label{figure_errors_e}}
\end{figure}

\begin{figure}
\epsscale{0.8}
% Comment in the following line to embed the postscript figure into the manuscript
% %\plotone{errors_all.ps}
\plotone{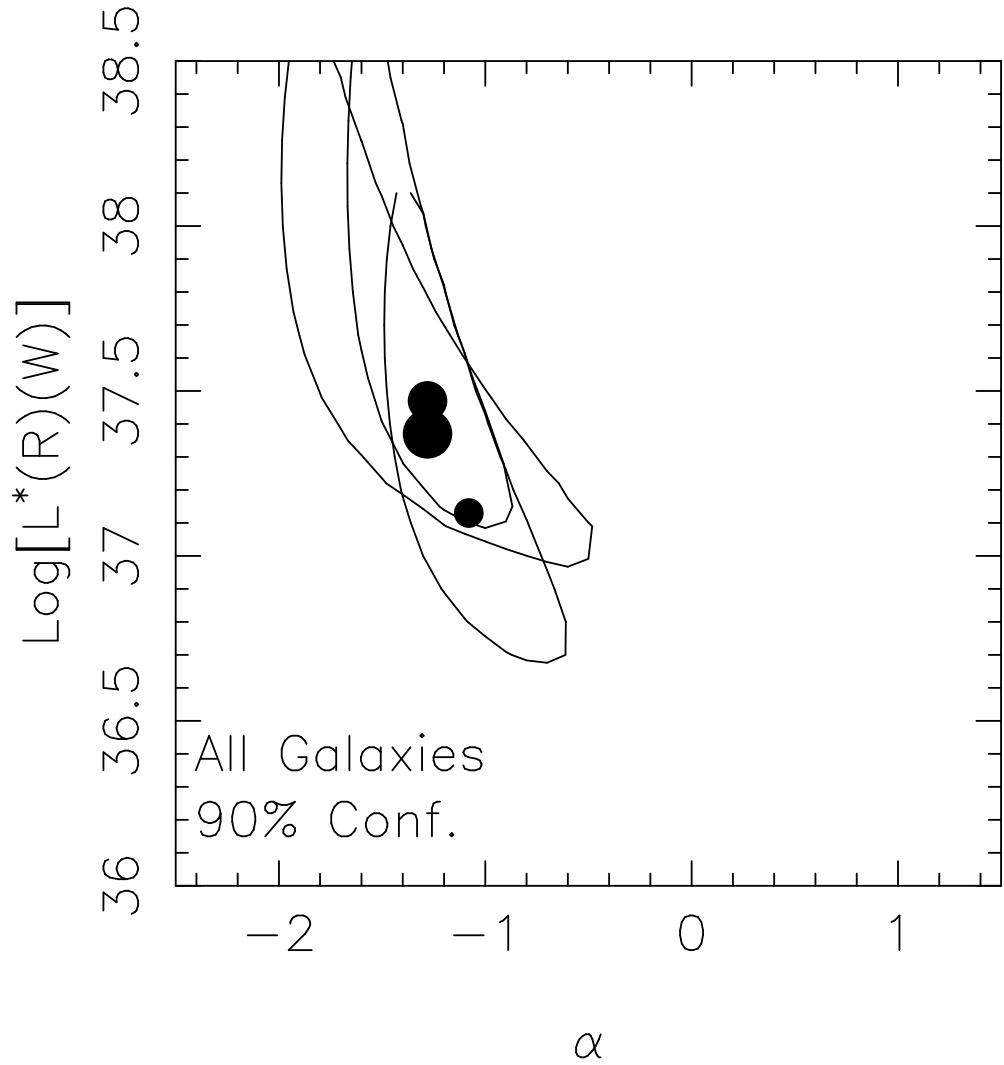}
\caption[figure5.ps]{The 90\% confidence level
contours for $\alpha$ and $L^*(R)$ for the all galaxies grouped together
(except AGNs)
is shown for the
three redshift bins between $z=0.25$ and 1.05.  The filled circle,
whose size increases with increasing redshift of the bin,
denotes the maximum likelihood solution. 
\label{figure_errors_all}}
\end{figure}

\begin{figure}
\epsscale{0.8}
% Comment in the following line to embed the postscript figure into the manuscript
% %\plotone{paper_figure7.ps}
\plotone{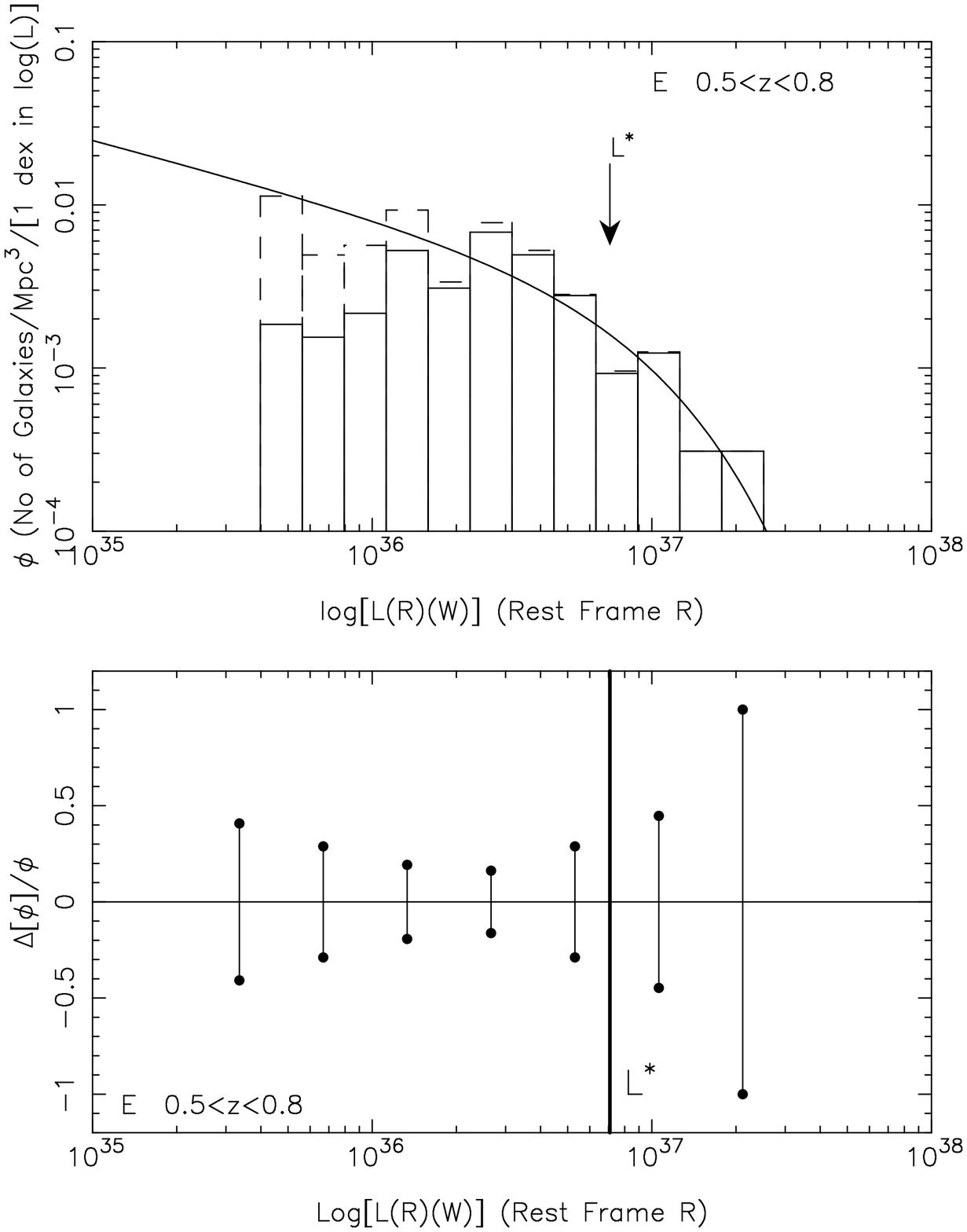}
\caption[figure5.ps]{The comparison between the observed
and best fit LF for $z=0.50$ to 0.80 for $\cal{E}$ galaxies is shown
in the left panel for rest frame $R$.   The solid lines denote the observed
sample, while the dashed lines include corrections for incompleteness in
the redshift survey.
In the right panel, we shown $\Delta\phi/\phi$,
an indication of the error arising from Poisson statistics, 
as a function of luminosity at rest frame $R$.
\label{figure_rlf_on_data_ie}}
\end{figure}

\begin{figure}
\epsscale{0.8}
% Comment in the following line to embed the postscript figure into the manuscript
% %\plotone{paper_figure8.ps}
\plotone{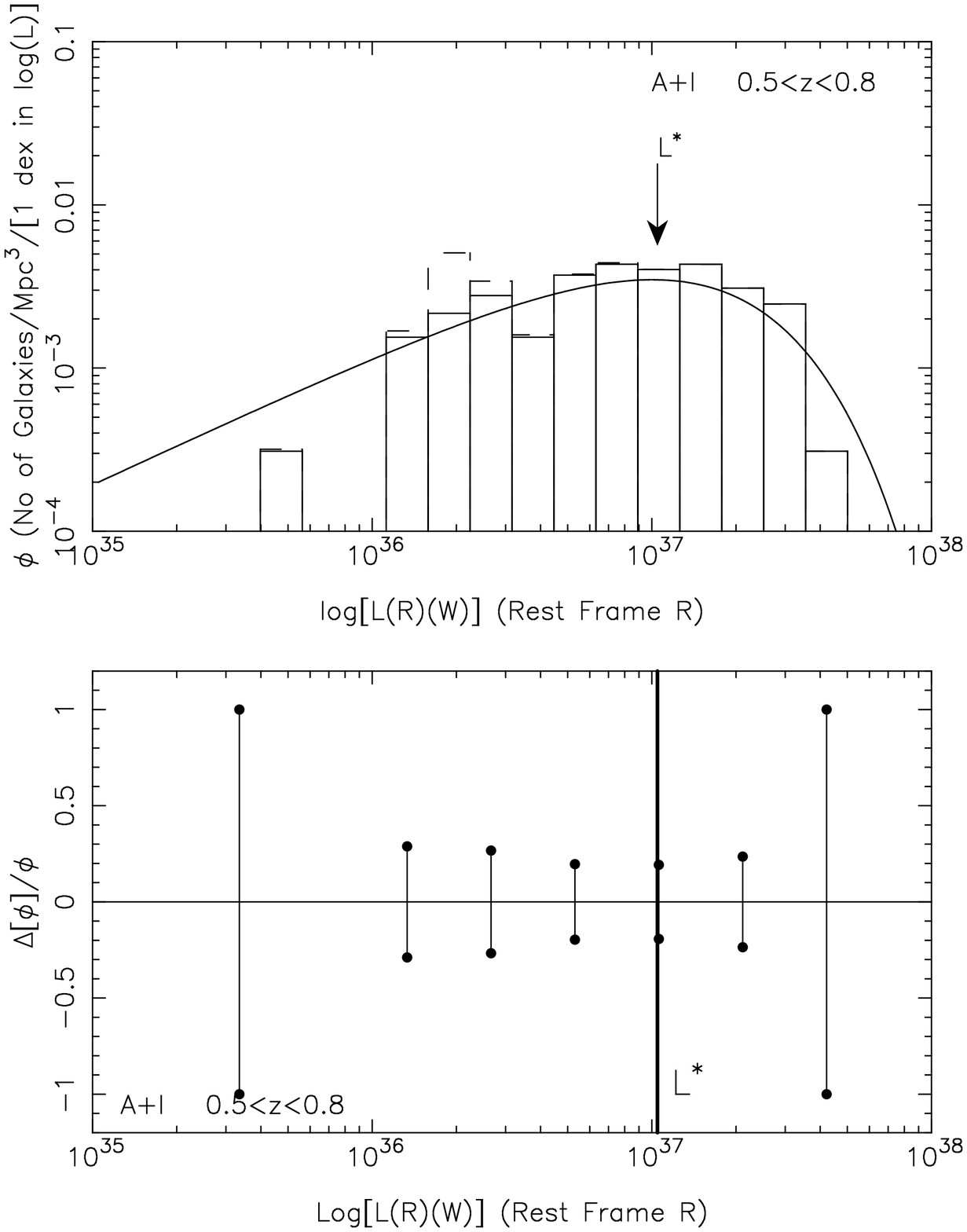}
\caption[figure5.ps]{The comparison between the observed
and best fit LF for $z=0.50$ to 0.80 for $\cal{A+I}$ galaxies is shown
in the left panel for rest frame $R$.  The solid lines denote the observed
sample, while the dashed lines include corrections for incompleteness in
the redshift survey.
In the right panel, we shown $\Delta\phi/\phi$,
an indication of the error  arising from Poisson statistics, 
as a function of luminosity at rest frame $R$.
\label{figure_rlf_on_data_a}}
\end{figure}

\begin{figure}
\epsscale{1.0}
% Comment in the following line to embed the postscript figure into the manuscript
% %\plotone{comparison_alpha.ps}
\plotone{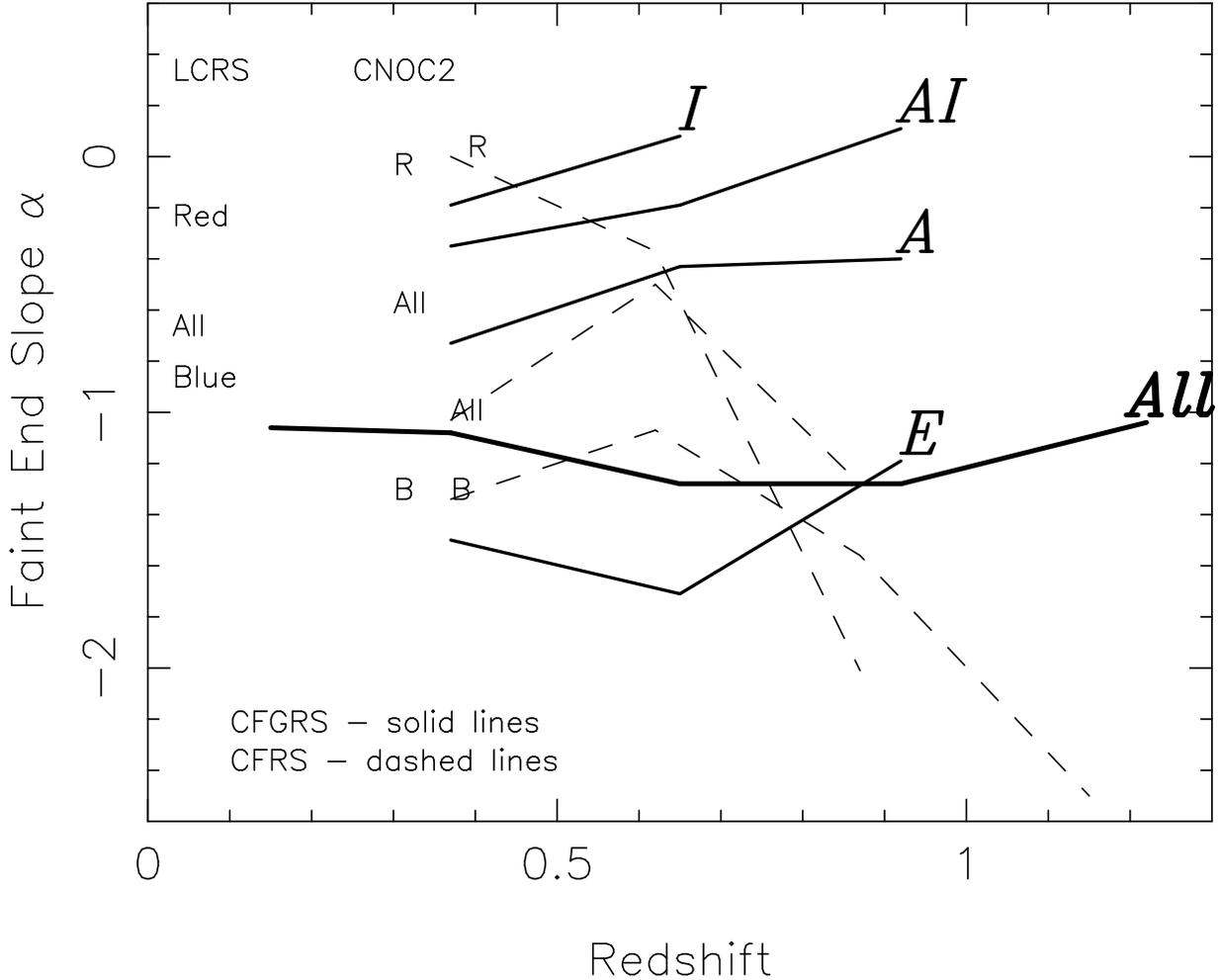}
\caption[figure6.ps]{The values we have
determined for $\alpha$ are shown as a function of redshift for
several galaxy spectral groupings by the thick solid curves.
These are from $L^*(R)$, except for the highest redshift bin, where
that of $L^*(K)$ is used.  For comparison, the faint end slopes
determined by Lin \etal\ (1996) for the blue galaxies, the red
galaxies, and the entire sample of the LCRS are shown, as are comparable
values for the CNOC2 survey (Lin \etal\ (1999).  The values 
of $\alpha$ deduced by Lilly \etal\ (1995) for the
CFRS are indicated as the thin dashed lines.
\label{figure_alpha_comp}}
\end{figure}

\begin{figure}
\epsscale{1.0}
% Comment in the following line to embed the postscript figure into the manuscript
% %\plotone{lstar_comp.ps}
\plotone{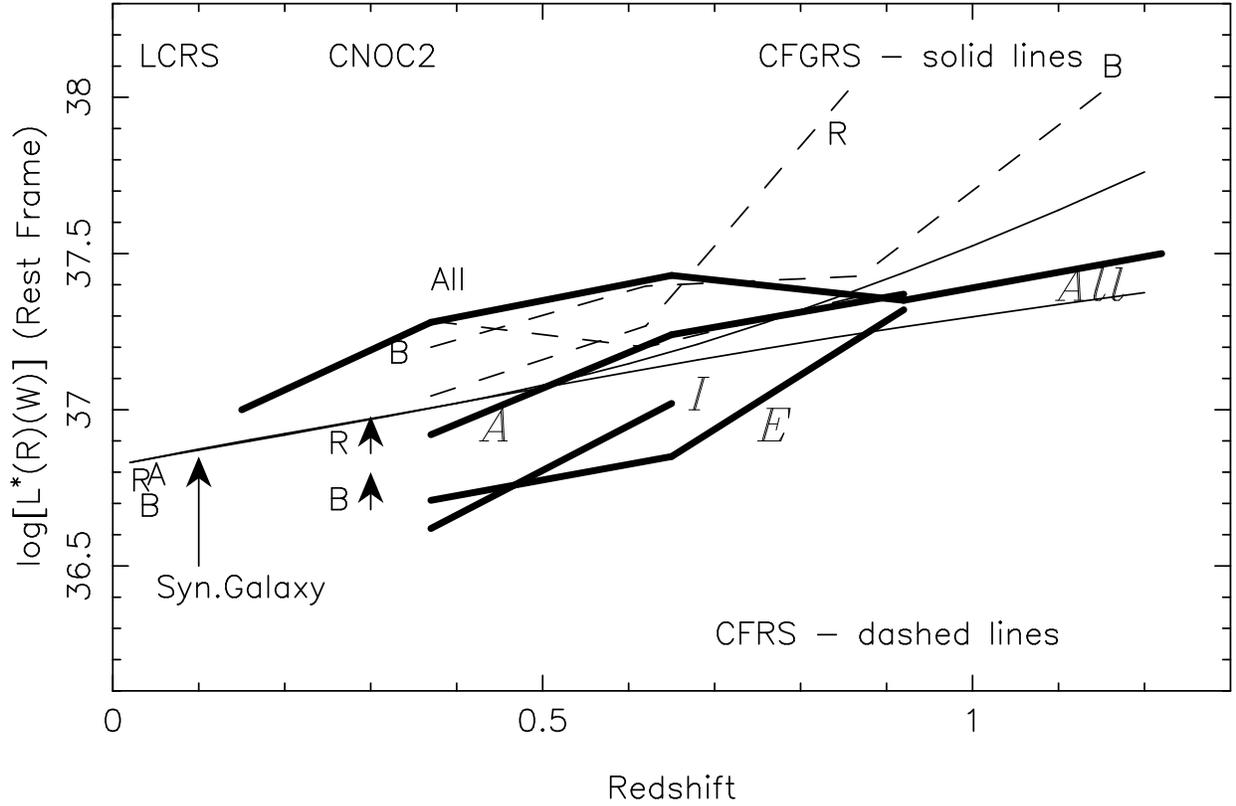}
\caption[figure6.ps]{The thick solid curves show
$L^*(R)$(W) as a function of
redshift derived here (see Table~\ref{table_fixa})
for the $\cal{A}$ and $\cal{E}$ galaxy spectral groupings,
as well as for our entire sample.  The values of $L^*(R)$ for the
complete sample, as well as that subdivided into
``blue'' and ``red'' (equivalently, ``emission'' and ''no emission'') galaxies,
from the LCRS (Lin \etal\ 1996), CNOC2 (Lin \etal 1999)
and the CFRS (Lilly \etal\ 1995) are shown for comparison.
The CFRS values are indicated by the dashed lines.
The thin solid lines denote the predicted evolution of a
elliptical and a Sc galaxy from the models of Poggianti (1997)
as a function of redshift.  
\label{figure_lstar_comp}}
\end{figure}

\begin{figure}
\epsscale{0.8}
% Comment in the following line to embed the postscript figure into the manuscript
% %\plotone{lum_density_ver2.ps}
\plotone{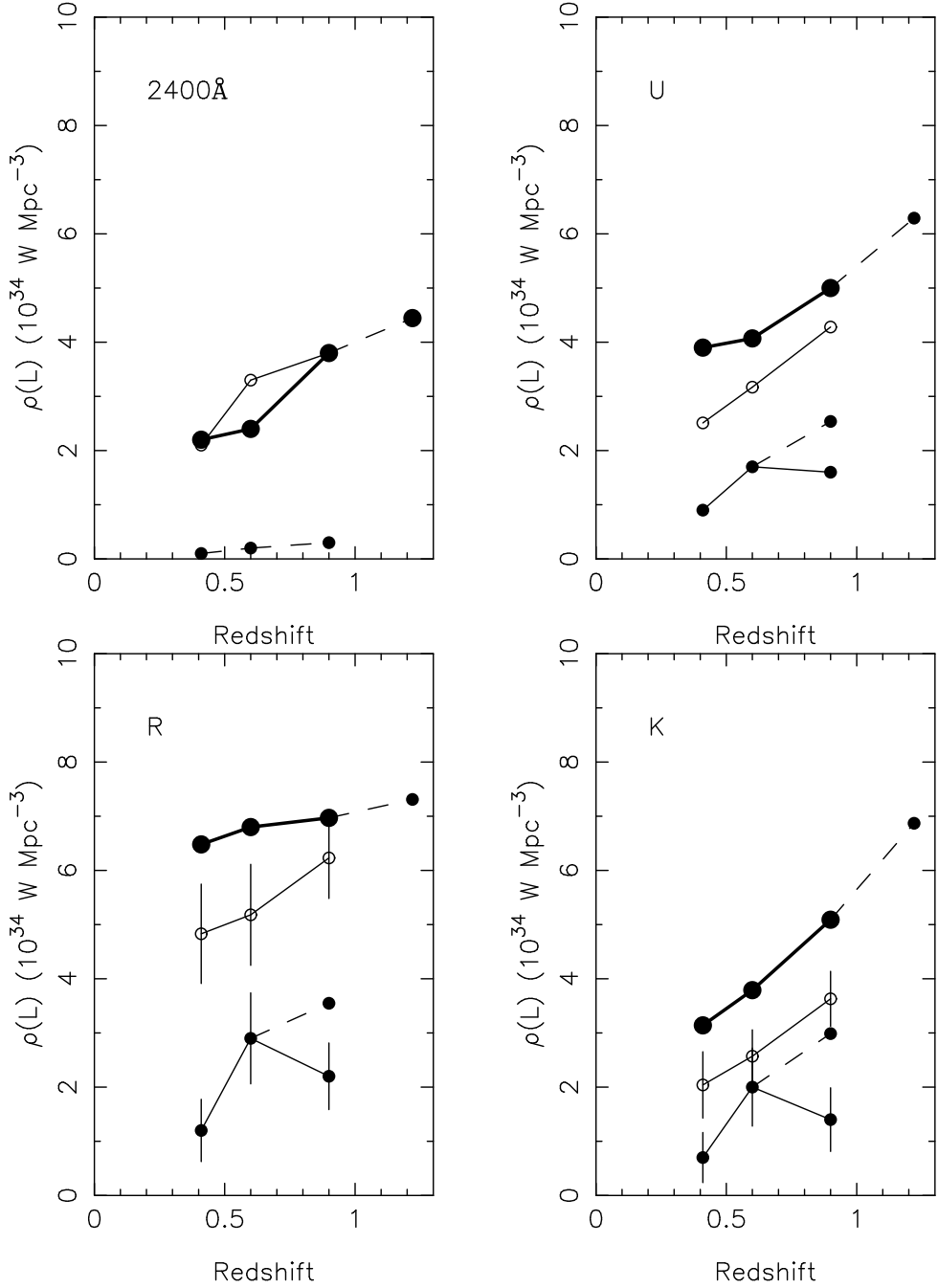}
\caption[figure8.ps]{
The evolution of the luminosity
density with redshift is shown for four rest frame
wavelengths from 0.24 to 2.2$\mu$  for the $\cal{A}$
galaxies (filled circles),
$\cal{I+E}$ galaxies (open circles), and for
the group of all galaxies, indicated
by the larger filled circles.  1$\sigma$ errors are shown
for $R$ and $K$ only. The three parameter solutions
have been used when available.  Corrections have been applied
for galaxies missing from the sample for the
highest redshift bin and for the $\cal{A}$ galaxy
grouping in the bin $0.8<z<1.05$ only.
Dashed lines connect
such corrected points.
\label{figure_lum_dens}}
\end{figure}

\begin{figure}
\epsscale{1.0}
% Comment in the following line to embed the postscript figure into the manuscript
%\plotone{lstar_urk_z.ps}
\plotone{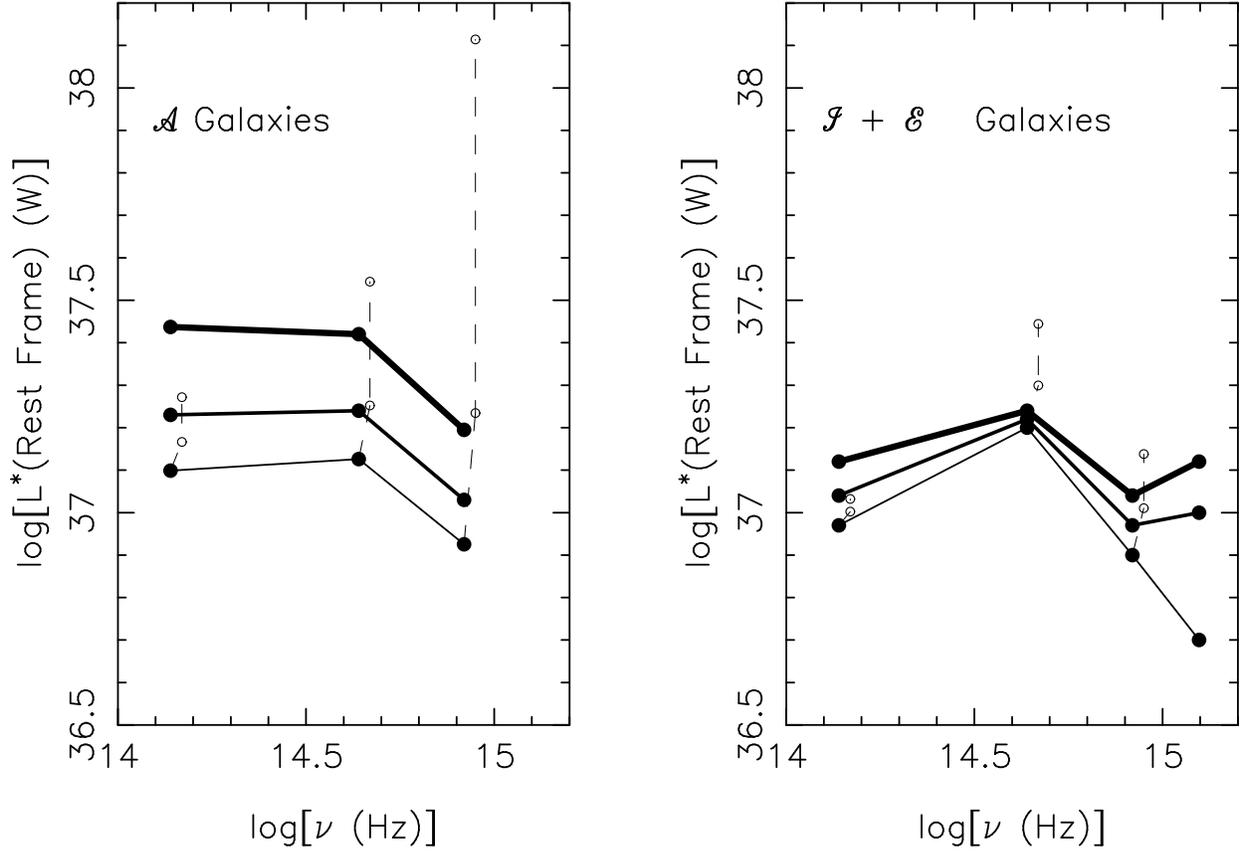}
\caption[figure8.ps]{The evolution of a $L^*$ galaxy in rest
frame $U$, $R$ and $K$ with redshift is shown.  The filled
circles represent the values computed from Table~\ref{table_q}
for absorption line galaxies (left panel) and emission line
galaxies (right panel).  The $\alpha$ determined from rest
frame $R$ is used throughout.  The results of the
three redshift bins covering
the range $0.25<z<1.05$ are shown, with the lines becoming thicker
as the redshift of the bin increases.
The results at rest frame 2400\AA\
are shown only for the emission line galaxies. 
The open circles (connected
by dashed lines) are the predicted evolution for an
elliptical with an initial star formation burst with a duration of 1 Gyr
and an age of 15 Gyr at present and for a Sc galaxy 
with a star formation rate dependent on gas density
as predicted by Poggianti (1997). 
\label{figure_lstar_urk_z}}
\end{figure}

We expect both solutions discussed here, that with $\alpha$ fixed 
given in the main body of this
paper, or that with $L^*$ fixed as given in this appendix,
to be satisfactory fits to the
data. We consider next
whether there is any way we can judge which of these two is more
likely to be valid.  The banana-shaped error contours in the 2D space 
with axes $\alpha,L^*$ (see Figures~\ref{figure_errors_a} to
\ref{figure_errors_all}) form the basis for our treatment.
We proceed to diagonalize the covariance
matrix by defining a new coordinate system which follows the long
major axis of the banana-shaped error contours with a second
axis perpendicular to that.   We measure the difference between the constrained
solution obtained by fixing $\alpha$ or alternatively $L^*$
(i.e. the solutions given in Tables~\ref{table_fixa} and \ref{table_fixlstar})
and the full two parameter solution given in Table~\ref{table2p}
in units of $\sigma$.   To accomplish this, we examine 
each of the error contours
in the $\alpha,L^*$ plane 
and measure $\sigma_l$ and $\sigma_s$ in these ``long'' and ``short''
coordinates.

\begin{figure}
\epsscale{0.7}
% Comment in the following line to embed the postscript figure into the manuscript
%\plotone{error_circles.ps}
\plotone{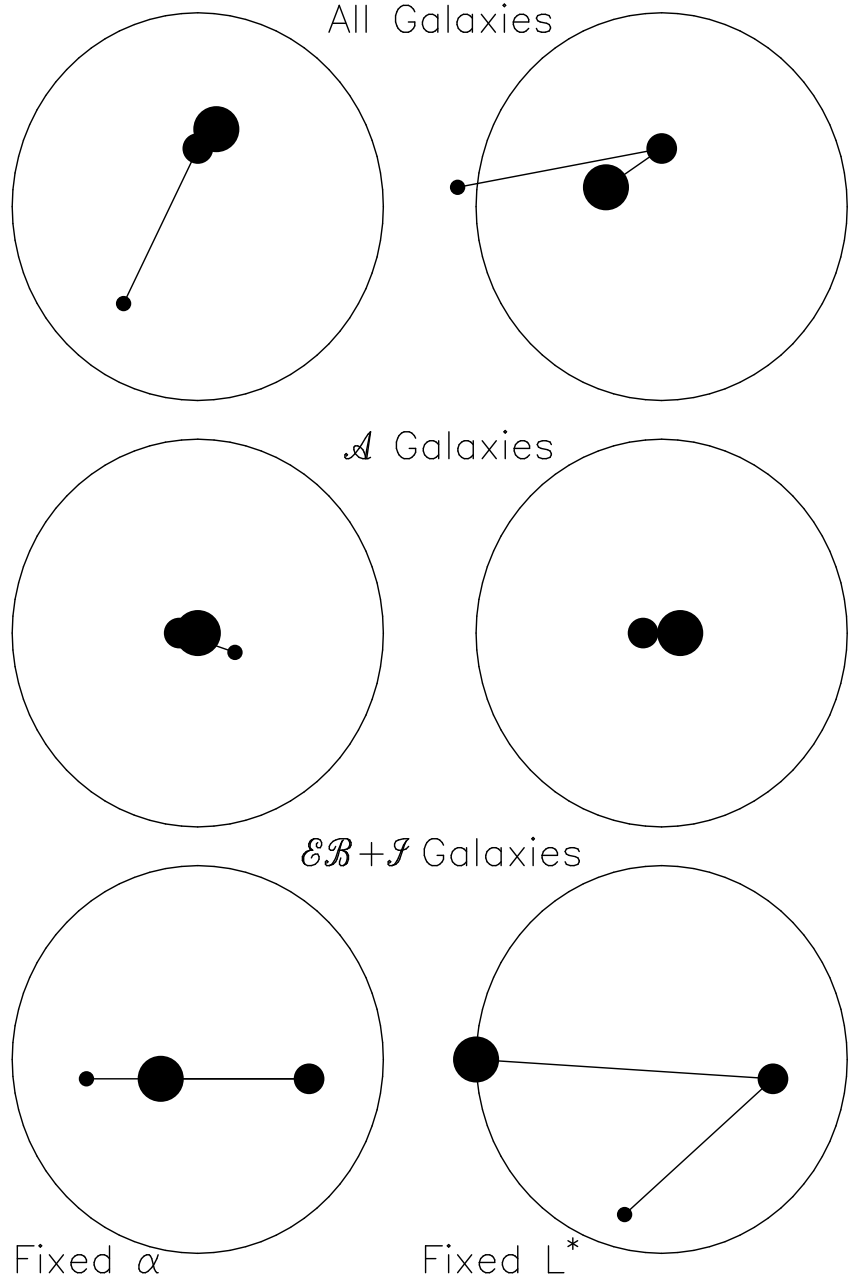}
\caption[figure8.ps]{The diagonalized errors are shown for the
two different assumptions adopted here in the LF solution. 
On the left side are
those obtained when $\alpha$ is fixed as a function of redshift
for each galaxy spectral grouping.  On the right side, those given
in the appendix, where $L^*$ is maintained fixed instead, are displayed.  
The difference between the $(\alpha,L^*)$ found in these solutions
and the values found with the full two parameter solutions of \S\ref{2pshec}
is expressed in units of 1$\sigma$ rms deviations. 
The large circles denotes a total rms difference of $1\sigma$.  
The results are shown for three redshift bins $0.25<z<0.5$,
$0.5<z<0.8$ and $0.8<z<1.05$, as well as for three galaxy spectral groupings. 
The symbol size increases
as the mean redshift of the bin increases.  See text for details.
\label{figure_error_circles}}
\end{figure}

Figure~\ref{figure_error_circles} presents the results for rest-frame $R$.
The large circles represent 1$\sigma$ rms difference.
The horizontal axis is $\sigma_l$, while the vertical axis is $\sigma_s$.
Three redshift bins were used, $0.25<z<0.5$, $0.5<z<0.8$ and
$0.8<z<1.05$, for each of three galaxy spectral groupings. 
The galaxy spectral groupings used are all galaxies, $\cal{A}$ galaxies
and $\cal{E+I}$ galaxies.  The latter two are chosen as they are
believed to be the pair which sum to the whole sample and
which are most likely to be stable in terms
of technical problems over the range of redshift considered
here (see \S\ref{2pshec} and
\ref{section_overall}). On the
left side of the figure are the differences for the solution where
$\alpha$ is fixed, while on the right are the solutions for the
case when $L^*$ is held fixed.

We see that both of these assumptions lead to reasonably good solutions, 
lying within  a difference of less than $1\sigma$ 
(i.e. within the large circles) in most cases.
For the $\cal{A}$ galaxies,
where the sample is very small and the errors are quite large
(see Table~\ref{table_fixa} and \ref{table_fixlstar} for the corresponding
uncertainties in the $\alpha,L^*$ plane), little distinction is
possible.    However, when one looks at the other two spectral
classes presented in this figure, where the samples are much larger,
the solution with fixed $L^*$ has slightly larger deviations
(appearing as a larger mean distance of the points from the center of
the circle) than does the solution with fixed $\alpha$.  We therefore
can have at least some degree of confidence that luminosity
evolution is in fact ocurring in our sample.  Only with a larger, deeper,
better sample will we be able to refine this analysis and this 
potential discriminant.

\clearpage

%
% Table 1
%
\begin{deluxetable}{crrrr}
\tablenum{1}
\tablewidth{0pt}
%\scriptsize
\tablecaption{Completeness at Observed $U$, $R$ and $K$}
\tablehead{\colhead{Mag Range} &  \colhead{N(Phot)(HDF)}
& \colhead{\%($z$)(HDF)} & \colhead{N(Phot)(FF)} & \colhead{\%($z$)(FF)} \nl
}
\startdata
% from /scr2/jlc/hdf_lf/completeness/rcompleteness.doc
% with objects from Dawson, Stern, Spinrad.. added
Obs. $U$ \nl
  17.00 --   17.25 &    0 &  ... &    2 &   100  \nl 
  17.25 --   17.50 &    0 &  ... &    0 &  ...  \nl 
  17.50 --   17.75 &    0 &  ... &    1 &  ...  \nl 
  17.75 --   18.00 &    0 &  ... &    1 &  ...  \nl 
  18.00 --   18.25 &    0 &  ... &    3 &   100  \nl 
  18.25 --   18.50 &    0 &  ... &    1 &  ...  \nl 
  18.50 --   18.75 &    0 &  ... &    1 &  100  \nl 
  18.75 --   19.00 &    0 &  ... &    1 &  100  \nl 
  19.00 --   19.25 &    1 &  100 &    0 &  ...  \nl 
  19.25 --   19.50 &    0 &  ... &    0 &  ...  \nl 
  19.50 --   19.75 &    1 &  100 &    3 &  100  \nl 
  19.75 --   20.00 &    2 &  100 &    1 &  100  \nl 
  20.00 --   20.25 &    1 &  100 &    2 &  100  \nl 
  20.25 --   20.50 &    0 &  ... &    2 &  100  \nl 
  20.50 --   20.75 &    1 &  100 &    3 &  100  \nl 
  20.75 --   21.00 &    0 &  ... &    6 &  100  \nl 
  21.00 --   21.25 &    0 &  ... &    5 &  100  \nl 
  21.25 --   21.50 &    2 &  100 &    3 &  100  \nl 
  21.50 --   21.75 &    2 &  100 &    8 &  100  \nl 
  21.75 --   22.00 &    4 &  100 &   19 &  100  \nl 
  22.00 --   22.25 &    1 &  100 &   17 &  100  \nl 
  22.25 --   22.50 &    5 &  100 &   34 &  100  \nl 
  22.50 --   22.75 &    3 &  100 &   39 &  100  \nl 
  22.75 --   23.00 &    4 &  100 &   45 &  100  \nl 
  23.00 --   23.25 &    9 &  100 &   49 &   95  \nl 
  23.25 --   23.50 &    6 &  100 &   67 &   91  \nl 
  23.50 --   23.75 &    8 &  100 &   57 &   75  \nl 
  23.75 --   24.00 &   20 &   90 &   84 &   66  \nl 
  24.00 --   24.25 &   14 &   78 &  103 &   29  \nl 
  24.25 --   24.50 &   17 &   64 &  135 &   21  \nl 
  24.50 --   24.75 &   32 &   40 &  170 &   15  \nl 
 & \nl
Obs. $R$ \nl
 16.50 --  16.75 & 0 & ... & 0 & ... \nl
 16.75 --  17.00 & 0 & ... & 2 & 100 \nl
 17.00 --  17.25 & 0 & ... & 1 & 100 \nl  
 17.25 --  17.50 & 0 & ... & 2 & 100 \nl 
 17.50 --  17.75 & 0 & ... & 2 & 100 \nl 
 17.75 --  18.00 & 1 & 100 & 1 & 100 \nl   
 18.00 --  18.25 & 0 & ... & 1 & 100 \nl
 18.25 --  18.50 & 0 & ... & 0 & ... \nl 
 18.50 --  18.75 & 2 & 100 & 1 & 100 \nl  
 18.75 --  19.00 & 0 & ... & 3 & 100 \nl 
 19.00 --  19.25 & 1 & 100 & 3 & 100 \nl 
 19.25 --  19.50 & 2 & 100 & 3 & 100 \nl  
 19.50 --  19.75 & 0 & ... & 4 & 100 \nl 
 19.75 --  20.00 & 1 & 100 & 5 & 100 \nl 
 20.00 --  20.25 & 1 & 100 & 6 & 100 \nl 
 20.25 --  20.50 & 3 & 100 & 11 & 100 \nl 
 20.50 --  20.75 & 3 & 100 & 29 & 100 \nl
 20.75 --  21.00 & 4 & 100 & 19 & 100 \nl   
 21.00 --  21.25 & 3 & 100 & 23 & 100 \nl 
 21.25 --  21.50 & 6 & 100 & 27 & 96 \nl 
 21.50 --  21.75 & 3 & 100 & 35 & 97 \nl 
 21.75 --  22.00 & 7 & 100 & 38 & 95 \nl 
 22.00 --  22.25 & 2 & 100 & 44 & 93 \nl  
 22.25 --  22.50 & 4 & 100 & 51 & 100 \nl  
 22.50 --  22.75 & 7 & 100 & 56 & 89 \nl  
 22.75 --  23.00 & 11 & 100 & 66 & 79 \nl  
 23.00 --  23.25 & 13 & 92 & 83 & 63 \nl    
 23.25 --  23.50 & 13 & 85 & 94 & 38 \nl  
 23.50 --  23.75 & 13 & 92 & 93 & 27 \nl 
 23.75 --  24.00 & 15 & 93 & 126 & 15 \nl
  &  & \nl
Obs. $K$ \nl
  14.75 --   15.00 &    0 & ...  &  2 & 100 \nl 
  15.00 --   15.25 &    0  & ... & 1 &  100 \nl 
  15.25 --   15.50 &     0 & ... & 2 & 100  \nl 
  15.50 --   15.75 &     1 & 100 & 0 & ... \nl 
  15.75 --   16.00 &     0 & ... &  2 & 100  \nl 
  16.00 --   16.25 &   0 & ... &  1 &  100  \nl 
  16.25 --   16.50 &     0 & ...  &  3 & 100  \nl 
  16.50 --   16.75 &     2 &   100 &  8 & 100  \nl 
  16.75 --   17.00 &     2 & 100 & 9 &   100 \nl 
  17.00 --   17.25 &    1 &  100 &  12 & 91 \nl 
  17.25 --   17.50 &   4 & 100 &   16 &  100  \nl 
  17.50 --   17.75 &    4 & 100 &   20 & 100 \nl 
  17.75 --   18.00 &    4 & 100 &  15 &  86  \nl 
  18.00 --   18.25 &    7  & 100 &  32 &   93  \nl 
  18.25 --   18.50 &     3 &  100 & 34 & 100  \nl 
  18.50 --   18.75 &    4 & 100 &  30 & 100  \nl 
  18.75 --   19.00 &     3 &  100 & 36 & 88  \nl 
  19.00 --   19.25 &      8 &  100 & 39 &  89  \nl 
  19.25 --   19.50 &      6 &  83\tablenotemark{a} & 48 & 100  \nl 
  19.50 --   19.75 &     1 &  100 & 47 & 68  \nl 
  19.75 --   20.00 &    11 &  100 & 54 &  72  \nl 
  20.00 --   20.25 &    8 &  100 &  76 & 47  \nl 
  20.25 --   20.50 &    10 & 100  & 106 &  41  \nl 
  20.50 --   20.75 &    13 &  100 & 120 & 25  \nl 
  20.75 --   21.00 &    13 &  100 & 110 & 19  \nl 
  21.00 --   21.25 &    15 &  40 &  ... & ... \nl 
  21.25 --   21.50 &     20 & 20 & ... & ... \nl 
\enddata
\tablenotetext{a}{This galaxy missing a spectroscopic redshift
is very red, though not red enough to
be considered an ERO, and is slightly fainter than the $R=24.0$ cutoff 
adopted for the HDF redshift survey.}
\label{table_complete}
\end{deluxetable}

%
% Table 2
%
\begin{deluxetable}{crr}
\tablenum{2}
\tablewidth{0pt}
%\scriptsize
\tablecaption{Properties of the Redshift Bins}
\tablehead{\colhead{$z$ Range} & \colhead{Mean $z$} & Co-moving Volume \nl
\colhead{} & \colhead{} & \colhead{(Mpc$^3$)} \nl
}
\startdata
0.01 -- 0.25 & 0.15 & 39 \nl
0.25 -- 0.5  &   0.41 & 183 \nl
0.5 -- 0.8 &     0.60 & 429 \nl
0.8 -- 1.05 &      0.90 & 487 \nl
1.05 -- 1.5  &     1.22 & 1206 \nl
\enddata
\label{table_zmean}
\end{deluxetable}

%
% Table 3
%
\begin{deluxetable}{crrrrr}
\tablenum{3}
\tablewidth{0pt}
%\scriptsize
\tablecaption{Solutions for $L^*(R)$ and $\alpha$}
\tablehead{\colhead{$z$ Range} & \colhead{Total No.} & \colhead{No. in HDF} &
\colhead{No. in} & \colhead{log[$L^*(R)$]} & \colhead{$\alpha$}  \nl
\colhead{} &   \colhead{} & \colhead{} & \colhead{Flanking Fields} & 
\colhead{(W)} \nl
}
\startdata
$\cal{A}$ Galaxies \nl
0.25 -- 0.5  &  17 & 1 & 16 & $37.15\pm0.45$ & $-0.73\pm0.60$ \nl
0.5 -- 0.8 &    36 & 6 & 30 & $37.19\pm0.30$ & $-0.43\pm0.50$ \nl
0.8 -- 1.05 &   19 & 7 & 12 & $37.32\pm0.50$ & $-0.40\pm1.00$ \nl
   &  \nl
$\cal{I}$ Galaxies \nl
0.25 -- 0.5  &  62 & 7 & 55 & $36.73\pm0.25$ & $-0.19\pm0.40$ \nl
0.5 -- 0.8 &    63 & 8 & 55 & $36.85\pm0.25$ & $+0.08\pm0.55$ \nl
0.8 -- 1.05 & 14 & 2 & 12 & ... & ... \nl
    &   \nl
$\cal{A}$+$\cal{I}$ Galaxies \nl
0.25 -- 0.5  &  79 & 8 & 71 & $36.83\pm0.25$ & $-0.35\pm0.35$ \nl
0.5 -- 0.8 &    99 & 14 & 85 & $37.00\pm0.20$ & $-0.19\pm0.40$ \nl
0.8 -- 1.05 &   33 & 9 & 24 & $37.16\pm0.45$ & $+0.11\pm1.00$ \nl

$\cal{I}$+$\cal{E}$ Galaxies \nl
0.25 -- 0.5  & 131 &  23 & 108 & $37.00\pm0.35$ & $-1.07\pm0.20$ \nl
0.5 -- 0.8 &   164 & 23 & 141 & $37.39\pm0.35$ & $-1.39\pm0.20$ \nl
0.8 -- 1.05 &  125 & 18 & 107 & $37.20\pm0.25$ & $-1.19\pm0.40$ \nl
   &  \nl
$\cal{E}$ Galaxies \nl
0.25 -- 0.5  &  69 & 16 & 53 & $36.75\pm0.40$ & $-1.50\pm0.25$ \nl
0.5 -- 0.8 &   101 & 15 & 86 & $37.19\pm0.45$ & $-1.71\pm0.25$ \nl
0.8 -- 1.05 &  111 & 16 & 95 & $37.05\pm0.20$ & $-1.09\pm0.60$ \nl
  &  \nl
All Galaxies \nl
0.01 -- 0.25 & 37 & 6 & 31 & $>36.70\pm0.60$\tablenotemark{a} & 
      $-1.06\pm0.25$ \nl
0.25 -- 0.5  & 148 & 24 & 124 & $37.13\pm0.25$ & $-1.08\pm0.15$ \nl
0.5 -- 0.8 &   200 & 29 & 171 & $37.47\pm0.25$ & $-1.28\pm0.25$ \nl
0.8 -- 1.05 &  144 & 25 & 119 & $37.37\pm0.25$ & $-1.28\pm0.30$ \nl
1.05 -- 1.50 & 18\tablenotemark{b} & 4 & 14 & $37.30\pm0.40$ & 
     $-0.45\pm1.30$ \nl
\enddata
\tablenotetext{a}{Assigned as a lower limit due to selection criteria for HDF.}
\tablenotetext{b}{Six galaxies in the HDF with $23.5 \le R \le 24.0$
had to be eliminated.  See text for details.}
\label{table2p}
\end{deluxetable}

%
% Table 4
%
\begin{deluxetable}{cr rrrr}
\tablenum{4}
\tablewidth{0pt}
%\scriptsize
\tablecaption{Mean $\alpha$ Adopted for Each Galaxy Spectral Grouping}
\tablehead{\colhead{Spectral Group} & \colhead{Typical} & 
\colhead{Mean $\alpha$} & \colhead{Mean $\alpha$} & \colhead{Mean $\alpha$}
& \colhead{Mean $\alpha$} \nl
\colhead{} & \colhead{N(Gal)} & 
 \colhead{(Rest Fr. $U$)} &
\colhead{(Rest Fr. $R$)} & \colhead{(Rest Fr. $K$)} &
\colhead{(Rest Fr. 2400\AA)}   \nl
\colhead{$z$-Range} & \colhead{($0.25-1.05$)} & \colhead{($0.25-1.05$)} &
\colhead{($0.25-1.05$)} & \colhead{($0.25-1.05$)}
& \colhead{($0.25-0.8$)}  \nl
}
\startdata
$\cal{A}$  &  70 &  $-0.60\pm0.45$ & $-0.50\pm0.40$  & $-0.10\pm0.30$ & ...  \nl
$\cal{I}$ &   140  & $+0.30\pm0.35$ & $-0.05\pm0.35$ & $-0.20\pm0.25$ 
         & ($-0.55\pm0.30$) \nl
$\cal{A+I}$ & 210 & $-0.25\pm0.25$ & $-0.20\pm0.30$ & $-0.30\pm0.20$ & 
        ($-0.60\pm0.25$) \nl
$\cal{I+E}$ & 385  & $-1.05\pm0.20$ &  $-1.25\pm0.15$ & $-1.20\pm0.15$ & 
         ($-0.60\pm0.20$) \nl
$\cal{E}$ & 245 &  $-1.30\pm0.25$ & $-1.45\pm0.25$ & $-1.35\pm0.20$ & 
         ($-0.60\pm0.25$) \nl
All & 495  & $-1.15\pm0.15$ & $-1.25\pm0.15$ & $-1.15\pm0.10$ & 
         ($-0.60\pm0.15$) \nl
\enddata
\label{table_meana}
\end{deluxetable}

%
% Table 5
%
\begin{deluxetable}{crrrrrr}
\tablenum{5}
\tablewidth{0pt}
%\scriptsize
\tablecaption{Rest Frame $R$ LF Solution For 
Fixed $\alpha$ For Each Galaxy Spectral Grouping}
\tablehead{\colhead{$z$ Range} & \colhead{Total No.} & \colhead{No.} &
\colhead{No.} & \colhead{log[$L^*(R)$]} & \colhead{$\alpha$} &
\colhead{$\phi^*$} \nl
\colhead{} & \colhead{} & \colhead{(in HDF)} & \colhead{(in Flanking Fields)} & 
\colhead{(log(W))} & \colhead{(fixed)} & \colhead{($Mpc^{-3}$)} \nl
}
\startdata
 $\cal{A}$ Galaxies \nl
0.25 -- 0.5  &  17 & 1 & 16 & 36.92 & $-$0.50 & 0.0028(1.63)\tablenotemark{b} \nl
0.5 -- 0.8 &    36 & 6 & 30 & 37.24 & $-$0.50 & 0.0033(1.38) \nl
0.8 -- 1.05 &   19 & 7 & 12 & 37.37 & $-$0.50 & 0.0019(1.43) \nl
   & \nl
$\cal{I}$ Galaxies \nl
0.25 -- 0.5  &  62 & 7 & 55 & 36.62 & $-$0.05 & 0.014(1.22) \nl
0.5 -- 0.8 &    63 & 8 & 55 & 36.86 & $-$0.05 & 0.0066(1.18) \nl
0.8 -- 1.05 & 14 & 2 & 12 & ... & ... \nl
   &  \nl
$\cal{A}$+$\cal{I}$ Galaxies \nl
0.25 -- 0.5  &  79 & 8 & 71 & 36.75 & $-$0.20 & 0.017(1.25) \nl
0.5 -- 0.8 &    99 & 14 & 85 & 37.02 & $-$0.20 & 0.011(1.33) \nl
0.8 -- 1.05 &   33 & 9 & 24 & 37.25 & $-$0.20 & 0.0036(1.33) \nl
  &  \nl
$\cal{I}$+$\cal{E}$ Galaxies \nl
0.25 -- 0.5  & 131 & 23 & 108 & 37.30 & $-$1.25 & 0.0066(1.47)\nl
0.5 -- 0.8 &   164 & 23 & 141 & 37.15 & $-$1.25 & 0.0066(1.54) \nl
0.8 -- 1.05 &  125 & 18 & 107 & 37.25 & $-$1.25 & 0.0095(1.54) \nl
  &  \nl
$\cal{E}$ Galaxies \nl
0.25 -- 0.5  &  69 & 16 & 53 & 36.71 & $-$1.45 & 0.0061(1.90) \nl
0.5 -- 0.8 &   101 & 15 & 86 & 36.85 & $-$1.45 & 0.0052(1.50) \nl
0.8 -- 1.05 &  111 & 16 & 95 & 37.32 & $-$1.45 & 0.0062(1.50) \nl
   &   \nl
All Galaxies & \nl
0.01 -- 0.25 & 37 & 6 & 31 & $>$36.9 &  $-$1.25 & $>$0.0044 \nl
0.25 -- 0.5  & 148 & 24 & 124 & 37.28 & $-$1.25 & 0.0073(1.50) \nl
0.5 -- 0.8 &   200 & 29 & 171 & 37.43 & $-$1.25 & 0.0058(1.54) \nl
0.8 -- 1.05 &  144 & 25 & 119 & 37.35 & $-$1.25 & 0.0094(1.50) \nl
1.05 -- 1.50 & 18 & 4 & 14 & 37.50\tablenotemark{a} & $-$1.25 & 0.00054 \nl
\tablenotetext{a}{$L^*$ could be anywhere between 37.3 and 37.7
depending on how the extrapolation for fixed $\alpha$ is made.}
\tablenotetext{b}{These uncertainties are given in the form a 
a factor by which $\phi*$ should be multiplied or divided to obtain
the 1$\sigma$ range for $\phi^*$.  The errors are calculated
allowing both $\alpha$ and $L^*$ to vary.}
\enddata
\label{table_fixa}
\end{deluxetable}

%
% Table 6
%
\begin{deluxetable}{crrrrr}
\tablenum{6}
\tablewidth{0pt}
%\scriptsize
\tablecaption{Solutions for $L^*(U)$ and $\alpha$}
\tablehead{\colhead{$z$ Range} & \colhead{Total No.} & \colhead{No. in HDF} &
\colhead{No. in} & \colhead{log[$L^*(U)$]} & \colhead{$\alpha$}  \nl
\colhead{} &   \colhead{} & \colhead{} & \colhead{Flanking Fields} & 
\colhead{(W)} \nl
}
\startdata
$\cal{A}$ Galaxies \nl
0.25 -- 0.5  &  17 & 1 & 16 & $37.19\pm0.60$ & $-0.84\pm0.50$ \nl
0.5 -- 0.8 &    36 & 6 & 30 & $37.15\pm0.40$ & $-0.66\pm0.50$ \nl
0.8 -- 1.05\ &   19 & 7 & 12 & $37.13\pm0.30$ & $-0.21\pm1.30$ \nl
   &  \nl
$\cal{I}$ Galaxies \nl
0.25 -- 0.5  &  62 & 7 & 55 & $36.42\pm0.25$ & $+0.30\pm0.50$ \nl
0.5 -- 0.8 &    63 & 8 & 55 & $36.60\pm0.20$ & $+0.24\pm0.55$ \nl
    &   \nl
$\cal{A}$+$\cal{I}$ Galaxies \nl
0.25 -- 0.5  &  79 & 8 & 71 & $36.62\pm0.20$ & $-0.24\pm0.35$ \nl
0.5 -- 0.8 &    99 & 14 & 85 & $36.85\pm0.15$ & $-0.30\pm0.40$ \nl
0.8 -- 1.05 &   33 & 9 & 24 & $37.02\pm0.35$ & $+0.00\pm0.80$ \nl

$\cal{I}$+$\cal{E}$ Galaxies \nl
0.25 -- 0.5  & 131 &  23 & 108 & $36.74\pm0.25$ & $-1.03\pm0.20$ \nl
0.5 -- 0.8 &   164 & 23 & 141 & $36.99\pm0.30$ & $-1.27\pm0.25$ \nl
0.8 -- 1.05 &  125 & 18 & 107 & $36.85\pm0.25$ & $-0.85\pm0.45$ \nl
   &  \nl
$\cal{E}$ Galaxies \nl
0.25 -- 0.5  &  69 & 16 & 53 & $36.70\pm0.45$ & $-1.51\pm0.25$ \nl
0.5 -- 0.8 &   101 & 15 & 86 & $37.03\pm0.65$ & $-1.71\pm0.20$ \nl
0.8 -- 1.05 &  111 & 16 & 95 & $36.76\pm0.25$ & $-0.85\pm0.55$ \nl
  &  \nl
All Galaxies \nl
0.01 -- 0.25 & 37 & 6 & 31 & $>36.58\pm0.40$\tablenotemark{a} & $-0.90\pm0.20$ \nl
0.25 -- 0.5  & 148 & 24 & 124 & $36.90\pm0.20$ & $-1.09\pm0.15$ \nl
0.5 -- 0.8 &   200 & 29 & 171 & $37.17\pm0.20$ & $-1.22\pm0.20$ \nl
0.8 -- 1.05 &  144 & 25 & 119 & $37.07\pm0.40$ & $-1.12\pm0.35$ \nl
1.05 -- 1.50 & 18\tablenotemark{b} & 4 & 14 & $37.41\pm0.50$ & $-0.73\pm1.20$ \nl
\enddata
\tablenotetext{a}{Assigned as a lower limit due to selection criteria for HDF.}
\tablenotetext{b}{Galaxies in the HDF with $23.5<R<24$ were eliminated.}
\label{table2p_u}
\end{deluxetable}

%
% Table 7
%
\begin{deluxetable}{crrrrr}
\tablenum{7}
\tablewidth{0pt}
%\scriptsize
\tablecaption{Solutions for $L^*(K)$ and $\alpha$}
\tablehead{\colhead{$z$ Range} & \colhead{Total No.} & \colhead{No. in} &
\colhead{No. in} & \colhead{log[$L^*(K)$]} & \colhead{$\alpha$}  \nl
\colhead{} &   \colhead{} & \colhead{HDF} & \colhead{Flanking Fields} & 
\colhead{(W)} \nl
}
\startdata
$\cal{A}$ Galaxies \nl
0.25 -- 0.5  &  15 & 0 & 15 & $37.00\pm0.50$\tablenotemark{a} & 
           $-0.57\pm0.40$\tablenotemark{a} \nl
0.5 -- 0.8 &    36 & 6 & 30 & $36.90\pm0.25$ & $+0.07\pm0.40$ \nl
0.8 -- 1.05 &   19 & 7 & 12 & $37.15\pm0.45$ & $-0.07\pm0.70$ \nl
   &  \nl
$\cal{I}$ Galaxies \nl
0.25 -- 0.5  &  60 & 7 & 53 & $36.75\pm0.25$ & $-0.60\pm0.25$ \nl
0.5 -- 0.8 &    62 & 7 & 55 & $36.70\pm0.25$ & $-0.21\pm0.40$ \nl
0.8 -- 1.05 & 16 & 2 & 14 & $36.70\pm0.50$ & $+1.20\pm1.20$ \nl
    &   \nl
$\cal{A}$+$\cal{I}$ Galaxies \nl
0.25 -- 0.5  &  75 & 7 & 68 & $36.80\pm0.25$ & $-0.60\pm0.25$ \nl
0.5 -- 0.8 &    98 & 13 & 85 & $36.85\pm0.20$ & $-0.21\pm0.30$ \nl
0.8 -- 1.05 &   35 & 9 & 26 & $37.00\pm0.30$ & $+0.22\pm0.60$ \nl

$\cal{I}$+$\cal{E}$ Galaxies \nl
0.25 -- 0.5  & 115 &  22 & 93 & $37.10\pm0.40$ & $-1.25\pm0.10$ \nl
0.5 -- 0.8 &   147 & 17 & 130 & $37.00\pm0.25$ & $-1.23\pm0.20$ \nl
0.8 -- 1.05 &  121 & 15 & 106 & $37.10\pm0.30$ & $-1.13\pm0.30$ \nl
   &  \nl
$\cal{E}$ Galaxies \nl
0.25 -- 0.5  &  55 & 15 & 40 & $37.00\pm0.60$ & $-1.68\pm0.20$ \nl
0.5 -- 0.8 &   85 & 10 & 75 & $36.40\pm0.40$ & $-1.26\pm0.35$ \nl
0.8 -- 1.05 &  105 & 13 & 92 & $37.00\pm0.35$ & $-1.20\pm0.30$ \nl
  &  \nl
All Galaxies \nl
0.01 -- 0.25 & 30 & 5 & 25 & $>36.40\pm0.40$\tablenotemark{b} & 
         $-0.81\pm0.15$ \nl
0.25 -- 0.5  & 130 & 22 & 108 & $37.10\pm0.30$ & $-1.21\pm0.15$ \nl
0.5 -- 0.8 &   183 & 23 & 160 & $37.20\pm0.25$ & $-1.16\pm0.15$ \nl
0.8 -- 1.05 &  140 & 22 & 118 & $37.30\pm0.30$ & $-1.14\pm0.25$ \nl
1.05 -- 1.50 & 27\tablenotemark{c} & 10 & 17 & $37.25\pm0.40$ & $-1.04\pm0.70$ \nl
\enddata
\tablenotetext{a}{These are 1 $\sigma$ one parameter errors throughout.  See
figures~\ref{figure_errors_a} through \ref{figure_errors_all} for the
two parameter error contours.}
\tablenotetext{b}{Assigned as a lower limit due to selection criteria for HDF.}
\tablenotetext{c}{One galaxy in the HDF with weight greater than 15
had to be eliminated.  See text for details.}
\label{table2p_k}
\end{deluxetable}

%
% Table 8
%
\begin{deluxetable}{crrrrrr}
\tablenum{8}
\tablewidth{0pt}
%\scriptsize
\tablecaption{Rest Frame $U$ LF Solution For 
Fixed $\alpha$ For Each Galaxy Spectral Grouping}
\tablehead{\colhead{$z$ Range} & \colhead{Total No.} & \colhead{No.} &
\colhead{No.} & \colhead{log[$L^*(U)$]} & \colhead{$\alpha$} &
\colhead{$\phi^*$} \nl
\colhead{} & \colhead{} & \colhead{(in HDF)} & \colhead{(in Flanking Fields)} & 
\colhead{(log(W))} & \colhead{(fixed)} & \colhead{($Mpc^{-3}$)} \nl
}
\startdata
 $\cal{A}$ Galaxies \nl
0.25 -- 0.5  &  17 & 1 & 16 & 36.89 & $-$0.60 & 0.241E-2 \nl
0.5 -- 0.8 &    36 & 6 & 30 & 37.10 & $-$0.60 & 0.282E-2 \nl
0.8 -- 1.05 &   19 & 7 & 12 & 37.30 & $-$0.60 & 0.164E-2 \nl
   & \nl
$\cal{I}$ Galaxies \nl
0.25 -- 0.5  &  62 & 7 & 55 & 36.42 & +0.30 & 0.144E-1 \nl
0.5 -- 0.8 &    63 & 8 & 55 & 36.56 & +0.30 & 0.655E-2 \nl
   &  \nl
$\cal{A}$+$\cal{I}$ Galaxies \nl
0.25 -- 0.5  &  79 & 8 & 71 & 36.62 & $-$0.25 & 0.159E-1\nl
0.5 -- 0.8 &    99 & 14 & 85 & 36.82 & $-$0.25 & 0.978E-2 \nl
0.8 -- 1.05 &   33 & 9 & 24 & 37.11 & $-$0.25 & 0.340E-2 \nl
  &  \nl
$\cal{I}$+$\cal{E}$ Galaxies \nl
0.25 -- 0.5  & 131 & 23 & 108 & 36.78 & $-$1.05 & 0.126E-1\nl
0.5 -- 0.8 &   164 & 23 & 141 & 36.78 & $-$1.05 & 0.107E-1 \nl
0.8 -- 1.05 &  125 & 18 & 107 & 36.92 & $-$1.05 & 0.130E-1 \nl
  &  \nl
$\cal{E}$ Galaxies \nl
0.25 -- 0.5  &  69 & 16 & 53 & 36.45 & $-$1.30 & 0.744E-2 \nl
0.5 -- 0.8 &   101 & 15 & 86 & 36.60 & $-$1.30 & 0.729E-2 \nl
0.8 -- 1.05 &  111 & 16 & 95 & 37.00 & $-$1.30 & 0.863E-2 \nl
   &   \nl
All Galaxies & \nl
0.01 -- 0.25 & 37 & 6 & 31 & $>$37.15 &  $-$1.15 & $>$0.568E-2 \nl
0.25 -- 0.5  & 148 & 24 & 124 & 37.03 & $-$1.15 & 0.929E-2 \nl
0.5 -- 0.8 &   200 & 29 & 171 & 37.05 & $-$1.15 & 0.834E-2 \nl
0.8 -- 1.05 &  144 & 25 & 119 & 37.05 & $-$1.15 & 0.118E-1 \nl
1.05 -- 1.50 & 18 & 4 & 14 & 37.65\tablenotemark{a} & $-$1.15 & 
0.445E-3\tablenotemark{a} \nl
\enddata
\tablenotetext{a}{Extremely uncertain values.}
\label{table_fixa_u}
\end{deluxetable}

%
% Table 9
%
\begin{deluxetable}{crrrrrr}
\tablenum{9}
\tablewidth{0pt}
%\scriptsize
\tablecaption{Rest Frame $K$ LF Solution For 
Fixed $\alpha$ For Each Galaxy Spectral Grouping}
\tablehead{\colhead{$z$ Range} & \colhead{Total No.} & \colhead{No.} &
\colhead{No.} & \colhead{log[$L^*(K)$]} & \colhead{$\alpha$} &
\colhead{$\phi^*$} \nl
\colhead{} & \colhead{} & \colhead{(in HDF)} & \colhead{(in Flanking Fields)} & 
\colhead{(log(W))} & \colhead{(fixed)} & \colhead{($Mpc^{-3}$)} \nl
}
\startdata
 $\cal{A}$ Galaxies \nl
0.25 -- 0.5  &  15 & 0 & 15 & 36.50 & $-$0.10 & 0.374E-2 \nl
0.5 -- 0.8 &    36 & 6 & 30 & 37.00 & $-$0.10 & 0.336E-2 \nl
0.8 -- 1.05 &   19 & 7 & 12 & 37.15 & $-$0.10 & 0.164E-2 \nl
   & \nl
$\cal{I}$ Galaxies \nl
0.25 -- 0.5  &  60 & 7 & 53 & 36.45 & $-$0.20 & 0.138E-1 \nl
0.5 -- 0.8 &    62 & 5 & 55 & 36.70 & $-$0.20 & 0.623E-2 \nl
0.8 -- 1.05 & 16 & 2 & 14 & 37.10 & $-$0.20 & 0.130E-2 \nl
   &  \nl
$\cal{A}$+$\cal{I}$ Galaxies \nl
0.25 -- 0.5  &  75 & 7 & 68 & 36.50 & $-$0.30 & 0.164E-1\nl
0.5 -- 0.8 &    98 & 13 & 85 & 36.90 & $-$0.30 & 0.875E-2 \nl
0.8 -- 1.05 &   35 & 9 & 26 & 37.25 & $-$0.30 & 0.256E-2 \nl
  &  \nl
$\cal{I}$+$\cal{E}$ Galaxies \nl
0.25 -- 0.5  & 115 & 22 & 93 & 36.95 & $-$1.20 & 0.698E-2\nl
0.5 -- 0.8 &   147 & 17 & 130 & 36.95 & $-$1.20 & 0.540E-2 \nl
0.8 -- 1.05 &  121 & 15 & 106 & 37.20 & $-$1.20 & 0.435E-2 \nl
  &  \nl
$\cal{E}$ Galaxies \nl
0.25 -- 0.5  &  55 & 15 & 40 & 36.30 & $-$1.35 & 0.640E-2 \nl
0.5 -- 0.8 &   85 & 10 & 75 & 36.50 & $-$1.35 & 0.496E-2 \nl
0.8 -- 1.05 &  105 & 13 & 92 & 37.20 & $-$1.35 & 0.299E-2 \nl
   &   \nl
All Galaxies & \nl
0.01 -- 0.25 & 30 & 5 & 25 & $>$37.0 &  $-$1.15 & $>$0.414E-2 \nl
0.25 -- 0.5  & 130 & 22 & 108 & 37.00 & $-$1.15 & 0.840E-2 \nl
0.5 -- 0.8 &   183 & 23 & 160 & 37.20 & $-$1.15 & 0.559E-2 \nl
0.8 -- 1.05 &  140 & 22 & 118 & 37.30 & $-$1.15 & 0.481E-2 \nl
1.05 -- 1.50 & 27 & 10 & 17 & 37.40 & $-$1.15 & 0.378E-3 \nl
\enddata
\label{table_fixa_k}
\end{deluxetable}

%
% Table 10
%
\begin{deluxetable}{crrrrr}
\tablenum{10}
\tablewidth{0pt}
%\scriptsize
\tablecaption{Solutions for $L^*(2400\AA)$ and $\alpha$}
\tablehead{\colhead{$z$ Range} & \colhead{Total No.} & \colhead{No. in} &
\colhead{No. in} & \colhead{log[$L^*(2400\AA)$]} & \colhead{$\alpha$}  \nl
\colhead{} &   \colhead{} & \colhead{HDF} & \colhead{Flanking Fields} & 
\colhead{(W)} \nl
}
\startdata
$\cal{A}$ Galaxies \nl
0.25 -- 0.5  &  8 & 1 & 7 & ... & ... \nl
0.5 -- 0.8 &   15 & 3 & 13 & $36.40\pm0.40$\tablenotemark{a} 
        & $-1.00\pm0.70$\tablenotemark{a}  \nl
0.8 -- 1.05 & 5 & 1 & 4 &  ... & ... \nl
   &  \nl
$\cal{I}$ Galaxies \nl
0.25 -- 0.5  &  56 & 8 & 48 & $36.12\pm0.25$ & $-0.28\pm0.45$ \nl
0.5 -- 0.8 &    51 & 8 & 44 & $36.65\pm0.30$ & $-0.87\pm0.35$ \nl
0.8 -- 1.05 & 10 & 1 & 9 & ... & ... \nl
    &   \nl
$\cal{A}$+$\cal{I}$ Galaxies \nl
0.25 -- 0.5  &  64 & 9 & 55 & $36.10\pm0.20$ & $-0.27\pm0.35$ \nl
0.5 -- 0.8 &    66 & 11 & 57 & $36.62\pm0.35$ & $-0.93\pm0.40$ \nl
0.8 -- 1.05 &   15 & 2 & 13 & $36.19\pm0.30$ & $+1.94\pm1.20$ \nl

$\cal{I}$+$\cal{E}$ Galaxies \nl
0.25 -- 0.5  & 121 &  25 & 96 & $36.05\pm0.15$ & $-0.33\pm0.30$ \nl
0.5 -- 0.8 &   140 & 23 & 120 & $36.51\pm0.20$ & $-0.76\pm0.30$ \nl
0.8 -- 1.05 &  110 & 18 & 95 & $36.25\pm0.20$ & $+1.08\pm0.65$ \nl
   &  \nl
$\cal{E}$ Galaxies \nl
0.25 -- 0.5  &  65 & 17 & 48 & $36.00\pm0.25$ & $-0.44\pm0.40$ \nl
0.5 -- 0.8 &   89 & 15 & 76 & $36.42\pm0.20$ & $-0.70\pm0.35$ \nl
0.8 -- 1.05 &  100 & 15 & 76 & $36.25\pm0.10$ & $+1.14\pm0.60$ \nl
  &  \nl
All Galaxies \nl
0.01 -- 0.25 & ...\tablenotemark{b} &  ... \tablenotemark{b} \nl
0.25 -- 0.5  & 129 & 26 & 103 & $35.86\pm0.15$ & $-0.43\pm0.25$ \nl
0.5 -- 0.8 &   155 & 26 & 133 & $36.49\pm0.15$ & $-0.77\pm0.20$ \nl
0.8 -- 1.05 &  115 & 19 & 99 & $36.25\pm0.10$ & $+1.14\pm0.70$ \nl
1.05 -- 1.50 & 22 & 12 & 12 & $36.55\pm0.35$ & 
     $-0.17\pm1.00$ \nl
\enddata
\tablenotetext{a}{These are 1 $\sigma$ one parameter errors throughout.  See
figures~\ref{figure_errors_a} through \ref{figure_errors_all} for the
two parameter error contours.}
\tablenotetext{b}{Not observable from the ground at $0.01<z<0.25$.}
\label{table2p_q}
\end{deluxetable}

%
% Table 11
%
\begin{deluxetable}{crrrrrr}
\tablenum{11}
\tablewidth{0pt}
%\scriptsize
\tablecaption{Rest Frame 2400\AA\ LF Solution For 
Fixed $\alpha$ For Each Galaxy Spectral Grouping}
\tablehead{\colhead{$z$ Range} & \colhead{Total No.} & \colhead{No.} &
\colhead{No.} & \colhead{log[$L^*(2400\AA)$]} & \colhead{$\alpha$} &
\colhead{$\phi^*$} \nl
\colhead{} & \colhead{} & \colhead{(in HDF)} & \colhead{(in Flanking Fields)} & 
\colhead{(log(W))} & \colhead{(fixed)} & \colhead{($Mpc^{-3}$)} \nl
}
\startdata
$\cal{I}$ Galaxies \nl
0.25 -- 0.5  &  56 & 8 & 48 & 36.28 & $-$0.55 & 0.118E-1 \nl
0.5 -- 0.8 &    51 & 8 & 44 & 36.44 & $-$0.55 & 0.624E-2 \nl
0.8 -- 1.05 & 12 & 1 & 12 & ... & ... \nl
   &  \nl
$\cal{A}$+$\cal{I}$ Galaxies \nl
0.25 -- 0.5  &  64 & 9 & 55 & 36.29 & $-$0.60 & 0.126E-1\nl
0.5 -- 0.8 &    66 & 11 & 57 & 36.40 & $-$0.60 & 0.842E-2 \nl
0.8 -- 1.05 &   15 & 2 & 14 & 36.85 & $-$0.60 & 0.124E-2 \nl
  &  \nl
$\cal{I}$+$\cal{E}$ Galaxies \nl
0.25 -- 0.5  & 121 & 25 & 96 & 36.17 & $-$0.55 & 0.291E-1\nl
0.5 -- 0.8 &  140 & 23 & 120 & 36.39 & $-$0.55 & 0.191E-1 \nl
0.8 -- 1.05 &  105 & 18 & 89 & 36.60 & $-$0.55 & 0.178E-1 \nl
  &  \nl
$\cal{E}$ Galaxies \nl
0.25 -- 0.5  &  65 & 17 & 48 & 36.10 & $-$0.60 & 0.166E-1 \nl
0.5 -- 0.8 &   89 & 15 & 76 & 36.36 & $-$0.60 & 0.126E-1 \nl
0.8 -- 1.05 &  100 & 17 & 85 & 36.60 & $-$0.60 & 0.165E-1 \nl
   &   \nl
All Galaxies & \nl
0.25 -- 0.5  & 129 & 26 & 103 & 36.18 & $-$0.60 & 0.292E-1 \nl
0.5 -- 0.8 &   155 & 26 & 133 & 36.37 & $-$0.60 & 0.211E-1 \nl
0.8 -- 1.05 &  115 & 19 & 99 & 36.60 & $-$0.60 & 0.180E-1 \nl
1.05 -- 1.50 & 22 & 12 & 12 & 36.65 & $-$0.60 & 0.360E-2 \nl
\enddata
\label{table_fixa_q}
\end{deluxetable}

%
% Table 12
%
\begin{deluxetable}{c rrrr}
\tablenum{12}
\tablewidth{0pt}
%\scriptsize
\tablecaption{Piecewise Calculation of LF Evolution from $z = 0$ to $z = 1$}
\tablehead{\colhead{Galaxy Types} & 
\colhead{$\Delta$[log($L^*(U)$)]\tablenotemark{a}} &
\colhead{$\Delta$[log($L^*(R)$)]\tablenotemark{a}} & 
\colhead{$\Delta$[log($L^*(K)$)]\tablenotemark{a}} &
\colhead{$\Delta$[log($L^*(2400\AA)$)]\tablenotemark{b}}  \nl
\colhead{ } & \colhead{(dex)} & \colhead{(dex)}  \nl
}
\startdata
$\cal{A}$ & 0.8 & 0.9 & 1.3 & ...\nl
$\cal{I}$\tablenotemark{c} & 0.3 & 1.0 & 1.3 & ... \nl
$\cal{A+I}$ & 1.1 & 1.0 & 1.5 & 0.4 \nl
$\cal{I+E}$ & 0.3 & 0.0 & 0.5 & 0.7 \nl
$\cal{E}$ & 1.1 & 1.2\tablenotemark{d} & 1.8\tablenotemark{d} & 0.9 \nl
All & 0.0 & 0.2 & 0.6 & 0.6 \nl
\enddata
\tablenotetext{a}{The regime $0.25<z<1.05$ is used for this calculation.}
\tablenotetext{b}{Only the regime $0.25<z<0.80$ is used for this calculation.}
\tablenotetext{c}{Quite uncertain due to low number of $\cal{I}$
galaxies in the $0.8 < z < 1.05$ bin.}
\tablenotetext{d}{The values listed for this entry are believed to
be spurious.   See the text for details.}
\label{table_qpiece}
\end{deluxetable}

%
% Table 13
%
\begin{deluxetable}{crrr rll}
\tablenum{13}
\tablewidth{0pt}
%\scriptsize
\tablecaption{Solution for $Q$, the LF Evolution from $z = 0$ to $z = 1$
\tablenotemark{a}}
\tablehead{\colhead{Galaxy Types} & \colhead{No. of} & 
\colhead{No.} & \colhead{No. in} & \colhead{log[$L^*(W)]$}
& \colhead{$\alpha$} & \colhead{$Q$} \nl
\colhead{ } & \colhead{Gal.} & \colhead{in HDF} & \colhead{Flanking}  & 
 (at $z=0.60$) & (fixed) \nl
\colhead{ } & \colhead{ } & \colhead{ } & \colhead{Fields} \nl
}
\startdata
Rest Fr. $U$ \nl
$\cal{A}$ & 72 & 14 & 58 & 37.10 & $-$0.60 & $+0.55\pm0.70$\tablenotemark{b} \nl
$\cal{I}$\tablenotemark{c} & 139 & 17 & 122 & 36.56 & +0.30 & $+1.45\pm0.45$ \nl 
$\cal{A+I}$ & 211 & 31 & 180 & 36.82 & $-$0.25 & $+1.20\pm0.45$ \nl
$\cal{I+E}$ & 420 & 64 & 356 & 36.78 & $-$1.05 & $+0.20\pm0.40$ \nl
$\cal{I+E}$(3p) & 420 & 64 & 356 & $36.68\pm0.30$ & $-1.15\pm0.15$ 
         & $+0.46\pm0.30$ \nl
 $\cal{E}$ & 281 & 47 & 234 & 36.60 & $-$1.30 & $+0.85\pm0.40$ \nl
All & 492 & 78 & 414 & 37.05 & $-$1.15 & $-0.15\pm0.40$ \nl
All (3p) & 492 & 78 & 414 & $36.88\pm0.25$ & $-1.20\pm0.15$ & $+0.30\pm0.25$ \nl
      & \nl
Rest Fr. $R$ \nl
$\cal{A}$ & 72 & 14 & 58 & 37.24 & $-$0.50 & $+0.62\pm0.70$ \nl
$\cal{I}$\tablenotemark{c} & 139 & 17 & 122 & 36.86 & $-$0.05 & $+1.25\pm0.50$ \nl
$\cal{A+I}$ & 211 & 31 & 180 & 37.02 & $-$0.20 & $+1.18\pm0.45$ \nl
$\cal{I+E}$ & 420 & 64 & 356 & 37.15 & $-$1.25 & $-0.02\pm0.40$ \nl
$\cal{I+E}$ (3p) & 420 & 64 & 356 & $37.16\pm0.35$ & $-1.30\pm0.15$ 
        & $+0.21\pm0.35$ \nl
$\cal{E}$ & 281 & 47 & 234 & 36.85 & $-$1.45 & $+0.90\pm0.40$ \nl
All & 492 & 78 & 414 & 37.43 & $-$1.25 & $-0.24\pm0.40$ \nl
All (3p) & 492 & 78 & 414 & $37.24\pm0.25$ & $-1.30\pm0.10$ & $+0.06\pm0.35$ \nl
      & \nl
Rest Fr. $K$ \nl
$\cal{A}$ & 71 & 14 & 57 & 37.00 & $-$0.10 & $+0.69\pm0.70$ \nl
$\cal{I}$\tablenotemark{c} & 138 & 16 & 122 & 36.70 & $-$0.20 & $+1.45\pm0.50$ \nl
$\cal{A+I}$ & 209 & 30 & 179 & 36.90 & $-$0.30 & $+1.39\pm0.50$ \nl
$\cal{I+E}$ & 387 & 58 & 329 & 36.95 & $-$1.20 & $+0.27\pm0.50$ \nl
$\cal{I+E}$ (3p)  & 387 & 58 & 329 & $36.90\pm0.40$ & $-1.30\pm0.20$ & 
       $+0.50\pm0.50$ \nl
$\cal{E}$ & 249 & 42 & 207 & 36.50 & $-$1.35 & $+1.32\pm0.50$ \nl
All & 458 & 62 & 386 & 37.20 & $-$1.15 & $-0.01\pm0.50$ \nl
All (3p) & 458 & 62 & 386 & $37.05\pm0.40$ & $-1.25\pm0.20$ &
           $+0.42\pm0.50$ \nl
      & \nl
Rest Fr. 2400\AA \nl
$\cal{I+E}$ &  371 & 66 & 311 & 36.37 & $-0.55$ & $+1.17\pm0.30$ \nl
$\cal{E}$ & 254 & 49 & 209 & 36.36 & $-$0.60 & $+1.31\pm0.30$ \nl
All & 399 & 71 & 335 & 36.37 & $-$0.60 & $+1.24\pm0.25$ \nl
\enddata
\tablenotetext{a}{The solution for $Q$ utilizes the regime
$0.25 < z < 1.05$.}
\tablenotetext{b}{These are single parameter 1$\sigma$ errors.}
\tablenotetext{c}{Quite uncertain due to low number of $\cal{I}$
galaxies in the $0.8 < z < 1.05$ bin.}
\label{table_q}
\end{deluxetable}

%
% Table 14
%
\begin{deluxetable}{c rr rr rr rr rr}
\tablenum{14}
\tablewidth{0pt}
%\scriptsize
\tablecaption{Predicted Versus Actual Galaxy Counts}
\tablehead{\colhead{Galaxy} & \multispan{2}{$0.01< z<0.25$} &
\multispan{2}{$0.25< z<0.5$}  
& \multispan{2}{$0.5<z<0.8$}
& \multispan{2}{$0.8<z<1.05$}
& \multispan{2}{$1.05<z<1.5$}  \nl
\colhead{Types} & \colhead{Obs.} & \colhead{Pred.}
& \colhead{Obs.} & \colhead{Pred.} & \colhead{Obs.} & \colhead{Pred.}
& \colhead{Obs.} & \colhead{Pred.}
& \colhead{Obs.} & \colhead{Pred.} \nl
}
\startdata
Rest Fr. $U$\tablenotemark{a} \nl
$\cal{A}$   & ... & ... & 17 & 18 & 36 & 46 & 19 & 24 & ... & ... \nl
$\cal{I}$   & ... & ... & 62 & 64 & 63 & 66 &  ... & ...  ... & ... \nl
$\cal{A+I}$ & ... & ... & 79 & 80 & 99 & 112 & 33 & 39 &  ... & ... \nl
$\cal{I+E}$ &  ... & ... & 131 & 146 & 164 & 205 & 125 & 162 &  ... & ... \nl
$\cal{E}$ & ... & ... & 69 & 78 & 101 & 131 & 111 & 143 &  ... & ... \nl
All & 37 & 45 & 148 & 145 & 200 & 250 & 144 & 193 & 
18\tablenotemark{b} & 29 \nl
 & \nl
Rest Fr. $R$\tablenotemark{a} \nl
$\cal{A}$   & ... & ... & 17 & 18 & 36 & 42 & 19 & 17 & ... & ... \nl
$\cal{I}$   & ... & ... & 62 & 59 & 63 & 62 &  ... & ...  ... & ... \nl
$\cal{A+I}$ & ... & ... & 79 & 83 & 99 & 118 & 33 & 32 &  ... & ... \nl
$\cal{I+E}$ &  ... & ... & 131 & 142 & 164 & 187 & 125 & 152 &  ... & ... \nl
$\cal{E}$ & ... & ... & 69 & 86 & 101 & 120 & 111 & 149 &  ... & ... \nl
All & 37 & 32 & 148 & 151 & 200 & 223 & 144 & 161 & 
18\tablenotemark{b} & 14 \nl
 & \nl
Rest Fr. $K$\tablenotemark{c} \nl
$\cal{A}$   & ... & ... &  15 & 17 & 36 & 37 & 19 & 19 & ... & ... \nl
$\cal{I}$   & ... & ... & 60 & 67 & 62 & 69 & 16 & 16 & ... & ... \nl
$\cal{A+I}$ & ... & ... & 75 & 84 & 98 & 107 & 35 & 35 &  ... & ... \nl
$\cal{I+E}$ &  ... & ... & 115 & 152 & 147 & 204 & 121 & 157 &  ... & ... \nl
$\cal{E}$ & ... & ... & 55 & 86 & 85 & 131 & 105 & 155 &  ... & ... \nl
All & 30 & 32 & 130 & 168 & 183 & 239 & 183 & 185 & 27 & 30 \nl
 & \nl
Rest Fr. 2400\AA\tablenotemark{c}  \nl
$\cal{I}$ & ... & ... & 56 & 59 & 51 & 57 & ... & ... & ... & ... \nl
$\cal{I+E}$ &  ... & ... & 129 & 135 & 140 & 171 & 105 & 129 & ... & ... \nl
$\cal{E}$ & ... & ... & 65 & 71 & 89 & 111 & 100 & 127 & ... & ... \nl
All & ... & ... & 129 & 138 & 155 & 184 & 115 & 123 & 22 & 28 \nl
\enddata
\tablenotetext{a}{The observed number of galaxies at rest frame
$R$ is calculated to an observed mag of $R=23.5$.}
\tablenotetext{b}{Six galaxies in the HDF with $23.5<R<24$ have been
eliminated.  See text for details.}
\tablenotetext{c}{The observed number of galaxies at rest frame
$K$ is calculated to an observed mag of $K=20.75$.}
\tablenotetext{d}{The observed number of galaxies at rest frame 2400\AA\
is calculated to an observed mag of $U=24$.0} 
\label{table_num}
\end{deluxetable}

%
% Table 15
%
\begin{deluxetable}{c rrr rr}
\tablenum{15}
\tablewidth{0pt}
%\scriptsize
\tablecaption{Mean Predicted Rest Frame Luminosity 
of Observed Sample\tablenotemark{a}}
\tablehead{\colhead{Galaxy Types} & \colhead{$z<0.25$} &
\colhead{$0.25< z<0.5$}  & \colhead{$0.5<z<0.8$} 
& \colhead{$0.8<z<1.05$} & \colhead{$1.05<z<1.5$}  \nl
\colhead{} & \colhead{$L(R)$ (10$^{36}$ W)} & 
\colhead{log[$L$ (10$^{36}$ W)]} &  \colhead{log[$L$ (10$^{36}$ W)]} &
\colhead{log[$L$ (10$^{36}$ W)]} & \colhead{log[$L$ (10$^{36}$ W)]} \nl
}
\startdata
Rest Fr. $U$ \nl
$\cal{A}$ & ...  & 4.8 & 8.2 & 16.1 & ... \nl
$\cal{I}$  & ...  & 4.2 & 6.0 &  ... & ... \nl
$\cal{A+I}$ & ... & 4.2 & 6.8 & 14.9 & ... \nl
$\cal{I+E}$ & ... & 2.4 & 3.3 & 6.8 & ... \nl
$\cal{E}$ & ...  &  1.3 & 2.3 & 6.5 & ... \nl
~ \nl
All (pred) & 2.0   & 3.6 & 4.2 & 7.5 & 20.9 \nl
All (obs) & 1.0 & 2.1 & 3.9 & 7.2 & 20.0 \nl
~ \nl
Rest Fr. $R$ \nl
$\cal{A}$ & ...  & 6.5 & 14.5 & 29.0 & ... \nl
$\cal{I}$  & ...  & 5.2 & 12.0 &  ... & ... \nl
$\cal{A+I}$ & ... & 6.0 & 12.0 & 25.5 & ... \nl
$\cal{I+E}$ & ... & 4.4 & 5.4 & 11.9 & ... \nl
$\cal{E}$ & ...  &  1.6 & 2.9 & 10.6 & ... \nl
~ \nl
All (pred) & 1.0   & 4.3 & 7.7 & 14.2 & 26.9 \nl
All (obs) & 1.3 & 3.7 & 7.1 & 11.6 & 17.8 \nl
  & \nl
Rest Fr. $K$ \nl
$\cal{A}$ & ...  &  3.6 & 11.3 & 16.5 & ... \nl
$\cal{I}$  & ...  & 3.0 & 5.4 & 13.5 & ... \nl
$\cal{A+I}$ & ... & 3.1 & 7.5 & 16.9  & ... \nl
$\cal{I+E}$ & ... & 2.1 & 2.8 & 3.7 & ... \nl
$\cal{E}$ & ...  & 0.8 & 1.4 & 4.4 & ... \nl
~ \nl
All (pred) & 1.4 & 2.5 & 4.3 & 6.7 & 10.0 \nl
All (obs) &  0.7 & 2.1 & 3.6 &  5.5 &  6.7  \nl
  & \nl
Rest Fr. 2400\AA \nl
$\cal{I}$  & ...  & 1.7 & 3.0 & ... & ... \nl
$\cal{A+I}$ & ... & 1.6 & 2.8 & 9.0  & ... \nl
$\cal{I+E}$ & ... & 1.4 & 2.7 & 5.5 & ... \nl
$\cal{E}$ & ...  & 1.3 & 2.5 & 5.2 & ... \nl
~ \nl
All (pred) & ... & 1.4 & 2.6 & 5.6 & 9.5 \nl
All (obs) &  ... & 1.0  & 2.3 &  4.7 &  7.3  \nl
\enddata
\tablenotetext{a}{This is the predicted mean luminosity of a sample
observed to our limiting magnitude, calculated 
to a cutoff of $R$(observed) = 23.5
and $K$(observed) = 20.75.  See Table ~\ref{table_lumdens} for 
the luminosity density.}
\label{table_lmean}
\end{deluxetable}

%
% Table 16
%
\begin{deluxetable}{c rrr rr}
\tablenum{16}
\tablewidth{0pt}
%\scriptsize
\tablecaption{Luminosity Density at Rest-Frame 2400\AA, $U$, $R$ and $K$ 
(10$L^*$ to $L^*/20$)}
\tablehead{\colhead{Galaxy Types} & \colhead{} & \colhead{} & \colhead{$z$ Range} \nl
\colhead{} & \colhead{$0.01-0.25$} &
\colhead{$0.25-0.5$}  & \colhead{$0.5-0.8$} 
& \colhead{$0.8-1.05$} & \colhead{$1.05-1.5$}  \nl
\colhead{} & \colhead{$\rho$} & \colhead{$\rho$} & \colhead{$\rho$} &
\colhead{$\rho$} & \colhead{$\rho$} \nl
\colhead{} & \colhead{} & \colhead{} & \colhead{(10$^{34}$ W Mpc$^{-3}$)} \nl
}
\startdata
Rest Fr. $U$ \nl
$\cal{A}$   &  ... & 0.9 & 1.7  & 1.6  & ... \nl
$\cal{I}$   & ... & 2.9 & 1.8 &  ... & ... \nl
$\cal{A+I}$ & ... & 3.6 & 3.5 & 2.4 & ... \nl
$\cal{I+E}$ &  ... & 3.8 & 3.3 & 5.5 & ... \nl
$\cal{E}$ & ... & 1.2 & 1.6 & 4.9  & ... \nl
All & $\ge$4.2 & 3.8  & 4.9 & 6.9 & 1.0\tablenotemark{a} \nl
 & \nl
Rest Fr. $R$ \nl
$\cal{A}$   &  ... & 1.2 ($\pm48$\%) & 2.9 ($\pm29$\%) & 
        2.2 ($\pm28$\%) & ... \nl
$\cal{I}$   & ... & 3.5 ($\pm26$\%) & 2.9 ($\pm29$\%) &  ... & ... \nl
$\cal{A+I}$ & ... & 5.2 ($\pm22$\%) & 6.6 ($\pm18$\%) & 
           3.5 ($\pm35$\%) & ... \nl
$\cal{I+E}$ &  ... & 7.2 ($\pm19$\%) & 5.1 ($\pm18$\%) & 
           9.3 ($\pm12$\%) & ... \nl
$\cal{E}$ & ... & 2.0 ($\pm29$\%) & 2.3 ($\pm$39\%) & 
           8.1 ($\pm10$\%) & ... \nl
All & $\ge$1.9 ($\pm55$\%) &   7.7 ($\pm17$\%) & 8.6 ($\pm14$\%) 
      & 11.5 ($\pm16$\%) & 0.9\tablenotemark{a} \nl
 & \nl
Rest Fr. $K$ \nl
$\cal{A}$   &  ... & 0.7 ($\pm66$\%) & 2.0 ($\pm36$\%)& 1.4 ($\pm42$\%) & ... \nl
$\cal{I}$   & ... & 2.2 ($\pm27$\%) & 1.7 ($\pm24$\%) & 0.9 ($\pm98$\%) & ... \nl
$\cal{A+I}$ & ... & 2.8 ($\pm23$\%) & 3.7 ($\pm18$\%) & 2.4 ($\pm36$\%) & ... \nl
$\cal{I+E}$ & ... &  3.3 ($\pm30$\%) & 2.6 ($\pm19$\%) & 3.7 ($\pm14$\%) & ... \nl
$\cal{E}$ & ... & 0.7($\pm43$\%)  & 0.9 ($\pm15$\%) & 2.8 ($\pm17$\%) & ... \nl
All & $\ge$2.2 ($\pm66$\%) & 4.4 ($\pm27$\%) & 4.6 ($\pm17$\%) & 
            5.0 ($\pm15$\%) & 0.5\tablenotemark{a} \nl
 & \nl
Rest Fr. 2400\AA \nl
$\cal{A}$   &  ... & ... & 0.2 & ... \nl
$\cal{I}$   & ... & 1.1 & 0.9 & ... & ... \nl
$\cal{A+I}$ & ... & 1.2 & 1.0 & 0.4 & ... \nl
$\cal{I+E}$ & ... & 2.1 & 2.3 & 3.5 & ... \nl
$\cal{E}$ & ... & 1.0 & 1.4 & 3.2 & ... \nl
All & ... & 2.2 & 2.4 & 3.5 & 0.8\tablenotemark{a} \nl
\enddata
\tablenotetext{a}{The values for the bin $1.05<z<1.5$ are suspected to
be underestimated by substantial factors whose approximate
values are given in the text.  See the text for details.}
\label{table_lumdens}
\end{deluxetable}

%
% Table 17
%
\begin{deluxetable}{c rrr }
\tablenum{17}
\tablewidth{0pt}
%\scriptsize
\tablecaption{Luminosity Density and Stellar Mass at Rest-Frame $K$ Extrapolated
to $z=0.0$}
\tablehead{\colhead{Galaxy Types} & \colhead{}  & \colhead{$z$ Range} \nl
\colhead{} & 
\colhead{$0.25-0.5$}  & \colhead{$0.5-0.8$} 
& \colhead{$0.8-1.05$}   \nl
\colhead{} &  \colhead{$\rho$\tablenotemark{a}} 
& \colhead{$\rho$\tablenotemark{a}} & \colhead{$\rho$\tablenotemark{a}}  \nl
\colhead{} & \colhead{} &  
\colhead{(10$^{34}$ W Mpc$^{-3}$)\tablenotemark{b}} \nl 
}
\startdata
$\cal{A}$   &  0.49 ($\pm65$\%) & 1.19 ($\pm39$\%) & 0.65 ($\pm44$\%) \nl
$\cal{I}$   & 1.31 ($\pm31$\%) & 0.93 ($\pm28$\%) & 0.38 ($\pm98$\%) \nl
$\cal{I+E}$ &  2.31 ($\pm36$\%) & 1.58 ($\pm22$\%) & 1.93 ($\pm20$\%) \nl
$\cal{E}$ & 0.56 ($\pm24$\%) & 0.64 ($\pm21$\%) & 1.82 ($\pm20$\%) \nl
 & \nl
Total ($\cal{A + (I+E)}$) & 2.80 ($\pm32$\%) & 2.77 ($\pm20$\%) & 2.58 ($\pm19$\%) \nl
\enddata
\tablenotetext{a}{Integrations go from 10$L^*(z)$ to $L^*(z)/20$.}
\tablenotetext{b}{Note that $1.0 \times 10^{34}$ W Mpc$^{-3} \equiv
1.3 \times 10^8 L$\subsun Mpc$^{-3} \equiv 
1.6 \times 10^8 M$\subsun Mpc$^{-3}$.}
\label{table_ml}
\end{deluxetable}

%
%   Table 18
%
\begin{deluxetable}{lc rrrrr }
\tablenum{18}
\tablewidth{0pt}
%\scriptsize
\tablecaption{Comparison With LF Parameters of Other Surveys}
\tablehead{\colhead{Reference} & \colhead{Galaxy Types} 
& \colhead{log[$L^*$ (W)]} &  \colhead{$\alpha$} &
\colhead{log[$L^*$ (W)]} & \colhead{$\alpha$} & \colhead{log[$L^*$ (W)]} \nl
\colhead{} & \colhead{} & \colhead{} & \colhead{} & 
\colhead{} & \colhead{(CFGRS)} 
& \colhead{(Adj.$\alpha$)\tablenotemark{a}}\nl
}
\startdata
Rest Fr. $R$ & Local \nl
LCRS & All & $36.76\pm0.02$ & $-0.70\pm0.03$ &  37.20 & $-1.30$
  &  36.65 \nl
        & Em. & $36.66\pm0.04$ & $-0.90\pm0.10$ &  36.71 & $-1.45$ & 36.15 \nl
        & No em. & $36.74\pm0.04$ & $-0.27\pm0.10$ & 36.92 & $-0.50$ & 36.60 \nl
SDSS & All &  $37.01\pm0.02$  &  $-1.20\pm0.03$ & 37.20 & $-1.30$ & 37.09 \nl
2dF & All & $37.00\pm0.02$ & $-1.09\pm0.09$ & 37.20 & $-1.30$ & 36.96 \nl
    & Em. & $36.70\pm0.04$ & $-1.70\pm0.10$ &  36.71 & $-1.45$ & 37.07 \nl
    & No. em. & $36.95\pm0.05$ & $-0.70\pm0.10$ & 36.92 & $-0.50$ & 37.13 \nl
 & \nl 
Rest Fr. $K$ & Local \nl
Cole 2001 & All & $36.89\pm0.02$ & $-0.96\pm0.05$ & 36.80 & $-$1.25 & 36.60 \nl
Kochanek 2001 & Early &  $36.92\pm0.03$ & $-0.90\pm0.10$ &  36.30 & $-0.10$  &  
   37.00\tablenotemark{c} \nl
          & Late & $36.70\pm0.03$ &  $-0.90\pm0.10$  &  36.10 & $-$1.35 &  35.60  \nl
 ~ \nl
Int. $z$ \nl
CNOC2 & $<z>=0.3$  \nl
Rest Fr. $R$    & Early & $36.84\pm0.05$ & $-0.06\pm0.14$ & 36.86 & $-0.50$ 
                  & 36.56 \nl
  & Late & $36.68\pm0.06$ & $-1.34\pm0.12$ & 36.65 & $-1.45$ & 36.52 \nl

Rest Fr. $U$     & Early & $35.93\pm0.05$ & $0.14\pm0.15$ & 36.23 & $-0.35$ &
                     36.03 \nl
  & Late &  $36.25\pm0.05$ & $-1.14\pm0.15$ & 36.66 & $-1.60$ & 36.26 \nl
~ \nl
CFRS & $<z>=0.56$ \nl
Rest Fr. $B$ & All & 36.62 & $-0.50$ & 37.15 & $-1.15$ & 
      36.35\tablenotemark{c} \nl
             & Red & 36.69 & $-0.37$ & 37.14 & $-0.50$ & 37.01 \nl
             & Blue & 36.82 & $-1.07$ & 36.79 & $-1.05$ & 36.79 \nl
High $z$ \nl
Steidel 1999 & $z\sim3$ \nl
Rest Fr. 1500\AA & LBG & $37.59 \pm0.06$ & $-1.60\pm0.13$ & 36.75\tablenotemark{b} 
         & $-0.60$ & 37.15\tablenotemark{c} \nl
\enddata
\tablenotetext{a}{Here the value of $L^*$ is adjusted so that the
value of $\alpha$(CFGRS) matches that of the comparison survey.}
\tablenotetext{b}{$L^*$ at rest frame 2400\AA\ for galaxies at $z\sim1.3$.}
\tablenotetext{c}{A very large extrapolation in $\alpha$ is required. The
resulting $L^*$ is quite uncertain.} 
\label{table_lfcomp}
\end{deluxetable}
% see file /scr2/jlc/hdf_lf/zsurvey_comparison_notes

%
% Table 19
%
\begin{deluxetable}{lll r rrrr}
\tablenum{19}
\tablewidth{0pt}
%\scriptsize
\tablecaption{SED Parameters for Additional Galaxies in the Flanking Fields of the HDF}
\tablehead{\colhead{RA} &  \colhead{Dec\tablenotemark{a}} &
\colhead{} &
\colhead{$z$\tablenotemark{a}} &
\colhead{log(L($\lambda_m$:{\rm{blue}})} & \colhead{$\alpha_{UV}$} &
\colhead{log(L($\lambda_m$:{\rm{red}})} & \colhead{$T$}  \nl
\colhead{ ($-12$h)} & 
\colhead{($-62^{\circ}$)} & \colhead{} &  \colhead{} &
\colhead{(W)\tablenotemark{b}} &   \colhead{(sBB)} & \colhead{(W)} & \colhead{(K)} \nl
}
\startdata
 36~~22.07 & 14 & 59.7 & 0.849 & 36.29 & 0.89 & 36.37 & 9155 \nl
 36~~24.16 & 15 & 14.4 & 0.796 & 36.39 & 1.72 & 36.42 & 15470 \nl
 36~~25.52 & 15 & 10.7 & 0.680 & 36.04 & 2.21 & 36.20 & 5305 \nl
 36~~33.99 & 16 & 04.6 & 0.834 & 36.58 & 1.08 & 36.64 & 6440 \nl
 36~~39.82 & 16 & 01.3 & 0.843 & 36.61 & 10.00 & 36.57 & 4825 \nl
 36~~45.24 & 11 & 08.7 & 0.512 & 35.59 & $-$0.27 & 35.73 & 8300 \nl 
\enddata
\tablenotetext{a}{From Dawson \etal\ (2001).}
\label{table_dawson}
\end{deluxetable}

%
% Table 20
%
\begin{deluxetable}{crclcr}
\tablenum{20}
\tablewidth{0pt}
%\scriptsize
\tablecaption{Rest Frame $R$ LF Solution For 
Fixed $L^*$ For Each Galaxy Spectral Grouping}
\tablehead{\colhead{$z$ Range} & \colhead{Total No.\tablenotemark{a}} &  
\colhead{log[$L^*(R)$]} & \colhead{$\alpha$} &
\colhead{$\phi^*$} & \colhead{$\rho(L(R))$} \nl
\colhead{} & \colhead{} &  
\colhead{(log(W)} & \colhead{} & \colhead{($Mpc^{-3}$)} &
\colhead{(10$^{34}$ W Mpc$^{-3}$)} \nl
\colhead{} & \colhead{} & \colhead{(fixed)} \nl
}
\startdata
 $\cal{A}$ Galaxies \nl
0.25 -- 0.5  &  17 & 37.24 & $-$0.78 & 0.0018 & 1.4 \nl
0.5 -- 0.8 &    36 & 37.24 & $-$0.50 & 0.0033 & 2.8 \nl
0.8 -- 1.05 &   19 & 37.24 & $-$0.08 & 0.0022 & 2.0 \nl
   & \nl
$\cal{I}$ Galaxies \nl
0.25 -- 0.5  &  62 &  36.86 & $-0.37$ & 0.116~ & 4.3 \nl
0.5 -- 0.8 &    63 &  36.86 & $-$0.05 & 0.0066 & 3.1 \nl
0.8 -- 1.05 & 14 &  ... & ... & ... & ... \nl
   &  \nl
$\cal{A}$+$\cal{I}$ Galaxies \nl
0.25 -- 0.5  &  79 &  37.02 & $-$0.55 & 0.012~ & 6.0 \nl
0.5 -- 0.8 &    99 &  37.02 & $-$0.20 & 0.011~ & 5.8 \nl
0.8 -- 1.05 &   33 &  37.02 & +0.63 & 0.0032 & 3.3 \nl
  &  \nl
$\cal{I}$+$\cal{E}$ Galaxies \nl
0.25 -- 0.5  & 131  & 37.15 & $-$1.27 & 0.0073 & 5.7 \nl
0.5 -- 0.8 &   164  & 37.15 & $-$1.25 & 0.0066 & 5.1 \nl
0.8 -- 1.05 &  125  & 37.15 & $-$1.35 & 0.0108 & 8.9 \nl
  &  \nl
$\cal{E}$ Galaxies \nl
0.25 -- 0.5  &  69 & 36.85 & $-$1.55 &  0.0040 & 1.9 \nl
0.5 -- 0.8 &   101 & 36.85 & $-$1.45 & 0.0052 & 2.3 \nl
0.8 -- 1.05 &  111 & 36.85 & $-0.82$ & 0.02~1 & 7.3 \nl
   &   \nl
All Galaxies & \nl
0.25 -- 0.5  & 148 & 37.43 & $-$1.33 & 0.0051 & 7.9 \nl
0.5 -- 0.8 &   200 & 37.43 & $-$1.25 & 0.0049  & 7.6 \nl
0.8 -- 1.05 &  144 & 37.43 & $-$1.45 & 0.0066 & 11.1 \nl
\tablenotetext{a}{The numbers of galaxies in the HDF and in the Flanking 
Fields respectively can be found from Table~\ref{table_fixa}.}
\enddata
\label{table_fixlstar}
\end{deluxetable}

\clearpage

% figure 1

% figure 2

% figure 3

% figure 4

% figure 5

% figure 6

% figure 7

% figure 8

% figure 9

% figure 10

% figure 11

% figure 12

% figure 13

\end{document}